\documentclass{LMCS}

\usepackage{url,cite,amssymb,stmaryrd,alltt,xypic,appendix,enumerate,hyperref}
\usepackage{color,alltt}

\definecolor{shade}{gray}{0.75}

\newcommand{\shadethis}[1]{\colorbox{shade}
{\protect\rule[-0.95mm]{0mm}{3.95mm}\hspace{0.3mm}#1\hspace{0.3mm}}}

\input{danvy-millikin-LMCS.mac}
\renewcommand{\synvarval}{\ensuremath{v}}

\def\doi{4 (4:12) 2008}
\lmcsheading%
{\doi}
{1--67}
{}
{}
{Dec.~16, 2006}
{Nov.~29, 2008}
{}   

\begin{document}

\title{
	A Rational Deconstruction \\
	of Landin's SECD Machine with the J Operator
}

\author[O.~Danvy]{Olivier Danvy\rsuper a}
\address{\lsuper a\protect\begin{tabular}[t]{@{}l}
         Department of Computer Science, Aarhus University
         \protect\\
         IT-parken, Aabogade 34, DK-8200 Aarhus N, Denmark
         \protect\end{tabular}}
\email{danvy@brics.dk}

\author[K.~Millikin]{Kevin Millikin\rsuper b}
\address{\lsuper b\protect\begin{tabular}[t]{@{}l}
         Google Inc.
         \protect\\
         Aabogade 15, DK-8200 Aarhus N, Denmark
         \protect\end{tabular}}
\email{kmillikin@google.com}


\begin{abstract}
Landin's SECD machine was the first abstract machine for applicative
expressions, \ie, functional programs.
Landin's J operator was the first control operator for functional
languages, and was specified by an extension of the SECD machine.
We present a family of
evaluation functions corresponding to this
extension of the SECD machine, using a series of
elementary transformations (transformation into continu\-ation-passing
style (CPS) and defunctionalization, chiefly) and their left inverses
(transformation into direct style and refunctionalization).  To this end,
we modernize the SECD machine into a bisimilar one that operates in
lockstep with the original one but that (1) does not use a data stack and (2)
uses the caller-save rather than the callee-save convention for
environments.  We also identify that the dump component of the SECD
machine is managed in a callee-save way.  The caller-save counterpart of
the modernized SECD machine precisely corresponds to Thielecke's
double-barrelled continuations and to Felleisen's encoding of J in
terms of call/cc.  We then variously characterize the J operator in terms
of CPS and in terms of
delimited-control operators in the CPS hierarchy.

As a byproduct, we also
present several reduction semantics for applicative expressions
with the J operator, based on Curien's original calculus of explicit
substitutions.  These reduction semantics mechanically correspond to the
modernized versions of the SECD machine and to the best of our knowledge,
they provide the first syntactic theories of applicative expressions with the J
operator.

The present work is concluded by a motivated wish to see Landin's name
added to the list of co-discoverers of continuations.
Methodologically, however, it mainly
illustrates the value of Reynolds's defunctionalization and of
refunctionalization as well as the expressive power of the CPS hierarchy
(1) to account for the first control operator and the first abstract machine
for functional languages and (2) to connect them to their successors.
Our work also illustrates the value of Danvy and Nielsen's refocusing technique
to connect environment-based abstract machines and syntactic theories in
the form of reduction semantics for calculi of explicit substitutions.
\end{abstract}

\maketitle

% \vfill

\clearpage

\section{Introduction}
\label{sec:intro}

\vspace{-0.5mm}

Forty years ago, Peter Landin
unveiled the first control operator, J, to a heretofore unsuspecting
world~\cite{Landin:TR65-Generalization, Landin:CACM65, Landin:CACM66}.
He did so to generalize the notion of jumps and labels for translating
Algol 60 programs into applicative expressions, using the J operator to
account for the meaning of an Algol label.  For a simple example,
consider the block

\[
  \mathbf{begin}\; s_1 \;;\; \mathbf{goto}\: L \;;\; L:\: s_2 \;\mathbf{end}
\]

\noindent
where the sequencing between the statements (`basic blocks,' in compiler
parlance~\cite{Aho-Sethi-Ullman:86}) $s_1$ and $s_2$ has been made
explicit with a label and a jump to this label.  This block
is translated into the applicative expression

\[
  \synlam{()}
         {\synhlet{L}
                  {\synapp{\rawJ}{s_2'}}
                  {\synhlet{()}
                           {\synapp{s_1'}{()}}
                           {\synapp{L}{()}}}}
\]

\noindent
where $s_1'$ and $s_2'$ respectively denote the translation of $s_1$ and
$s_2$.
The occurrence of J captures the continuation of the outer let expression and
yields a `program closure' that is bound to $L$.
Then, $s_1'$ is applied to $()$.
If this application completes, the program closure bound to $L$ is applied:
(1) $s_2'$ is applied to $()$ and then,
if this application completes, (2)
the captured continuation is resumed, thereby completing the execution of
the block.

Landin also showed that the notion of program closure makes sense not
just in an imperative setting, but also in a functional one.  He
specified the J operator by extending the SECD machine~\cite{Landin:CJ64,
Landin:Fox66}.

\subsection{The SECD machine}
\label{subsec:the-SECD-machine}

Over the years, the SECD machine has been the topic of considerable
study: it provides an unwavering support for operational
semantics~\cite{Abramsky:TCS92, Banerjee:PhD, Bierman:TR97,
Birtwistle-Graham:90, Chung:MS, Clark:PhD, Felleisen-Friedman:FDPC3,
Hannan-Miller:MSCS92-one, Hardin-al:JFP98, Jones:ICALP81, Neergaard:PhD,
Plotkin:TCS75, Vasconcelos:JFP05}, compilation~\cite{Ager-al:Jones02,
Bayer:Scheme01, Burge:75, Cardelli:Polymorphism83, Georgeff:TOPLAS84,
Gomard-Sestoft:PEPM91, Graham:92, Nielson-Nielson:TOPLAS86, Paulson:PhD,
Pleban:PLDI84, Sestoft:PhD}, and parallelism~\cite{Abramsky-Sykes:FPCA85,
Blelloch-Greiner:FPCA95}, and it lends itself readily to
variations~\cite{Davie:92, Davie-McNally:TENCON89, Felleisen:CL87,
Hannan:PEPM91, Ramsdell:JAR99, Runciman-Toyn:NGC86, Thielecke:HOSC02} and
generalizations~\cite{McGowan:STOC70, Meijer:TR88, Xi:PLILP97}.  In
short, it is standard textbook material~\cite{Field-Harrison:88,
Glaser-al:84, Henderson:80, Henson:87, Kluge:05, Spivey:LN-SECD}, even though its
architecture is generally agreed to be on the `baroque' side, since most
subsequent abstract machines have no data stack and only one control
stack instead of two.  Nobody, however, seems to question its existence
as a distinct artifact (\ie, man-made construct) mediating between
applicative expressions (\ie, functional programs) and traditional
sequential imperative implementations.

Indeed abstract machines provide
a natural meeting ground for theoretically-minded and
experimentally-minded computer scientists: they are as close to an actual
implementation as most theoreticians will ever get, and to an actual
formalization as most experimentalists will ever go.  For example,
Plotkin~\cite{Plotkin:TCS75} proved the correctness of the SECD machine
in reference to a definitional interpreter due to
Morris~\cite{FLMorris:LaSC93-one} and a variety of implementations take
the SECD machine as their starting point~\cite{Bayer:Scheme01,
Blelloch-Greiner:FPCA95, Cardelli:Polymorphism83, McGowan:STOC70}.

\subsection{The authors' thesis}

Is there, however, such a gap between applicative expressions and
abstract machines?  The thrust of Steele's MSc thesis~\cite{Steele:MS}
was that after CPS transformation,\footnote{%
`CPS' stands for `Continuation-Passing Style;'
this term is due to Steele~\cite{Steele:MS}.  In a CPS program, all
calls are tail calls and functions thread a functional accumulator, the
continuation, that represents `the rest of the
computation'~\cite{Strachey-Wadsworth:TR74}.  CPS programs are either
written directly or the result of a CPS
transformation~\cite{Danvy-Filinski:MSCS92-one, Plotkin:TCS75}.
(See Appendix~\ref{subsec:fib-cps}.)
The left inverse of the CPS transformation is the direct-style
transformation~\cite{Danvy:SCP94, Danvy-Lawall:LFP92}.
} a $\lambda$-abstraction
can be seen as a label and a tail call as a machine jump with the machine
registers holding the actual parameters.  Furthermore the point of
Reynolds's defunctionalization~\cite{Reynolds:72} is that higher-order
programs can be given an equivalent first-order
representation.\footnote{%
In the early 1970's~\cite{Reynolds:72},
John Reynolds introduced defunctionalization as a variation of
Landin's `function closures'~\cite{Landin:CJ64}, where a term is paired
together with its environment.  In a defunctionalized program, what is
paired with an environment is not a term, but a tag that determines this
term uniquely.  In ML, the tagged environments are grouped into data
types, and auxiliary apply functions dispatch over the tags.
(See Appendix~\ref{subsec:fib-cps-defunct}.)
The left inverse of defunctionalization is
`refunctionalization'~\cite{Danvy-Nielsen:PPDP01, Danvy-Millikin:SCP0?}.
}

It nevertheless took 40 years for the SECD machine to be `rationally
deconstructed' into a compositional evaluation
function~\cite{Danvy:IFL04}, the point being that

\begin{enumerate}[(1)]

\item
the SECD machine is essentially in defunctionalized form, and

\item
its refunctionalized counterpart is an evaluation function in CPS, which
turns out to be compositional.

\end{enumerate}

\noindent
This deconstruction laid the ground for a functional correspondence
between evaluators and abstract machines~\cite{Ager:PhD, Ager-al:PPDP03,
Ager-al:IPL04, Ager-al:TCS05, Biernacka-al:LMCS05, Biernacki:PhD,
Danvy:DSc, Danvy:ICFP08, Johannsen:MS, Midtgaard:PhD, Millikin:PhD, Munk:MS}.

It is therefore the authors' thesis~\cite{Danvy:DSc, Millikin:PhD} that
the gap between abstract machines and applicative expressions is bridged
by Reynolds's defunctionalization.

Our goal here is to show that the functional correspondence between
evaluators and abstract machines also applies to the SECD machine with
the J operator, which, as we show, also can be deconstructed into a compositional
evaluation function.  As a corollary, we present several new simulations of
the J operator, and the first syntactic theories for applicative
expressions with the J operator.

\subsection{Deconstruction of the SECD machine with the J operator}

Let us outline our deconstruction of the SECD machine before
substantiating it
in the next sections.  We
follow the order of the first deconstruction~\cite{Danvy:IFL04}, though with
a twist: for simplicity and without loss of generality,
in the middle of the derivation, we first abandon the
stack-threading, callee-save features of the SECD machine,
which are non-standard,
for the more
familiar---and therefore less `baroque'---stackless,
caller-save features of traditional
definitional interpreters~\cite{Friedman-Wand:08,
FLMorris:LaSC93-one, Reynolds:72, Steele-Sussman:TR78-AoI}.
(These concepts are reviewed in the appendices.  The point here is that the
SECD machine manages the environment in a callee-save fashion.)
We then identify that the dump too is managed in a callee-save fashion and we
present the corresponding caller-save counterpart.

The SECD machine is defined as the
iteration of
a state-transition function operating over a
quadruple---a data stack (of type \stt{S}) containing intermediate
values, an environment (of type \stt{E}), a control stack (of type
\stt{C}), and a dump (of type \stt{D}) and yielding a value (of type
\stt{value}):

%

% \begin{quote}
\begin{verbatim}
  run : S * E * C * D -> value
\end{verbatim}
% \end{quote}

%

\noindent
The first deconstruction~\cite{Danvy:IFL04}
showed that together the C and D components
represent the current continuation and that the D component represents
the continuation of the current caller, if there is one.
As already pointed out in Section~\ref{subsec:the-SECD-machine},
since Landin's
work,
the C and D components of his abstract machine have been unified into one
component;
reflecting this unification,
control operators capture both what used to be C and D instead of only what
used to be D.

\subsubsection{Disentangling and refunctionalization
               (Section~\protect\ref{sec:deconstruction-1})}
\label{subsubsec:d-and-r}

The above definition of
\stt{run} looks complicated
because it has
several induction variables, \ie, it dispatches over several components of
the quadruple.  Our deconstruction proceeds as follows:
\begin{enumerate}[$\bullet$]

\item
We disentangle
\stt{run} into four mutually recursive transition functions,
each of which has one induction variable, \ie,
dispatches over one
component of the quadruple (boxed in the signature below):

%

%

%\begin{quote}
\begin{small}
\begin{alltt}
  run_c :                 S * E * \ourframebox{C} * D -> value
  run_d :                     value * \ourframebox{D} -> value
  run_t :          \ourframebox{term} * S * E * C * D -> value
  run_a : \ourframebox{value} * value * S * E * C * D -> value
\end{alltt}
\end{small}
%\end{quote}

%

\noindent
The first function, \stt{run\_c}, dispatches towards
\stt{run\_d} if the control stack is empty, \stt{run\_t} if the top of
the control stack contains a term, and \stt{run\_a} if the top of the
control stack contains an apply directive.
This disentangled specification, as it were, is
in defunctionalized form~\cite{Danvy-Millikin:SCP0?, Danvy-Nielsen:PPDP01,
Reynolds:72}:
the control stack
and the dump are defunctionalized data types, and \stt{run\_c} and
\stt{run\_d} are the corresponding apply functions.

\item
Refunctionalization eliminates the two apply functions:

%

%\begin{quote}
\begin{small}
\begin{alltt}
  run_t :          term * S * E * C * D -> value
  run_a : value * value * S * E * C * D -> value
  where C = \shadethis{S * E * D -> value} and D = \shadethis{value -> value}
\end{alltt}
\end{small}
%\end{quote}

%

\noindent
\stt{C} and \stt{D} are now function types.
As identified in
the first rational deconstruction~\cite{Danvy:IFL04},
the resulting
program is a continuation-pass\-ing interpreter.  This interpreter
threads a data stack to hold intermediate results and uses a callee-save
convention for environments to process subterms.
(For information and comparison,
Appendix~\ref{app:caller-save-stackless-evaluator} illustrates an 
interpreter with no data stack for intermediate results and a caller-save
convention for environments,
Appendix~\ref{app:callee-save-stackless-evaluator} illustrates an 
interpreter with no data stack for intermediate results and a callee-save
convention for environments, and
Appendix~\ref{app:caller-save-stack-threading-evaluator} illustrates an 
interpreter with a data stack for intermediate results and a caller-save
convention for environments.) 

\end{enumerate}

\noindent
At this point, we could continue as in the first
deconstruction~\cite{Danvy:IFL04} and exhibit the direct-style
counterpart of this interpreter.  The result, however, would be less
simple and less telling than first making do without the data stack
(Section~\ref{subsubsec:eliminating-the-data-stack}) and
second adopting the more familiar caller-save convention for
environments
(Section~\ref{subsubsec:from-callee-save-to-caller-save-environments})
before continuing the deconstruction towards a compositional
interpreter in direct style
(Section~\ref{subsubsec:continuing-the-deconstruction}).

\subsubsection{A first modernization: eliminating the data stack
               (Section~\protect\ref{sec:deconstruction-2})}
\label{subsubsec:eliminating-the-data-stack}

In order to focus on the nature of the J operator, we first eliminate the
data stack:

%

% \begin{quote}
\begin{small}
\begin{alltt}
  run_t :          term * E * C * D -> value
  run_a : value * value * E * C * D -> value
  where C = value * E * D -> value and D = value -> value
\end{alltt}
\end{small}
% \end{quote}

%

\noindent
(Two simpler interpreters are presented and contrasted in
Appendices~\ref{app:caller-save-stackless-evaluator}
and~\ref{app:caller-save-stack-threading-evaluator}.  The first, in
Appendix~\ref{app:caller-save-stackless-evaluator}, has no data stack for
intermediate results, and the second, in
Appendix~\ref{app:caller-save-stack-threading-evaluator}, has one.)

\subsubsection{A second modernization:
               from callee-save to caller-save environments
               (Section~\protect\ref{sec:deconstruction-2})}
\label{subsubsec:from-callee-save-to-caller-save-environments}

Again, in order to focus on the nature of the J operator, we
adopt the more familiar caller-save
convention for environments.  In passing, we also
rename \stt{run\_t} as \stt{eval} and \stt{run\_a}
as \stt{apply}:

%

% \begin{quote}
\begin{verbatim}
   eval :      term * E * C * D -> value
  apply : value * value * C * D -> value
  where C = value * D -> value and D = value -> value
\end{verbatim}
% \end{quote}

%

\noindent
(Two simpler interpreters are presented and contrasted in
Appendices~\ref{app:caller-save-stackless-evaluator}
and~\ref{app:callee-save-stackless-evaluator}.  The first, in
Appendix~\ref{app:caller-save-stackless-evaluator}, uses a caller-save
convention for environments, and the second, in
Appendix~\ref{app:callee-save-stackless-evaluator}, uses a callee-save
convention.)

\subsubsection{Continuing the deconstruction:
               towards a compositional interpreter in direct style}
\label{subsubsec:continuing-the-deconstruction}

\begin{enumerate}[$\bullet$]

\item
A direct-style transformation eliminates the dump continuation:

%

% \begin{quote}
\begin{verbatim}
   eval :      term * E * C -> value
  apply : value * value * C -> value
  where C = value -> value
\end{verbatim}
% \end{quote}

%

\noindent
The clause
for the J operator and the main evaluation function
are expressed using the delimited-control operators shift and
reset~\cite{Danvy-Filinski:LFP90}.\footnote{%
\label{foo:CPS-hierarchy}
Delimited continuations represent part of
the rest of the computation: the control operator reset delimits
control and the control operator shift captures the current delimited
continuation~\cite{Danvy-Filinski:LFP90}.
These two control operators provide a direct-style handle
for programs with two layers of continuations.
This programming pattern is also used for `success' and
`failure' continuations in the functional-programming
approach to backtracking.
Programs that have been CPS-transformed twice exhibit two such layers
of continuations.
Here, \stt{C} is the first layer and \stt{D}
is the second.
Iterating a CPS transformation gives rise to a CPS
hierarchy~\cite{Biernacka-al:LMCS05, Danvy-Filinski:LFP90,
Kameyama:HOSC07, Murthy:CW92}.
}
The resulting
interpreter still
threads an explicit continuation, even though it is not tail-recursive.

\item
Another direct-style transformation eliminates the control continuation:

%

% \begin{quote}
\begin{verbatim}
   eval :      term * E -> value
  apply : value * value -> value
\end{verbatim}
% \end{quote}

%

\noindent
The clauses catering for the non-tail-recursive uses of the control
continuation are expressed using the delimited-control operators
shift$_1$, reset$_1$, shift$_2$, and
reset$_2$~\cite{Biernacka-al:LMCS05, Danvy-Filinski:LFP90,
Danvy-Yang:ESOP99, Kameyama:CSL04 ,Murthy:CW92}.  The
resulting evaluator is in direct style.  It is also in closure-converted
form: the applicable values are a defunctionalized data type
and
\stt{apply} is the corresponding apply function.

\item
Refunctionalization eliminates the apply function:

%

% \begin{quote}
\begin{verbatim}
  eval : term * E -> value
\end{verbatim}
% \end{quote}

%

\noindent
The resulting
evaluation function is compositional, and the corresponding
syntax-directed encoding gives rise to new simulations of the J
operator.

\end{enumerate}

\subsubsection{A variant: from callee-save to caller-save dumps
               (Section~\protect\ref{sec:deconstruction-3})}
\label{subsubsec:from-callee-save-to-caller-save-dumps}

In Section~\ref{subsubsec:from-callee-save-to-caller-save-environments},
we kept the dump component because it is part of the SECD machine semantics
of the J operator.  We observe, however, that the dump is managed in a
callee-save way.  We therefore change gear and
consider the caller-save counterpart of
the interpreter:

% \begin{quote}
\begin{verbatim}
   eval :      term * E * C * D -> value
  apply : value * value * C     -> value
  where C = value -> value and D = value -> value
\end{verbatim}
% \end{quote}

%

\noindent
This caller-save interpreter is still
in CPS.  We can write its direct-style counterpart and refunctionalize
its applicable values, which yields another compositional evaluation
function in direct style.  This compositional evaluation function gives
rise to new simulations of the J operator, some of which had already been
invented independently.

\subsubsection{Assessment}

As
illustrated in
Sections~\ref{subsubsec:eliminating-the-data-stack},
\ref{subsubsec:from-callee-save-to-caller-save-environments},
and~\ref{subsubsec:from-callee-save-to-caller-save-dumps},
there is plenty of room for variation in the present deconstruction.
Each
step is reversible: one can
CPS-trans\-form and defunctionalize an evaluator and (re)construct an
abstract machine~\cite{Ager:PhD, Ager-al:PPDP03, Ager-al:IPL04,
Ager-al:TCS05, Biernacka-al:LMCS05, Biernacki:PhD, Danvy:IFL04,
Danvy:DSc, Danvy:ICFP08}.

\subsection{Syntactic theories of applicative expressions with the J operator}

Let us outline our syntactic theories of applicative expressions
substantiating them
in the next sections.

\subsubsection{Explicit, callee-save dumps
               (Section~\ref{sec:syntactic-theory-callee-save-dumps})}

We present a reduction semantics for Curien's calculus of closures
extended with the J operator, and we derivationally link it to the
caller-save, stackless SECD machine of
Section~\ref{subsec:transmogrified-SECD-machine}.

\subsubsection{Implicit, caller-save dumps
               (Section~\ref{sec:syntactic-theory-implicit-caller-save-dumps})}

We present another reduction semantics for Curien's calculus of closures
extended with the J operator, and we derivationally link it to a version
of the SECD machine which is not in defunctionalized form.

\subsubsection{Explicit, caller-save dumps
               (Section~\ref{sec:syntactic-theory-explicit-caller-save-dumps})}

We outline a third reduction semantics for Curien's calculus of closures
extended with the J operator, and we show how it leads towards
Thielecke's double-barrelled continuations.

\subsubsection{Inheriting the dump through the environment
               (Section~\ref{sec:syntactic-theory-environment})}

We present a fourth reduction semantics for Curien's calculus of closures
extended with the J operator, and we derivationally link it to a version
of the CEK machine that reflects Felleisen's simulation of the J
operator.

\subsection{Prerequisites and domain of discourse:
            the functional correspondence}
\label{subsec:prerequisites-and-domain-of-discourse-fc}

We mostly use pure ML as a meta-language.  We assume a basic familiarity with
Standard ML and with reasoning about pure ML programs as well as an elementary
understanding of defunctionalization~\cite{Danvy-Millikin:SCP0?, Danvy-Nielsen:PPDP01,
Reynolds:72} and its left inverse, refunctionalization;
of the CPS transformation~\cite{Danvy-Filinski:LFP90,
Danvy-Lawall:LFP92,
Friedman-Wand:08, FLMorris:LaSC93-one, Reynolds:72,
Steele:MS} and its left inverse, the direct-style transformation;
and of delimited continuations~\cite{Biernacka-al:LMCS05,
Danvy-Filinski:LFP90, Danvy-Yang:ESOP99, Filinski:POPL94,
Kameyama:CSL04}.  From
Section~\ref{subsec:SECD-refunctionalized-and-dump-less}, we use pure ML
with delimited-control operators as a meta-language.

\paragraph{The source language of the SECD machine.}

The source language is the $\lambda$-calculus, extended with literals (as
observables) and the J operator.  Except for the variables in the initial
environment of the SECD machine, a program is a closed term:

%

% \begin{quote}
\begin{verbatim}
  datatype term = LIT of int
                | VAR of string
                | LAM of string * term
                | APP of term * term
                | J
  type program = term
\end{verbatim}
% \end{quote}

%

%

\paragraph{The control directives.}  The control component of the SECD
machine is a list of control directives, where a directive is a term or the
tag \stt{APPLY}:

%

% \begin{quote}
\begin{verbatim}
  datatype directive = TERM of term | APPLY
\end{verbatim}
% \end{quote}

%

%

%
%
%

\paragraph{The environment.}
We use a structure \stt{Env} with the following signature:

%

% \begin{quote}
\begin{verbatim}
  signature ENV = sig
                    type 'a env
                    val empty : 'a env
                    val extend : string * 'a * 'a env -> 'a env
                    val lookup : string * 'a env -> 'a
                  end
\end{verbatim}
% \end{quote}

%

\noindent
The empty environment is denoted by \stt{Env.empty}.
The function extending an environment with a new binding is denoted by
\stt{Env.extend}.
The function fetching the value of an identifier from an environment is
denoted by \stt{Env.lookup}.
These functions are pure and total and therefore throughout, we call them
without passing them any continuation, \ie,
in direct style~\cite{Danvy-Hatcliff:MFPS93}.

\paragraph{Values.}

There are five kinds of values: integers, the successor function,
function closures,
``state appenders''~\cite[page~84]{Burge:75}, and program closures:

%

% \begin{quote}
\begin{verbatim}
  datatype value = INT of int
                 | SUCC
                 | FUNCLO of E * string * term
                 | STATE_APPENDER of D
                 | PGMCLO of value * D
  withtype S = value list                                        (* data stack *)
       and E = value Env.env                                    (* environment *)
       and C = directive list                                       (* control *)
       and D = (S * E * C) list                                        (* dump *)
\end{verbatim}
% \end{quote}

%

\noindent
A function closure pairs a $\lambda$-abstraction (\ie, its formal parameter
and its body) and
its lexical environment.
A
state appender
is an intermediate value; applying it
yields a program
closure.
A program closure is a first-class continuation.\footnote{%
The terms `function closures' and `program closures' are due to
Landin~\cite{Landin:TR65-Generalization}.  The term `state appender' is
due to Burge~\cite{Burge:75}.  The term `continuation' is due to
Wadsworth~\cite{Wadsworth:HOSC00}.  The term `first-class' is due to
Strachey~\cite{Strachey:67}.  The term `first-class continuation' is due
to Friedman and Haynes~\cite{Friedman-Haynes:POPL85}.
}

\paragraph{The initial environment.}

The initial environment binds the successor function:
%
%
%
%
%
%
%
%
%
%
%

%

% \begin{quote}
\begin{verbatim}
  val e_init = Env.extend ("succ", SUCC, Env.empty)
\end{verbatim}
% \end{quote}

%

%

\paragraph{The starting specification:}
Several formulations of the SECD machine with the J operator have been
published~\cite{Burge:75, Felleisen:CL87, Landin:TR65-Generalization}.
We take the most recent one, \ie,
Felleisen's~\cite{Felleisen:CL87}, as our starting point,
and we consider the others in Section~\ref{sec:rw}:

%

%

% \begin{quote}
\begin{verbatim}
  (*  run : S * E * C * D -> value                                             *)
  fun run (v :: s, e, nil, nil)
      = v
    | run (v :: s', e', nil, (s, e, c) :: d)
      = run (v :: s, e, c, d)
    | run (s, e, (TERM (LIT n)) :: c, d)
      = run ((INT n) :: s, e, c, d)
    | run (s, e, (TERM (VAR x)) :: c, d)
      = run ((Env.lookup (x, e)) :: s, e, c, d)
    | run (s, e, (TERM (LAM (x, t))) :: c, d)
      = run ((FUNCLO (e, x, t)) :: s, e, c, d)
    | run (s, e, (TERM (APP (t0, t1))) :: c, d)
      = run (s, e, (TERM t1) :: (TERM t0) :: APPLY :: c, d)
    | run (s, e, (TERM J) :: c, d)                                        (* 1 *)
      = run ((STATE_APPENDER d) :: s, e, c, d)
    | run (SUCC :: (INT n) :: s, e, APPLY :: c, d)
      = run ((INT (n+1)) :: s, e, c, d)
    | run ((FUNCLO (e', x, t)) :: v :: s, e, APPLY :: c, d)
      = run (nil, Env.extend (x, v, e'), (TERM t) :: nil, (s, e, c) :: d)
    | run ((STATE_APPENDER d') :: v :: s, e, APPLY :: c, d)               (* 2 *)
      = run ((PGMCLO (v, d')) :: s, e, c, d)
    | run ((PGMCLO (v, d')) :: v' :: s, e, APPLY :: c, d)                 (* 3 *)
      = run (v :: v' :: nil, e_init, APPLY :: nil, d')
\end{verbatim}
% \end{quote}

% \begin{quote}
\begin{verbatim}
  fun evaluate0 t                            (*  evaluate0 : program -> value  *)
      = run (nil, e_init, (TERM t) :: nil, nil)
\end{verbatim}
% \end{quote}

%

\noindent
The function \stt{run} implements the
iteration of a transition
function for the SECD machine:
\stt{(s, e, c, d)}
is a state of the machine and each clause of the definition of
\stt{run} specifies a state transition.

The SECD machine is deterministic.  It terminates if it reaches a state with
an empty control stack and an empty dump; in that case, it produces a value
on top of the data stack.  It does not terminate for divergent source terms.
It becomes stuck if it attempts to apply an integer or attempts to apply the
successor function to a non-integer value, in that case an ML
pattern-matching error is raised (alternatively, the codomain of \stt{run}
could be made \stt{value option} and a fallthrough \stt{else} clause could
be added).
The clause marked ``\stt{1}'' specifies that the J operator, at any
point, denotes the current dump; evaluating it captures this dump and
yields
a state appender
that, when applied (in the clause marked
``\stt{2}''), yields a program closure.  Applying a program closure (in
the clause marked ``\stt{3}'') restores the captured dump.

\subsection{Prerequisites and domain of discourse:
            the syntactic correspondence}
\label{subsec:prerequisites-and-domain-of-discourse-sc}

We assume a basic familiarity with reduction semantics as can be gathered
in Felleisen's PhD thesis~\cite{Felleisen:PhD} and undergraduate lecture
notes~\cite{Felleisen-Flatt:LN} and with Curien's original calculus of
closures~\cite{Biernacka-Danvy:TOCL07, Curien:TCS91}, which is the
ancestor of calculi of explicit substitutions.
% In addition, we review
We also review
the syntactic correspondence between reduction semantics and abstract
machines in Section~\ref{app:from-RS-to-AM} by deriving the CEK machine
from Curien's calculus of closures for
% left-to-right 
applicative order.

\subsection{Overview}
\label{subsec:overview}

We first disentangle and refunctionalize Felleisen's version of the SECD
machine (Section~\ref{sec:deconstruction-1}).  We then modernize it,
eliminating its data stack and making go from callee-save to caller-save
environments, and deconstruct the resulting specification into a
compositional evaluator in direct style; we then analyze the J operator
(Section~\ref{sec:deconstruction-2}).  Identifying that dumps are managed
in a callee-save way in the modernized SECD machine, we also present a
variant where they are managed in a caller-save way, and we deconstruct
the resulting specification into another compositional evaluator in
direct style; we then analyze the J operator
(Section~\ref{sec:deconstruction-3}).  Overall, the deconstruction takes
the form of a series of elementary transformations.  The correctness of
each step is very simple: most of the time, it is a corollary of the
correctness of the transformation itself.

We then review related work (Section~\ref{sec:rw})
and
outline the deconstruction of the original version of the SECD machine,
which is due to Burge (Section~\ref{sec:alt-deconstruction}).

We then
present a reduction semantics for the J operator that corresponds to the
specification of Section~\ref{sec:deconstruction-2}
(Section~\ref{sec:syntactic-theory-callee-save-dumps}).
We further present a syntactic theory of applicative expressions with the J
operator using delimiters
(Section~\ref{sec:syntactic-theory-implicit-caller-save-dumps}), and we
show how this syntactic theory specializes to a reduction semantics that
yields the abstract machine of Section~\ref{sec:deconstruction-3}
(Section~\ref{sec:syntactic-theory-explicit-caller-save-dumps}) and to
another reduction semantics that embodies Felleisen's embedding of J into
Scheme described in Section~\ref{subsec:Felleisen-s-embedding}
(Section~\ref{sec:syntactic-theory-environment}).

We then conclude (Sections~\ref{sec:concl}
and~\ref{sec:on-the-origin-of-continuations}).

\section{Deconstruction of the SECD machine with the J operator:
  \texorpdfstring{\\}{} 
  disentangling and refunctionalization}
\label{sec:deconstruction-1}

\subsection{A disentangled specification}
\label{subsec:SECD-description-disentangled}

In the starting specification of
Section~\ref{subsec:prerequisites-and-domain-of-discourse-fc}, all
the possible transitions are meshed together in one recursive function,
\stt{run}.  As in
the first rational deconstruction~\cite{Danvy:IFL04},
we factor
\stt{run} into
four mutually recursive functions, each with
one induction variable.
In this disentangled definition, \stt{run\_c} dispatches to the three
other transition functions, which all dispatch back to \stt{run\_c}:

\begin{enumerate}[$\bullet$]

\item
\stt{run\_c} interprets the list of control directives, \ie, it specifies
which transition to take
according to whether
the list is empty, starts with a term, or
starts with an apply directive.  If the list is empty, it calls
\stt{run\_d}.  If the list starts with a term, it calls \stt{run\_t},
caching the term in an extra component (the first parameter of
\stt{run\_t}).  If the list starts with an apply directive, it calls
\stt{run\_a}.

\item
\stt{run\_d} interprets the dump, \ie, it specifies which transition to
take
according to whether
the dump is empty or non-empty, given a valid data stack;
\stt{run\_t} interprets the top term in the list of control directives;
and
\stt{run\_a} interprets the top value in the
% current 
data stack.

\end{enumerate}

\noindent
Graphically:

{\let\labelstyle=\textstyle
\spreaddiagramrows{5mm}
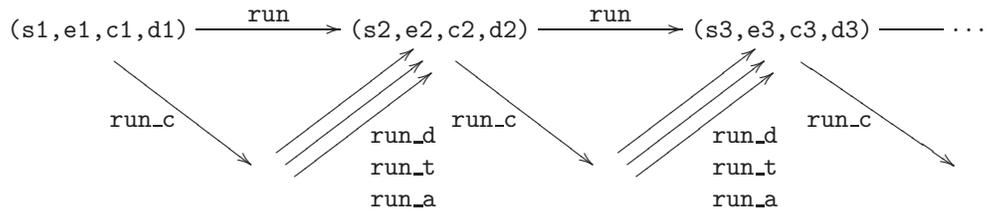
 $$
 \diagram
 {\Text{\stt{(s1,e1,c1,d1)}}}
 \ar@<+0.0cm>@{->}[rr]^{\Text{\stt{run}}}
 \ar@<-0.2cm>@{->}[dr]_{\Text{\stt{run\_c}}}
 &&
 {\Text{\stt{(s2,e2,c2,d2)}}}
 \ar@<+0.0cm>@{->}[rr]^{\Text{\stt{run}}}
 \ar@<-0.2cm>@{->}[dr]_{\Text{\stt{run\_c}}}
 &&
 {\Text{\stt{(s3,e3,c3,d3)}}}
 \ar@<+0.0cm>@{-}[r]
 \ar@<-0.2cm>@{->}[dr]_{\Text{\stt{run\_c}}}
 &
 \text{\ldots}
 \\
 &
 {\Text{\stt{\ }}}
 \ar@{->}[ur]
 \ar@<-0.2cm>@{->}[ur]
 \ar@<-0.4cm>@{->}[ur]_{\Text{\stt{run\_d} \\
                              \stt{run\_t} \\
                              \stt{run\_a}}}
 &&
 {\Text{\stt{\ }}}
 \ar@{->}[ur]
 \ar@<-0.2cm>@{->}[ur]
 \ar@<-0.4cm>@{->}[ur]_{\Text{\stt{run\_d} \\
                              \stt{run\_t} \\
                              \stt{run\_a}}}
 &&
 \enddiagram
 $$
}

%

%

% \begin{quote}
\begin{verbatim}
  (*  run_c :                 S * E * C * D -> value                           *)
  (*  run_d :                     value * D -> value                           *)
  (*  run_t :          term * S * E * C * D -> value                           *)
  (*  run_a : value * value * S * E * C * D -> value                           *)
  fun run_c (v :: s, e, nil, d)
      = run_d (v, d)
    | run_c (s, e, (TERM t) :: c, d)
      = run_t (t, s, e, c, d)
    | run_c (v0 :: v1 :: s, e, APPLY :: c, d)
      = run_a (v0, v1, s, e, c, d)
  and run_d (v, nil)
      = v
    | run_d (v, (s, e, c) :: d)
      = run_c (v :: s, e, c, d)
  and run_t (LIT n, s, e, c, d)
      = run_c ((INT n) :: s, e, c, d)
    | run_t (VAR x, s, e, c, d)
      = run_c ((Env.lookup (x, e)) :: s, e, c, d)
    | run_t (LAM (x, t), s, e, c, d)
      = run_c ((FUNCLO (e, x, t)) :: s, e, c, d)
    | run_t (APP (t0, t1), s, e, c, d)
      = run_c (s, e, (TERM t1) :: (TERM t0) :: APPLY :: c, d)
    | run_t (J, s, e, c, d)
      = run_c ((STATE_APPENDER d) :: s, e, c, d)
\end{verbatim}
% \end{quote}

% \begin{quote}
\begin{verbatim}
  and run_a (SUCC, INT n, s, e, c, d)
      = run_c ((INT (n+1)) :: s, e, c, d)
    | run_a (FUNCLO (e', x, t), v, s, e, c, d)
      = run_c (nil, Env.extend (x, v, e'), (TERM t) :: nil, (s, e, c) :: d)
    | run_a (STATE_APPENDER d', v, s, e, c, d)
      = run_c ((PGMCLO (v, d')) :: s, e, c, d)
    | run_a (PGMCLO (v, d'), v', s, e, c, d)
      = run_c (v :: v' :: nil, e_init, APPLY :: nil, d')
\end{verbatim}
% \end{quote}

% \begin{quote}
\begin{verbatim}
  fun evaluate1 t                            (*  evaluate1 : program -> value  *)
      = run_c (nil, e_init, (TERM t) :: nil, nil)
\end{verbatim}
% \end{quote}

\noindent
By construction, the two machines operate in
lockstep, with each transition of the original machine corresponding to two
transitions of the disentangled machine.
Since the two machines start in the same initial state, the
correctness of the disentangled machine is a corollary of them
operating in lockstep:

\begin{prop}[full correctness]
Given a program, \stt{evaluate0} and \stt{evaluate1} either both
diverge 
or both yield values that are structurally equal.
\end{prop}

In the rest of this section, we only consider programs that yield an
integer value, if any.  Indeed we are going to modify the data type of
the values as we go from abstract machine to evaluator, and we want a
simple, comparable characterization of the results they yield.

Furthermore, again for simplicity, we short-circuit four state
transitions in the abstract machine above:

% \begin{quote}
\begin{verbatim}
  ...
    | run_t (APP (t0, t1), s, e, c, d)
      = run_t (t1, s, e, (TERM t0) :: APPLY :: c, d)
  ...   
    | run_a (FUNCLO (e', x, t), v, s, e, c, d)
      = run_t (t, nil, Env.extend (x, v, e'), nil, (s, e, c) :: d)
  ...
    | run_a (PGMCLO (v, d'), v', s, e, c, d)
      = run_a (v, v', nil, e_init, nil, d')
  ...
\end{verbatim}
% \end{quote}

% \begin{quote}
\begin{verbatim}
  fun evaluate1 t
      = run_t (t, nil, e_init, nil, nil)
\end{verbatim}
% \end{quote}

%

%
%
%
%
%
%
%
%
%
%
%
%
%
%
%
%
%
%
%
%
%
%
%
%
%
%
%
%
%
%
%
%
%
%
%
%
%
%
%
%
%
%
%
%
%
%
%
%
%
%
%
%
%
%
%
%
%
%
%
%
%
%
%

\vspace{-1mm}

\subsection{A higher-order counterpart}
\label{subsec:SECD-refunctionalized}

In the disentangled definition of
Section~\ref{subsec:SECD-description-disentangled},
there are two possible ways to construct a dump---nil and consing a triple---and three
possible ways to construct a list of control direc\-tives---nil, consing a
term, and consing an apply directive.  One could phrase these
constructions as two specialized data types rather than as two lists.

These data types, together with \stt{run\_d} and \stt{run\_c} as their
apply functions, are in
the image of defunctionalization.  
After
refunctionalization,
the higher-order
evaluator reads as follows;\footnote{%
Had we not short-circuited the four state transitions at the end of
Section~\ref{subsec:SECD-description-disentangled}, the resulting
higher-order evaluator would contain four $\beta_v$-redexes.  Contracting
these redexes corresponds to short-circuiting these transitions.
}
it is higher-order because \stt{c} and
\stt{d} now denote functions:

%

% \begin{quote}
\begin{verbatim}
  datatype value = INT of int
                 | SUCC
                 | FUNCLO of E * string * term
                 | STATE_APPENDER of D
                 | PGMCLO of value * D
  withtype S = value list                                        (* data stack *)
       and E = value Env.env                                    (* environment *)
       and D = value -> value                             (* dump continuation *)
       and C = S * E * D -> value                      (* control continuation *)
\end{verbatim}
% \end{quote}

%

% \begin{quote}
\begin{verbatim}
  val e_init = Env.extend ("succ", SUCC, Env.empty)
\end{verbatim}
% \end{quote}

%

% \begin{quote}
\begin{verbatim}
  (*  run_t :          term * S * E * C * D -> value                           *)
  (*  run_a : value * value * S * E * C * D -> value                           *)
  fun run_t (LIT n, s, e, c, d)
      = c ((INT n) :: s, e, d)
    | run_t (VAR x, s, e, c, d)
      = c ((Env.lookup (x, e)) :: s, e, d)
    | run_t (LAM (x, t), s, e, c, d)
      = c ((FUNCLO (e, x, t)) :: s, e, d)
    | run_t (APP (t0, t1), s, e, c, d)
      = run_t (t1, s, e, fn (s, e, d) =>
          run_t (t0, s, e, fn (v0 :: v1 :: s, e, d) =>
            run_a (v0, v1, s, e, c, d), d), d)
    | run_t (J, s, e, c, d)
      = c ((STATE_APPENDER d) :: s, e, d)
  and run_a (SUCC, INT n, s, e, c, d)
      = c ((INT (n+1)) :: s, e, d)
    | run_a (FUNCLO (e', x, t), v, s, e, c, d)
      = run_t (t, nil, Env.extend (x, v, e'), fn (v :: s, e, d) => d v,
               fn v => c (v :: s, e, d))
\end{verbatim}
% \end{quote}

% \begin{quote}
\begin{verbatim}
    | run_a (STATE_APPENDER d', v, s, e, c, d)
      = c ((PGMCLO (v, d')) :: s, e, d)
    | run_a (PGMCLO (v, d'), v', s, e, c, d)
      = run_a (v, v', nil, e_init, fn (v :: s, e, d) => d v, d')
\end{verbatim}
% \end{quote}

%

% \begin{quote}
\begin{verbatim}
  fun evaluate2 t                            (*  evaluate2 : program -> value  *)
      = run_t (t, nil, e_init, fn (v :: s, e, d) => d v, fn v => v)
\end{verbatim}
% \end{quote}

%

\noindent
The resulting evaluator is in
CPS,
with two
layered continuations \stt{c} and \stt{d}.
It threads a stack of intermediate results (\stt{s}),
an environment (\stt{e}), a control continuation (\stt{c}), and a dump
continuation (\stt{d}).
Except for the environment being callee-save, the evaluator
follows a traditional eval--apply schema:
\stt{run\_t} is eval and \stt{run\_a} is apply.  Defunctionalizing it
yields the definition of
Section~\ref{subsec:SECD-description-disentangled} and as illustrated in
Appendix~\ref{app:fib}, by construction, \stt{run\_t} and \stt{run\_a} in
the defunctionalized version operate in lockstep with \stt{run\_t} and
\stt{run\_a} in the refunctionalized version:

\begin{prop}[full correctness]
\label{prop:evaluate1-and-evaluate2}
Given a program, \stt{evaluate1} and \stt{evaluate2} either both
diverge 
or both yield values; and if these values have an integer type, they are
the same integer.
\end{prop}

\section{Deconstruction of the SECD machine with the J operator: 
  \texorpdfstring{\\}{}
  no data stack and caller-save environments}
\label{sec:deconstruction-2}

We want to focus on J, and the non-standard aspects of the evaluator of
Section~\ref{subsec:SECD-refunctionalized} (the
callee-save environment and the data stack) are a
distraction.  We therefore
modernize this evaluator into a more
familiar caller-save, stackless
form~\cite{Friedman-Wand:08, FLMorris:LaSC93-one, Reynolds:72,
Steele-Sussman:TR78-AoI}.
Let us describe this
modernization in two steps: first we transform the
evaluator to use a caller-save convention for environments
(as outlined in Section~\ref{subsubsec:eliminating-the-data-stack}
and illustrated in
Appendices~\ref{app:caller-save-stackless-evaluator}
and~\ref{app:callee-save-stackless-evaluator}),
and second we
transform it to not use a data stack
(as outlined in Section~\ref{subsubsec:from-callee-save-to-caller-save-environments}
and illustrated in
Appendices~\ref{app:caller-save-stackless-evaluator}
and~\ref{app:caller-save-stack-threading-evaluator}).

The environments of the evaluator of
Section~\ref{subsec:SECD-refunctionalized} are callee-save because the apply
function \stt{run\_a} receives an environment \stt{e} as an argument and
``returns'' one to its continuation
\stt{c}~\cite[pages~404--408]{Aho-Sethi-Ullman:86}.  Inspecting the
evaluator shows that whenever \stt{run\_a} is passed a control directive
\stt{c} and an environment \stt{e}
and applies \stt{c}, then the environment \stt{e} is passed to \stt{c}.
Thus, the environment is
passed to \stt{run\_a} only in order to thread it to the control
continuation.  The control continuations created in \stt{run\_a} and
\stt{evaluate2} ignore their environment argument, and the control
continuations created in \stt{run\_t} are passed an environment that is
already in their lexical scope.  Therefore, neither the apply function
\stt{run\_a} nor the control continuations need to be passed an environment
at all.

Turning
to the data stack,
we first observe that the control continuations of the evaluator in
Section~\ref{subsec:SECD-refunctionalized} are always applied to a data
stack with at least one element.  Therefore, we can pass the top element of
the data stack as a separate argument, changing the type of control
continuations from \stt{S * E * D -> value} to \stt{value * S * E * D ->
value}.
We can thus eliminate the data stack following
an argument similar to the one for environments in the previous paragraph:
the \stt{run\_a} function merely threads its data
stack along to its control continuation; the control continuations created
in \stt{run\_a} and \stt{evaluate2} ignore their data-stack argument, and
the control continuations created in \stt{run\_t} are passed a data stack
that is already in their lexical scope.  Therefore, neither the apply
function \stt{run\_a}, the eval function \stt{run\_t}, nor the control
continuations need to be passed a data stack at all.

\subsection{A specification with no data stack and caller-save environments}
\label{subsec:a-specification-with-no-data-stack-and-caller-save-environments}

The caller-save, stackless counterpart of the evaluator of
Section~\ref{subsec:SECD-refunctionalized} reads as follows,
renaming \stt{run\_t} as \stt{eval} and \stt{run\_a}
as \stt{apply} in passing:

%

% \begin{quote}
\begin{verbatim}
  datatype value = INT of int
                 | SUCC
                 | FUNCLO of E * string * term
                 | STATE_APPENDER of D
                 | PGMCLO of value * D
  withtype E = value Env.env                                    (* environment *)
       and D = value -> value                             (* dump continuation *)
       and C = value * D -> value                      (* control continuation *)
\end{verbatim}
% \end{quote}

%

% \begin{quote}
\begin{verbatim}
  val e_init = Env.extend ("succ", SUCC, Env.empty)
\end{verbatim}
% \end{quote}

%

% \begin{quote}
\begin{verbatim}
  (*  eval  :      term * E * C * D -> value                                   *)
  (*  apply : value * value * C * D -> value                                   *)
  fun eval (LIT n, e, c, d)
      = c (INT n, d)
    | eval (VAR x, e, c, d)
      = c (Env.lookup (x, e), d)
    | eval (LAM (x, t), e, c, d)
      = c (FUNCLO (e, x, t), d)
    | eval (APP (t0, t1), e, c, d)
      = eval (t1, e, fn (v1, d) =>
          eval (t0, e, fn (v0, d) =>
            apply (v0, v1, c, d), d), d)
    | eval (J, e, c, d)
      = c (STATE_APPENDER d, d)
  and apply (SUCC, INT n, c, d)
      = c (INT (n+1), d)
    | apply (FUNCLO (e', x, t), v, c, d)
      = eval (t, Env.extend (x, v, e'), fn (v, d) => d v,
              fn v => c (v, d))
    | apply (STATE_APPENDER d', v, c, d)
      = c (PGMCLO (v, d'), d)
    | apply (PGMCLO (v, d'), v', c, d)
      = apply (v, v', fn (v, d) => d v, d')
\end{verbatim}
% \end{quote}

%

% \begin{quote}
\begin{verbatim}
  fun evaluate2' t                          (*  evaluate2' : program -> value  *)
      = eval (t, e_init, fn (v, d) => d v, fn v => v)
\end{verbatim}
% \end{quote}

%

\noindent
The new evaluator is still in
CPS, 
with two
layered continuations.
In order to
justify it formally,
we consider the corresponding abstract machine as obtained by
defunctionalization (shown in Section~\ref{sec:syntactic-theory-callee-save-dumps}; the
ML code for \stt{evaluate1'} is not shown here).  This abstract machine and
the disentangled abstract machine of
Section~\ref{subsec:SECD-description-disentangled} operate in lockstep and
we establish a bisimulation between them.  The full details of this formal
justification are found in the second author's PhD
dissertation~\cite[Section~4.4]{Millikin:PhD}.  Graphically:

{\let\labelstyle=\textstyle
 \spreaddiagramcolumns{2.1cm}
 \spreaddiagramrows{1.2cm}
 $$
 \diagram
 {\Text{\stt{evaluate0}}}
 \ar@{->}[r]^{\Text{disentangling}}
 &
 {\Text{\stt{evaluate1}}}
 \ar@<0.15cm>@{->}[r]^{\Text{refunctionalization}}
 \ar@{<..>}[d]_{\Text{bisimulation}}
 &
 {\Text{\stt{evaluate2}}}
 \ar@<0.15cm>@{->}[l]
 \ar@{-->}[d]^{\Text{`modernization': \myphantom{aaaaaa}
                     \\
                     no data stack and \myphantom{aaaaa}
                     \\
                     caller-save environments}}
 \\
 &
 {\Text{\stt{evaluate1'}}}
 \ar@<0.15cm>@{->}[r]
 &
 {\Text{\stt{evaluate2'}}}
 \ar@<0.15cm>@{->}[l]^{\Text{defunctionalization}}
 \enddiagram
 $$
}

\noindent
The following proposition follows as a corollary of the bisimulation and
of the correctness of defunctionalization:

\begin{prop}[full correctness]
Given a program, \stt{evaluate2} and \stt{evaluate2'} either both
diverge 
or both yield values; and if these values have an integer type, they are
the same integer.
\end{prop}

\subsection{A dump-less direct-style counterpart}
\label{subsec:SECD-refunctionalized-and-dump-less}

The evaluator of
Section~\ref{subsec:a-specification-with-no-data-stack-and-caller-save-environments}
is in continu\-ation-passing style, and therefore it is in the image of the
CPS transformation.  In order to highlight the control effect of the J
operator, we now present the direct-style counterpart of this evaluator.

The clause for J captures the current continuation (\ie,
the dump) in a state appender, and therefore its direct-style counterpart
naturally uses the undelimited control operator
call/cc~\cite{Danvy-Lawall:LFP92}.  With an eye on our next
step, we do not, however, use call/cc but its delimited cousins shift and
reset~\cite{Biernacka-al:LMCS05,
Danvy-Filinski:LFP90, Danvy-Yang:ESOP99} to write the
direct-style counterpart.

Concretely, we use an ML functor to obtain an instance of shift
and reset with \stt{value} as the type of intermediate
answers~\cite{Danvy-Yang:ESOP99, Filinski:POPL94}:
reset delimits
the (now implicit) dump continuation in \stt{eval}, and corresponds
to its initialization with the identity function; and shift
captures it in the clauses where J is evaluated and where a program
closure is applied.  There is one non-tail call to \stt{eval}, to evaluate
the body of a $\lambda$-abstraction; this context is captured by
\stt{shift} when J is evaluated:

% \begin{quote}
\begin{verbatim}
  datatype value = INT of int
                 | SUCC
                 | FUNCLO of E * string * term
                 | STATE_APPENDER of D
                 | PGMCLO of value * D
  withtype E = value Env.env                                    (* environment *)
       and C = value -> value                          (* control continuation *)
       and D = value -> value                 (* first-class dump continuation *)
\end{verbatim}
% \end{quote}

%

% \begin{quote}
\begin{verbatim}
  val e_init = Env.extend ("succ", SUCC, Env.empty)
\end{verbatim}
% \end{quote}

%

% \begin{quote}
\begin{verbatim}
  structure SR = make_Shift_and_Reset (type intermediate_answer = value)
\end{verbatim}
% \end{quote}

%

%

% \begin{quote}
\begin{small}
\begin{alltt}
  (*  eval  :      term * E * C -> value                                       *)
  (*  apply : value * value * C -> value                                       *)
  fun eval (LIT n, e, c)
      = c (INT n)
    | eval (VAR x, e, c)
      = c (Env.lookup (x, e))
    | eval (LAM (x, t), e, c)
      = c (FUNCLO (e, x, t))
    | eval (APP (t0, t1), e, c)
      = eval (t1, e, fn v1 => eval (t0, e, fn v0 => apply (v0, v1, c)))
    | eval (J, e, c)
      = SR.\shadethis{shift} (fn d => d (c (STATE_APPENDER d)))                       (* * *)
  and apply (SUCC, INT n, c)
      = c (INT (n+1))
    | apply (FUNCLO (e', x, t), v, c)
      = c (eval (t, Env.extend (x, v, e'), fn v => v))                    (* * *)
    | apply (STATE_APPENDER d, v, c)
      = c (PGMCLO (v, d))
    | apply (PGMCLO (v, d), v', c)
      = SR.\shadethis{shift} (fn d' => d (apply (v, v', fn v => v)))                  (* * *)
\end{alltt}
\end{small}
% \end{quote}

% \begin{quote}
\begin{small}
\begin{alltt}
  fun evaluate3' t                          (*  evaluate3' : program -> value  *)
      = SR.\shadethis{reset} (fn () => eval (t, e_init, fn v => v))
\end{alltt}
\end{small}
% \end{quote}

\noindent
The dump continuation is now implicit and is accessed using shift.
The first occurrence of shift captures the current dump when J is
evaluated.  The second occurrence is used to discard the current dump
when a program closure is applied.
CPS-trans\-form\-ing this evaluator yields the evaluator of
Section~\ref{subsec:a-specification-with-no-data-stack-and-caller-save-environments}:

\begin{prop}[full correctness]
Given a program, \stt{evaluate2'} and \stt{evaluate3'} either both
diverge 
or both yield values; and if these values have an integer type, they are
the same integer.
\end{prop}

\subsection{A control-less direct-style counterpart}
\label{subsec:SECD-refunctionalized-and-control-less}

The evaluator of Section~\ref{subsec:SECD-refunctionalized-and-dump-less}
still threads an explicit continuation, the control continuation.
It however is not in
continuation-passing style because of the non-tail calls to
\stt{c}, \stt{eval}, and \stt{apply} (in the clauses marked ``\stt{*}''
above)
and the occurrences of shift and reset.  This pattern
of control is characteristic of the CPS
hierarchy~\cite{Biernacka-al:LMCS05, Danvy-Filinski:LFP90,
Danvy-Yang:ESOP99, Kameyama:CSL04} (see also
Footnote~\ref{foo:CPS-hierarchy}, page~\pageref{foo:CPS-hierarchy}).
We therefore use the
delimited-control operators shift$_1$, reset$_1$,
shift$_2$, and reset$_2$ to write the direct-style
counterpart of this evaluator
(shift$_2$ and reset$_2$ are the direct-style
counterparts of shift$_1$ and reset$_1$, and shift$_1$
and reset$_1$ are synonyms for shift and reset).

Concretely, we use two ML functors to obtain
layered instances of shift and reset with \stt{value} as the
type of intermediate answers~\cite{Danvy-Yang:ESOP99, Filinski:POPL94}:
reset$_2$
delimits the (now twice implicit) dump continuation
in \stt{eval};
shift$_2$
captures it in the clauses where
J is evaluated and where a program closure is applied; reset$_1$
delimits the (now implicit) control continuation in \stt{eval} and in
\stt{apply}, and corresponds to its initialization with the identity
function; and shift$_1$ captures it in the clause where J is
evaluated:

%

% \begin{quote}
\begin{verbatim}
  datatype value = INT of int
                 | SUCC
                 | FUNCLO of E * string * term
                 | STATE_APPENDER of D
                 | PGMCLO of value * D
\end{verbatim}
% \end{quote}

% \begin{quote}
\begin{verbatim}
  withtype E = value Env.env                                    (* environment *)
       and D = value -> value                 (* first-class dump continuation *)
\end{verbatim}
% \end{quote}

%

% \begin{quote}
\begin{verbatim}
  val e_init = Env.extend ("succ", SUCC, Env.empty)
\end{verbatim}
% \end{quote}

%

% \begin{quote}
\begin{verbatim}
  structure SR1 = make_Shift_and_Reset (type intermediate_answer = value)
\end{verbatim}
% \end{quote}

%

% \begin{quote}
\begin{verbatim}
  structure SR2 = make_Shift_and_Reset_next (type intermediate_answer = value
                                             structure over = SR1)
\end{verbatim}
% \end{quote}

%

% \begin{quote}
\begin{verbatim}
  (*  eval  :      term * E -> value                                           *)
  (*  apply : value * value -> value                                           *)
  fun eval (LIT n, e)
      = INT n
    | eval (VAR x, e)
      = Env.lookup (x, e)
    | eval (LAM (x, t), e)
      = FUNCLO (e, x, t)
    | eval (APP (t0, t1), e)
      = let val v1 = eval (t1, e)
            val v0 = eval (t0, e)
        in apply (v0, v1) end
    | eval (J, e)
      = SR1.shift (fn c => SR2.shift (fn d => d (c (STATE_APPENDER d))))
  and apply (SUCC, INT n)
      = INT (n+1)
    | apply (FUNCLO (e', x, t), v)
      = SR1.reset (fn () => eval (t, Env.extend (x, v, e')))
    | apply (STATE_APPENDER d, v)
      = PGMCLO (v, d)
    | apply (PGMCLO (v, d), v')
      = SR1.shift (fn c' => SR2.shift (fn d' =>
          d (SR1.reset (fn () => apply (v, v')))))
\end{verbatim}
% \end{quote}

% \begin{quote}
\begin{verbatim}
  fun evaluate4' t                          (*  evaluate4' : program -> value  *)
      = SR2.reset (fn () => SR1.reset (fn () => eval (t, e_init)))
\end{verbatim}
% \end{quote}

\noindent
The control continuation is now implicit and is accessed using
shift$_1$.  The dump continuation is still implicit and is accessed
using shift$_2$.
CPS-transforming this evaluator yields the evaluator of
Section~\ref{subsec:SECD-refunctionalized-and-dump-less}:

\begin{prop}[full correctness]
Given a program, \stt{evaluate3'} and \stt{evaluate4'} either both
diverge 
or both yield values; and if these values have an integer type, they are
the same integer.
\end{prop}

\subsection{A compositional counterpart}
\label{subsec:SECD-compositional}

We now turn to the data flow of the evaluator of
Section~\ref{subsec:SECD-refunctionalized-and-control-less}.
As for the SECD machine without J~\cite{Danvy:IFL04},
this evaluator is in defunctionalized form: each of the values
constructed with \stt{SUCC}, \stt{FUNCLO}, \stt{PGMCLO}, and
\stt{STATE\_APPENDER}
is constructed at exactly one place and consumed at
exactly one other (the \stt{apply}
function).  We therefore refunctionalize them into the function space
\stt{value -> value}, which is shaded below:

%

% \begin{quote}
\begin{small}
\begin{alltt}
  datatype value = INT of int
                 | FUN of \shadethis{value -> value}
  withtype E = value Env.env
\end{alltt}
\end{small}
% \end{quote}

%

%
%

% \begin{quote}
\begin{verbatim}
  val e_init = Env.extend ("succ", FUN (fn (INT n) => INT (n+1)), Env.empty)
\end{verbatim}
% \end{quote}

%

% \begin{quote}
\begin{verbatim}
  structure SR1 = make_Shift_and_Reset (type intermediate_answer = value)
\end{verbatim}
% \end{quote}

%

% \begin{quote}
\begin{verbatim}
  structure SR2 = make_Shift_and_Reset_next (type intermediate_answer = value
                                             structure over = SR1)
\end{verbatim}
% \end{quote}

%

% \begin{quote}
\begin{verbatim}
  (*  eval  : term * E -> value                                                *)
  (*  where E = value Env.env                                                  *)
  fun eval (LIT n, e)
      = INT n
    | eval (VAR x, e)
      = Env.lookup (x, e)
    | eval (LAM (x, t), e)
      = FUN (fn v => SR1.reset (fn () => eval (t, Env.extend (x, v, e))))
    | eval (APP (t0, t1), e)
      = let val      v1 = eval (t1, e)
            val (FUN f) = eval (t0, e)
        in f v1 end
    | eval (J, e)
      = SR1.shift (fn c => SR2.shift (fn d =>
          d (c (FUN (fn v =>
                  FUN (fn v' => SR1.shift (fn c' =>
                                  SR2.shift (fn d' =>
                                    d (SR1.reset (fn () => let val (FUN f) = v
                                                           in f v' end))))))))))
\end{verbatim}
% \end{quote}

% \begin{quote}
\begin{verbatim}
  fun evaluate4'' t                        (*  evaluate4'' : program -> value  *)
      = SR2.reset (fn () => SR1.reset (fn () => eval (t, e_init)))
\end{verbatim}
% \end{quote}

\noindent
Unlike all the abstract machines and evaluators before,
this evaluation function is compositional: all the recursive calls on the
right-hand side are over proper sub-parts of the corresponding expression
on the left-hand side.
Defunctionalizing this evaluation function yields the evaluator of
Section~\ref{subsec:SECD-refunctionalized-and-control-less}:

\begin{prop}[full correctness]
Given a program,
\stt{evaluate4'} and \stt{evaluate4''} either both
diverge 
or both yield values; and if these values have an integer type, they are
the same integer.
\end{prop}

\subsection{Assessment}

 From Section~\ref{subsec:a-specification-with-no-data-stack-and-caller-save-environments}
to Section~\ref{subsec:SECD-compositional},
we have
modernized the SECD machine into a
stackless machine with a caller-save convention for environments,
and then deconstructed
the modernized version of this machine into a series of equivalent
specifications, starting (essentially) from a relation between states and
ending with an evaluation function.  The diagram below graphically
summarizes
the deconstruction.  The evaluators in
the top row are the defunctionalized counterparts of the evaluators in
the bottom row.  (The ML code for \stt{evaluate2''} and \stt{evaluate3''}
is not shown here.)

{\let\labelstyle=\textstyle
 \spreaddiagramcolumns{1.3cm}
 \spreaddiagramrows{1.2cm}
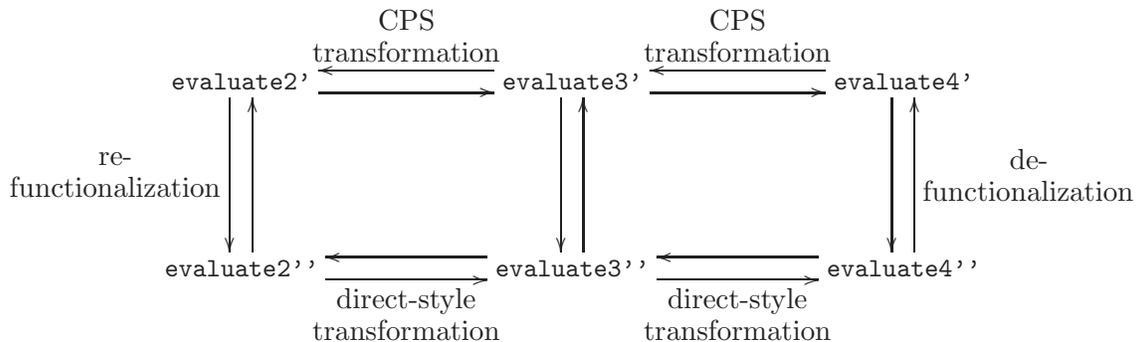
 $$
 \diagram
 {\Text{\stt{evaluate2'}}}
 \ar@<-0.15cm>@{->}[r]
 \ar@<-0.15cm>@{->}[d]_{\Text{re- \\functionalization}}
 &
 {\Text{\stt{evaluate3'}}}
 \ar@<-0.15cm>@{->}[r]
 \ar@<-0.15cm>@{->}[l]_{\Text{CPS \\ transformation}}
 \ar@<-0.15cm>@{->}[d]
 &
 {\Text{\stt{evaluate4'}}}
 \ar@<-0.15cm>@{->}[l]_{\Text{CPS \\ transformation}}
 \ar@<-0.15cm>@{->}[d]
 \\
 {\Text{\stt{evaluate2''}}}
 \ar@<-0.15cm>@{->}[r]_{\Text{direct-style \\ transformation}}
 \ar@<-0.15cm>@{->}[u]
 &
 {\Text{\stt{evaluate3''}}}
 \ar@<-0.15cm>@{->}[l]
 \ar@<-0.15cm>@{->}[r]_{\Text{direct-style \\ transformation}}
 \ar@<-0.15cm>@{->}[u]
 &
 {\Text{\stt{evaluate4''}}}
 \ar@<-0.15cm>@{->}[l]
 \ar@<-0.15cm>@{->}[u]_{\Text{de- \\ functionalization}}
 \enddiagram
 $$
}

\noindent
Using the tracing technique of Appendix~\ref{app:fib}, we can show that
\stt{evaluate2'} and \stt{evaluate2''} operate in lockstep.  We have
however not proved this lockstep property
for \stt{evaluate3'} and \stt{evaluate3''} and for
\stt{evaluate4'} and \stt{evaluate4''}, satisfying ourselves with
Plotkin's Simulation theorem~\cite{Plotkin:TCS75}, suitably extended for
shift and reset~\cite{Kameyama:HOSC07, Kameyama-Hasegawa:ICFP03}.

\subsection{On the J operator}
\label{subsec:J}

We now reap the fruits of
the modernization and the reconstruction, and present a
series of simulations of the J operator
(Sections~\ref{subsubsec:three-simulations-of-J},
\ref{subsubsec:C-and-the-hierarchy}, and
\ref{subsubsec:callcc-and-the-hierarchy}).  We then put the J operator into
perspective (Section~\ref{subsubsec:on-the-design-of-control-operators}).

\subsubsection{Three simulations of the J operator}
\label{subsubsec:three-simulations-of-J}

The evaluator of Section~\ref{subsec:SECD-compositional}
(\stt{evaluate4''}) and the
refunctionalized counterparts of the evaluators of
Sections~\ref{subsec:SECD-refunctionalized-and-dump-less} and
\ref{subsec:a-specification-with-no-data-stack-and-caller-save-environments} (\stt{evaluate3''}
and \stt{evaluate2''}) are
compositional.  They can be viewed as
syntax-directed encodings into their meta-language, as embodied in the first
Futamura projection~\cite{Futamura:71}
and the original approach to denotational semantics~\cite{Stoy:77}.
Below, we state these encodings
as three simulations of J: one in direct style, one in CPS with one layer of
continuations, and one in CPS with two layers of continuations.

We assume a call-by-value meta-language with right-to-left evaluation.
\begin{enumerate}[$\bullet$]

\item
In direct style, using shift$_2$ ($\rawshift_2$), reset$_2$
($\rawreset_2$), shift$_1$ ($\rawshift_1$), and reset$_1$
($\rawreset_1$), based on the compositional evaluator \stt{evaluate4''}
in direct style:
\[
  \begin{array}{r@{\ }c@{\ }l@{\hspace{3.25cm}}}
  \dmszero{\synlit}
  & = &
  \synlit
  \\
  \dmszero{\synvar}
  & = &
  \synvar
  \\
  \dmszero{\synapp{\synterm_0}
                  {\synterm_1}}
  & = &
  \synapp{\dmszero{\synterm_0}}{\dmszero{\synterm_1}}
  \\
  \dmszero{\synlam{\synvar}
                  {\synterm}}
  & = &
  \synlam{\synvar}
         {\synresetn{1}
                    {\dmszero{\synterm}}}
  \\
  \dmszero{\rawJ}
  & = &
  \synshiftn{1}{\synvarcont}
   {\synshiftn{2}{\synvarmcont}
     {\synapp{\synvarmcont}
             {\synappp{\synvarcont}
                      {\synlam{\synvarval}
                        {\ourbox{\synlam{\synvarval'}
                          {\synshiftn{1}{\synvarcont'}
                            {\synshiftn{2}{\synvarmcont'}
                              {\synapp{\synvarmcont}
                                      {\synresetn{1}
                                        {\synapp{\synvarval}
                                                {\synvarval'}}}}}}}}}}}}
  \end{array}
\]

\noindent
A
program $p$ is translated as
$\synresetn{2}{\synresetn{1}{\dmszero{p}}}$.

\item
In CPS with one layer of continuations, using shift ($\rawshift$)
and reset ($\rawreset$), based on the compositional evaluator
\stt{evaluate3''} in CPS with one layer of continuations:
\[
  \begin{array}{r@{\ }c@{\ }l@{\hspace{2.65cm}}}
  \dmsone{\synlit}
  & = &
  \synlam{\synvarcont}
         {\synapp{\synvarcont}
                 {\synlit}}
  \\
  \dmsone{\synvar}
  & = &
  \synlam{\synvarcont}
         {\synapp{\synvarcont}
                 {\synvar}}
  \\
  \dmsone{\synapp{\synterm_0}{\synterm_1}}
  & = &
  \synlam{\synvarcont}
         {\synapp{\dmsone{\synterm_1}}
                 {\synlam{\synvarval_1}
                         {\synapp{\dmsone{\synterm_0}}
                                 {\synlam{\synvarval_0}
                                         {\synapp{\synapp{\synvarval_0}
                                                         {\synvarval_1}}
                                                 {\synvarcont}}}}}}
  \\
  \dmsone{\synlam{\synvar}
                 {\synterm}}
  & = &
  \synlam{\synvarcont}
   {\synapp{\synvarcont}
           {\synlam{\synvar}
             {\synlam{\synvarcont}
               {\synapp{\synvarcont}
                       {\synappp{\dmsone{t}}
                                {\synlam{\synvarval}
                                  {\synvarval}}}}}}}
  \\
  \dmsone{\rawJ}
  & = &
  \synlam{\synvarcont}
    {\synshift{\synvarmcont}
      {\synapp
        {\synvarmcont}
        {\synappp{\synvarcont}
                 {\synlam{\synvarval}
                   {\synlam{\synvarcont}
                     {\synapp{\synvarcont}
                             {\ourbox{\synlam{\synvarval'}
                                     {\synlam{\synvarcont'}
                                       {\synshift{\synvarmcont'}
                                         {\synapp{\synvarmcont}
                                                 {\synappp{\synapp{\synvarval}
                                                                  {\synvarval'}}
                                                          {\synlam{\synvarval''}
                                                            {\synvarval''}}}}}}}}}}}}}}
  \end{array}
\]

\noindent
A
program $p$ is translated as
$\synreset{\synapp{\dmsone{p}}{\synlam{\synvarval}{\synvarval}}}$.

\item
In CPS with two layers of continuations (the outer continuation, \ie, the
dump continuation, can be $\eta$-reduced in
the first three clauses), based on the compositional evaluator
\stt{evaluate2''} in CPS with two layers of continuations:
\[
  \begin{array}{r@{\ }c@{\ }l}
  \dmstwo{\synlit}
  & = &
  \synlam{\synvarcont}
         {\synlam{\synvarmcont}
                 {\synapp{\synapp{\synvarcont}
                                 {\synlit}}
                         {\synvarmcont}}}
  \\
  \dmstwo{\synvar}
  & = &
  \synlam{\synvarcont}
         {\synlam{\synvarmcont}
                 {\synapp{\synapp{\synvarcont}
                                 {\synvar}}
                         {\synvarmcont}}}
  \\
  \dmstwo{\synapp{\synterm_0}{\synterm_1}}
  & = &
  \synlam{\synvarcont}
    {\synlam{\synvarmcont}
      {\synapp{\synapp{\dmstwo{\synterm_1}}
                      {\synlamp{\synvarval_1}
                        {\synlam{\synvarmcont}
                          {\synapp{\synapp{\dmstwo{\synterm_0}}
                                          {\synlamp{\synvarval_0}
                                            {\synlam{\synvarmcont}
                                              {\synapp{\synapp{\synapp{\synvarval_0}
                                                                      {\synvarval_1}}
                                                              {\synvarcont}}
                                                      {\synvarmcont}}}}}
                                  {\synvarmcont}}}}}
              {\synvarmcont}}}
  \\
  \dmstwo{\synlam{\synvar}{\synterm}}
  & = &
  \synlam{\synvarcont}
    {\synlam{\synvarmcont}
      {\synapp{\synapp{\synvarcont}
                      {\synlamp{\synvar}
                        {\synlam{\synvarcont}
                          {\synlam{\synvarmcont}
                            {\synapp{\synapp{\dmstwo{\synterm}}
                                            {\synlamp{\synvarval}
                                              {\synlam{\synvarmcont}
                                                {\synapp{\synvarmcont}
                                                        {\synvarval}}}}}
                                    {\synlam{\synvarval}
                                      {\synapp{\synapp{\synvarcont}
                                                      {\synvarval}}
                                              {\synvarmcont}}}}}}}}
              {\synvarmcont}}}
  \\
  \dmstwo{\rawJ}
  & = &
  \synlam{\synvarcont}
    {\synlam{\synvarmcont}
      {\synapp{\synapp{\synvarcont}
                      {\synlamp{\synvarval}
                        {\synlam{\synvarcont}
                          {\synlam{\synvarmcont'''}
                            {\synapp{\synapp{\synvarcont}
                                            {\ourbox{\synlamp{\synvarval'}
                                                    {\synlam{\synvarcont'}
                                                      {\synlam{\synvarmcont'}
                                                        {\synapp{\synapp{\synapp{\synvarval}
                                                                                {\synvarval'}}
                                                                        {\synlamp{\synvarval''}
                                                                          {\synlam{\synvarmcont''}
                                                                            {\synapp{\synvarmcont''}
                                                                                    {\synvarval''}}}}}
                                                                {\synvarmcont}}}}}}}
                                    {\synvarmcont'''}}}}}}
              {\synvarmcont}}}
  \end{array}
\]

\noindent
A
program $p$ is translated as
$\synapp{\synapp{\dmstwo{p}}{\synlamp{\synvarval}{\synlam{\synvarmcont}{\synapp{\synvarmcont}{\synvarval}}}}}{\synlam{\synvarval}{\synvarval}}$.

\end{enumerate}

\paragraph{Analysis:}
The simulation of literals, variables, and applications is
standard.
The control continuation of the body of each
$\lambda$-abstraction is delimited, corresponding to it being evaluated with
an empty control stack in the SECD machine.  The J operator abstracts the
control continuation and the dump continuation and immediately restores
them, resuming the computation with a
state appender which holds the abstracted dump continuation captive.
Applying this state appender to a
value $v$ yields a program closure (boxed in the three simulations above).
Applying this program closure to a value $v'$ has the effect of
discarding
both the current control continuation and the current dump continuation,
applying $v$ to $v'$, and resuming the
captured dump continuation with the result.

\paragraph{Assessment:}
The first rational deconstruction~\cite{Danvy:IFL04} already characterized the
SECD machine in terms of the CPS hierarchy: the control stack is the first
continuation, the dump is the second one (\ie, the meta-continuation), and
abstraction bodies are evaluated within a control delimiter (\ie, an empty
control stack).  Our work further characterizes the J operator as capturing (a
copy of) the meta-continuation.

\subsubsection{The $\rawC$ operator and the CPS hierarchy}
\label{subsubsec:C-and-the-hierarchy}

In the terminology of reflective towers~\cite{Danvy-Malmkjaer:LFP88},
continuations captured with shift are ``pushy''---at their point
of invocation, they compose with the current continuation by ``pushing''
it on the meta-continuation.
In the second encoding of J in Section~\ref{subsubsec:three-simulations-of-J}, the
term ${\synshift{\synvarmcont'} {\synapp{\synvarmcont}
{\synappp{\synapp{\synvarval} {\synvarval'}} {\synlam{\synvarval''}
{\synvarval''}}}}}$ serves to discard the current continuation $d'$ before
applying the captured continuation $d$.
Because of this use of shift to discard $d'$, the continuation $d$ is composed
with the identity continuation.

In contrast, still using the terminology of reflective towers,
continuations captured with call/cc~\cite{Clinger-Friedman-Wand:85-one}
or with Felleisen's $\rawC$ operator~\cite{Felleisen:PhD} are
``jumpy''---at their point
of invocation, they discard the current continuation.
If the continuation $d$ were captured with $\rawC$, then the
term $ {\synapp{\synvarmcont} {\synappp{\synapp{\synvarval} {\synvarval'}}
{\synlam{\synvarval''} {\synvarval''}}}} $ would suffice to discard the
current continuation.

The first encoding of J in Section~\ref{subsubsec:three-simulations-of-J} uses the
pushy control operators $\rawshift_1$
(\ie, $\rawshift$) and $\rawshift_2$.
Murthy~\cite{Murthy:CW92} and Kameyama~\cite{Kameyama:CSL04} have investigated
their jumpy counterparts in the CPS hierarchy, $\rawC_1$
(\ie, $\rawC$) and $\rawC_2$.  Jumpy
continuations therefore suggest 
two new simulations of the J operator.  We show only the
clauses for J, which are the only ones that change compared to
Section~\ref{subsubsec:three-simulations-of-J}.  As before, we assume a
call-by-value meta-language with right-to-left evaluation.
\begin{enumerate}[$\bullet$]

\item
In direct style, using $\rawC_2$, reset$_2$ ($\rawreset_2$), $\rawC_1$, and
reset$_1$ ($\rawreset_1$):

\[
\begin{array}{r@{\ }c@{\ }l@{\hspace{1.7cm}}}
\dmszero{\rawJ}
& = &
\synCn{1}{\synvarcont}
  {\synCn{2}{\synvarmcont}
    {\synapp{\synvarmcont}
      {\synappp{\synvarcont}
        {\synlam{\synvarval}
          {\ourbox
          {\synlam{\synvarval'}
            {\synapp{\synvarmcont}
              {\synresetn{1}
                {\synapp{\synvarval}
                  {\synvarval'}}}}}
          }
  }}}}
\end{array}
\]

\noindent
This simulation provides a new example of programming in the CPS hierarchy with
jumpy delimited continuations.

\item
In CPS with one layer of continuations, using $\rawC$ and reset ($\rawreset$):

\[
\begin{array}{r@{\ }c@{\ }l@{\hspace{0.45cm}}}
\dmsone{\rawJ}
& = &
\synlam{\synvarcont}
  {\synC{\synvarmcont}
    {\synapp
      {\synvarmcont}
      {\synappp{\synvarcont}
        {\synlam{\synvarval}
          {\synlam{\synvarcont}
            {\synapp{\synvarcont}
              {\ourbox
              {\synlam{\synvarval'}
                {\synlam{\synvarcont'}
                  {\synapp{\synvarmcont}
                    {\synappp{\synapp{\synvarval}
                        {\synvarval'}}
                      {\synlam{\synvarval''}
                        {\synvarval''}}}}}}
              }
  }}}}}}
\end{array}
\]

\end{enumerate}

\noindent
The corresponding CPS simulation of J with two layers of continuations
coincides with the one in Section~\ref{subsubsec:three-simulations-of-J}.

\subsubsection{The call/cc
operator and the CPS hierarchy}
\label{subsubsec:callcc-and-the-hierarchy}

Like
shift and $\rawC$,
call/cc takes a
snapshot of the current context.
However, unlike shift and $\rawC$, in so doing call/cc leaves the current
context in place.  So for example,
$1 + \synappp{\syncallcc}
             {\synlam{\synvarkont}
                     {10}}$
yields 11 because call/cc leaves the context $1 + [\:]$ in place,
whereas both
$1 + \synshiftp{\synvarkont}{10}$
and
$1 + \synCp{\synvarkont}{10}$
yield 10 because the context $1 + [\:]$ is tossed away.

Therefore J can be simulated in CPS with one layer of
continuations, using call/cc and exploiting its non-abortive behavior:

\[
\begin{array}{r@{\ }c@{\ }l}
\dmsone{\rawJ}
& = &
\synlam{\synvarcont}
  {\synapp{\rawcallcc}
          {\synlam{\synvarmcont}
      {\synapp{\synvarcont}
        {\synlam{\synvarval}
          {\synlam{\synvarcont}
            {\synapp{\synvarcont}
              {\ourbox
              {\synlam{\synvarval'}
                {\synlam{\synvarcont'}
                  {\synapp{\synvarmcont}
                    {\synappp{\synapp{\synvarval}
                        {\synvarval'}}
                      {\synlam{\synvarval''}
                        {\synvarval''}}}}}}
              }
  }}}}}}
\end{array}
\]

The obvious generalization of call/cc to the CPS hierarchy does not work,
however.  One needs an abort operator as well in order for call/cc$_2$ to
capture the initial continuation and the current meta-continuation.
We leave the rest of this train of thought to the imagination of the
reader.

\subsubsection{On the design of control operators}
\label{subsubsec:on-the-design-of-control-operators}

We note that replacing $\rawC$ with $\rawshift$ in
Section~\ref{subsubsec:C-and-the-hierarchy} (resp.~$\rawC_1$ with $\rawshift_1$
and $\rawC_2$ with $\rawshift_2$) yields a pushy counterpart for J, \ie,
program closures returning to their point of activation.  (Similarly,
replacing $\rawC$ with $\rawshift$ in the specification of call/cc in terms
of $\rawC$ yields a pushy version of call/cc, assuming a global control
delimiter.)  One can also
envision an abortive version of J that tosses away the context it
abstracts.
In that sense, control operators are easy to invent,
though not always easy to implement efficiently.
Nowadays, however, the
litmus test for a new control
operator lies elsewhere, for example:
\begin{enumerate}[(1)]

\item
Which programming idiom does this control operator
reflect~\cite{Clinger-Friedman-Wand:85-one, Danvy-Filinski:LFP90,
Danvy-Lawall:LFP92, Reynolds:72, Shan:HOSC07}?
\item
What is the logical content of this control
operator~\cite{Griffin:POPL90, Parigot:ICLPAR92}?

\end{enumerate}

\noindent
Even though it was the first control operator ever, J
passes
this
litmus test.
As pointed out by Thielecke,

\begin{enumerate}

\item
besides reflecting Algol jumps and labels~\cite{Landin:CACM65},
J
provides a generalized return~\cite[Section~2.1]{Thielecke:HOSC98}, and

\item
the type of $\synapp{\rawJ}{\synlam{\synvarval}{\synvarval}}$
is the law of the excluded middle~\cite[Section~5.2]{Thielecke:HOSC02}.

\end{enumerate}

\noindent
On the other hand, despite
their
remarkable fit to Algol labels and
jumps (as illustrated in the beginning of Section~\ref{sec:intro}),
the state appenders denoted by J are
unintuitive to use.  For
example, if a let expression is the syntactic sugar of a beta-redex (and
$\synvar_1$ is fresh), the observational equivalence

\[
  \begin{array}{r@{\hspace{2mm}}c@{\hspace{2mm}}l}
  \synapp{\synterm_0}{\synterm_1}
  & \cong &
  \synhlet{\synvar_1}{\synterm_1}{\synapp{\synterm_0}{\synvar_1}}
  \end{array}
\]

\noindent
does \emph{not} hold in the presence of J due to the non-standard
translation of abstractions, even though it does hold in the
presence of call/cc, $\rawC$, and shift for right-to-left evaluation.
For example, given
$C[\:]
 = \synapp{\synlamp{\synvar_2}
                   {\synapp{succ}{[\:]}}}
          {10}$,
$\synterm_0
 = \synapp{\synapp{\rawJ}
                  {\synlamp{k}{k}}}
          {0}$,
and
$\synterm_1
 = 100$,
$C[\synapp{\synterm_0}{\synterm_1}]$
yields $0$ whereas
$C[\synhlet{\synvar_1}{\synterm_1}{\synapp{\synterm_0}{\synvar_1}}]$
yields $1$.

\section{Deconstruction of the SECD machine with the J operator: 
  \texorpdfstring{\\}{}
  caller-save dumps}
\label{sec:deconstruction-3}

In Section~\ref{sec:deconstruction-2}, we
modernized the SECD machine by removing the intermediate data stack and
by managing the environment in a caller-save rather than
callee-save fashion.  We left the `non-modern' feature of the dump
continuation alone because it was part of the SECD-machine semantics of the
J operator.  In this section, we turn our attention to this dump
continuation, and we identify that like the environment in the original
SECD machine, \emph{the dump continuation is managed in a callee-save
fashion}.  Indeed the apply
function receives a dump continuation from its caller and passes it in turn
to the control continuation.

\vspace{-1mm}

\subsection{A specification with caller-save dump continuations}
\label{subsubsec:a-specification-with-caller-save-dump-continuations}

Let us modernize the SECD machine further by managing dump continuation
in a caller-save fashion.  Our reasoning is similar to that used in
Section~\ref{sec:deconstruction-2} for the
environment.  Inspecting
the evaluator \stt{evaluate2'} shows that when either
\stt{eval} or \stt{apply} receives a control continuation \stt{c} and a dump
continuation \stt{d} as arguments and applies \stt{c}, the dump
continuation \stt{d} is passed to \stt{c}.  Therefore, when the control
continuation passed to \stt{eval} or \stt{apply} is \stt{fn (v, d) => d v}
and the dump continuation is some
\stt{d'}, \stt{d'} can
be substituted for \stt{d} in the body of the control continuation.  After
this change, inspecting the control continuations reveals that the ones
created in \stt{apply} and \stt{evaluate2'} ignore their dump-continuation
arguments, and the ones created in \stt{eval} are passed a dump continuation
that is already in their lexical scope.  Therefore, the control
continuations do not need to be passed a dump continuation.  Since the dump
continuation was passed to \stt{apply} solely for the purpose of threading
it to the control continuation, \stt{apply} does not need to be passed a dump
continuation either.

The evaluator of
Section~\ref{subsec:a-specification-with-no-data-stack-and-caller-save-environments} with
caller-save dump continuations reads as follows:
% \begin{quote}
\begin{verbatim}
  datatype value = INT of int
                 | SUCC
                 | FUNCLO of E * string * term
                 | STATE_APPENDER of D
                 | PGMCLO of value * D
  withtype E = value Env.env                                    (* environment *)
       and D = value -> value                             (* dump continuation *)
       and C = value -> value                          (* control continuation *)
\end{verbatim}
% \end{quote}

% \begin{quote}
\begin{verbatim}  
  val e_init = Env.extend ("succ", SUCC, Env.empty)
\end{verbatim}
% \end{quote}

% \begin{quote}
\begin{verbatim}
  (*  eval  :  term * E * C * D -> value                                       *)
  (*  apply : value * value * C -> value                                       *)
  fun eval (LIT n, e, c, d)
      = c (INT n)
    | eval (VAR x, e, c, d)
      = c (Env.lookup (x, e))
    | eval (LAM (x, t), e, c, d)
      = c (FUNCLO (e, x, t))
    | eval (APP (t0, t1), e, c, d)
      = eval (t1, e, fn v1 =>
          eval (t0, e, fn v0 =>
            apply (v0, v1, c), d), d)
    | eval (J, e, c, d)
      = c (STATE_APPENDER d)
\end{verbatim}
% \end{quote}

% \begin{quote}
\begin{verbatim}
  and apply (SUCC, INT n, c)
      = c (INT (n + 1))
    | apply (FUNCLO (e', x, t), v, c)
      = eval (t, Env.extend (x, v, e'), c, c)
    | apply (STATE_APPENDER d', v, c)
      = c (PGMCLO (v, d'))
    | apply (PGMCLO (v, d'), v', c)
      = apply (v, v', d')
\end{verbatim}
% \end{quote}

% \begin{quote}
\begin{verbatim}  
  fun evaluate2'_alt t                   (*  evaluate2'_alt : program -> value *)
      = eval (t, e_init, fn v => v, fn v => v)
\end{verbatim}
% \end{quote}
%

\noindent
This evaluator still passes two continuations to \stt{eval}.  However, the
dump continuation is no longer passed as an argument to the control
continuation.  Thus, the two continuations have the same type.  The dump
continuation is
a snapshot of the control
continuation of the caller.  It is reset to be the continuation of the
caller when evaluating the body of a function closure and it is captured in
a state appender by the J operator.  Applying a program closure discards the
current control continuation in favor of the captured dump continuation.

As in
Section~\ref{subsec:a-specification-with-no-data-stack-and-caller-save-environments},
the abstract machine corresponding to \stt{evaluate2'\_alt} (obtained by
defunctionalization and displayed in
Section~\ref{sec:syntactic-theory-implicit-caller-save-dumps})
operates in lockstep
with the abstract machine corresponding to \stt{evaluate2'} (obtained by
defunctionalization and displayed in
Section~\ref{sec:syntactic-theory-callee-save-dumps}).  The following
proposition is a corollary of this bisimulation and the correctness of
defunctionalization:

\begin{prop}[full correctness]
Given a program, \stt{evaluate2'} and \stt{evaluate2'\_alt} either both diverge
or both yield values; and if these values have an integer type, they are the
same integer.
\end{prop}

\vspace{-1mm}

\subsection{The rest of the rational deconstruction}
\label{subsec:the-rest-of-the-rational-deconstruction}

The evaluator of
Section~\ref{subsubsec:a-specification-with-caller-save-dump-continuations}
can be transformed exactly as the higher-order evaluators of
Sections~\ref{subsec:SECD-refunctionalized}
and~\ref{subsec:a-specification-with-no-data-stack-and-caller-save-environments}:

\begin{enumerate}

\item
A
direct-style transformation with respect to the control continuation
yields an evaluator in direct style.

\item
Refunctionalizing the applicable values yields a compositional,
higher-order evaluator in direct style.

\end{enumerate}

\noindent
Graphically:

\vspace{-5mm}

{\let\labelstyle=\textstyle
 \spreaddiagramcolumns{2.0cm}
 \spreaddiagramrows{0.9cm}
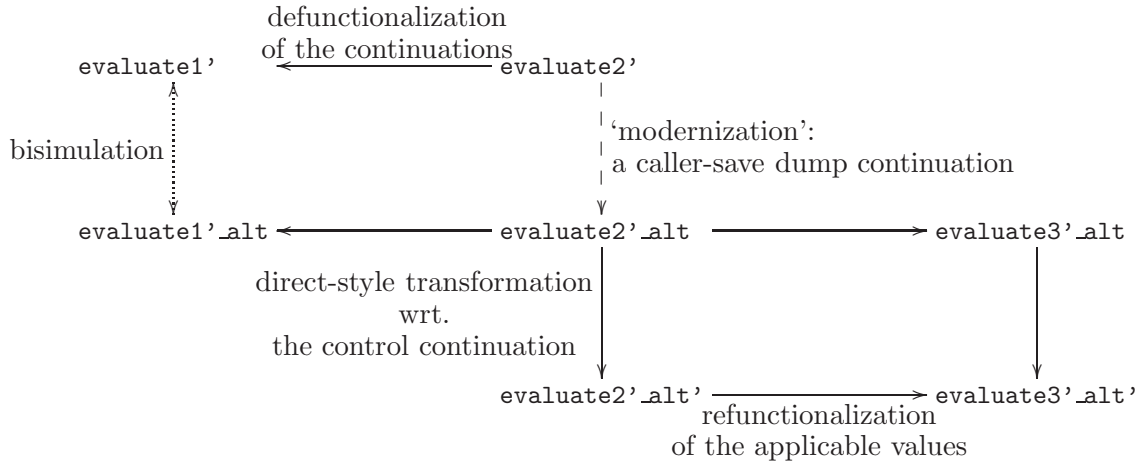
 $$
 \diagram
 {\Text{\stt{evaluate1'}\myphantom{\stt{\_alt}}}}
 \ar@{<..>}[d]_{\Text{bisimulation}}
 &
 {\Text{\stt{evaluate2'}\myphantom{\stt{\_alt'}}}}
 \ar@{->}[l]_{\Text{defunctionalization \\ of the continuations}}
 \ar@{-->}[d]^{\Text{`modernization': \myphantom{aaaaaaaaaaaaa}
                     \\
                     a caller-save dump continuation}}
 \\
 {\Text{\stt{evaluate1'\_alt}}}
 &
 {\Text{\stt{evaluate2'\_alt}\myphantom{\stt{'}}}}
 \ar@{->}[l]
 \ar@{->}[r]
 \ar@{->}[d]_{\Text{direct-style transformation \\ wrt.\\ the control continuation}}
 &
 {\Text{\stt{evaluate3'\_alt}\myphantom{\stt{'}}}}
 \ar@{->}[d]
 \\
 &
 {\Text{\stt{evaluate2'\_alt'}}}
 \ar@{->}[r]_{\Text{refunctionalization \\ of the applicable values}}
 &
 {\Text{\stt{evaluate3'\_alt'}}}
 \enddiagram
 $$
}

\clearpage

\subsection{Two other simulations of the J operator}
\label{subsec:two-other-simulations-of-J}

As in Section~\ref{subsubsec:three-simulations-of-J}, the compositional
evaluators of
Section~\ref{subsec:the-rest-of-the-rational-deconstruction} can be
viewed as syntax-directed translations into their meta-language.  Below,
we state these encodings as two further simulations of the J operator:
one in CPS with an additional return continuation, and one in
direct-style with a return continuation.

\begin{enumerate}[$\bullet$]

\item
In CPS with an additional return continuation, based on \stt{evaluate3'\_alt}:

\[
\begin{array}{r@{\ }c@{\ }l}
\dmsone{\synlit}
& = &
\synlam{\synvarcont}
  {\synlam{\synvardump}
    {\synapp{\synvarcont}{\synlit}}}
\\
\dmsone{\synvar}
& = &
\synlam{\synvarcont}
  {\synlam{\synvardump}
    {\synapp{\synvarcont}{\synvar}}}
\\
\dmsone{\synapp{\synterm_0}{\synterm_1}}
& = &
\synlam{\synvarcont}
  {\synlam{\synvardump}
    {\synapp
      {\synapp
        {\dmsone{\synterm_1}}
        {\synlamp{\synvarval_1}
          {\synapp
            {\synapp
              {\dmsone{\synterm_0}}
              {\synlamp{\synvarval_0}
                {\synapp{\synapp{\synvarval_0}
                                {\synvarval_1}}
                        {\synvarcont}}}}
            {\synvardump}}}}
      {\synvardump}}}
\\
\dmsone{\synlam{\synvar}{\synterm}}
& = &
\synlam{\synvarcont}
  {\synlam{\synvardump}
    {\synapp
      {\synvarcont}
      {\synlam{\synvarval}
        {\synlam{\synvarcont}
          {\synapp{\synapp{\dmsone{\synterm}}
                          {\synvarcont}}
                          {\synvarcont}}}}}}
\\
\dmsone{\rawJ}
& = &
\synlam{\synvarcont}
  {\synlam{\synvardump}
    {\synapp
      {\synvarcont}
      {\synlam{\synvarval_0}
        {\synlam{\synvarcont}
          {\synapp
            {\synvarcont}
            {\ourbox{\synlam{\synvarval_1}
              {\synlam{\synvarcont}
                {\synapp{\synapp{\synvarval_0}
                                {\synvarval_1}}
                                {\synvardump}}}}}}}}}}
\end{array}
\]
A program $p$ is translated as
$\synapp{\synapp{\dmsone{p}}
                {\synlamp{\synvarval}{\synvarval}}}
                {\synlam{\synvarval}{\synvarval}}$.

\item
In direct style with a return continuation, based on \stt{evaluate3'\_alt'}:
\[
\begin{array}{r@{\ }c@{\ }l@{\hspace{2.05cm}}}
\dmszero{\synlit}
& = &
\synlam{\synvardump}
  {\synlit}
\\
\dmszero{\synvar}
& = &
\synlam{\synvardump}
  {\synvar}
\\
\dmszero{\synapp{\synterm_0}{\synterm_1}}
& = &
\synlam{\synvardump}
  {\synapp
    {\synapp
      {\dmszero{\synterm_0}}
      {\synvardump}}
    {\synappp
      {\dmszero{\synterm_1}}
      {\synvardump}}}
\\
\dmszero{\synlam{\synvar}{\synterm}}
& = &
\synlam{\synvardump}
  {\synlam{\synvar}
    {\synshift{\synvarcont}
      {\synreset{\synapp
        {\synvarcont}
        {\synappp
          {\dmszero{\synterm}}
          {\synvarcont}}}}}}
\\
\dmszero{\rawJ}
& = &
\synlam{\synvardump}
  {\synlam{\synvarval_0}
    {\ourbox{\synlam{\synvarval_1}
      {\synshift{\synvarcont}
        {\synreset{\synapp{\synvardump}{\synappp{\synvarval_0}{\synvarval_1}}}}}}}}
\end{array}
\]
A program $p$ is translated as
$\synreset{\synapp{\dmszero{p}}{\synlam{\synvarval}{\synvarval}}}$.

NB. Operationally, the two occurrences of reset surrounding the body of
the shift-expression are unnecessary.  They could be omitted.

\end{enumerate}

\paragraph{Assessment:}
Transformed terms are passed a pair of
continuations, the usual continuation of the call-by-value CPS
transform and a return continuation.  Abstractions set the return
continuation to be the continuation at their point of invocation,
\ie, the continuation of their caller.  The
J operator captures the current return continuation in a program closure
(boxed above).

\subsection{Thielecke}

In his work on comparing control constructs~\cite{Thielecke:HOSC02},
Thielecke introduced a `double-barrelled' CPS transformation, where
terms are passed an additional `jump continuation' in addition to the
usual continuation of the call-by-value CPS transformation.  By
varying the transformation of abstractions, he was able to account for
first-class continuations, exceptions, and jumping.  His
double-barrelled CPS transformation, including a clause for his JI
operator (\ie, $\synJ{\synvar}{\synvar}$) and modified for
right-to-left evaluation, reads as follows:
\[
\begin{array}{r@{\ }c@{\ }l}
\htszero{\synvar}
& = &
\synlam{\synvarcont}
  {\synlam{\synvarjumpcont}
    {\synapp{\synvarcont}
            {\synvar}}}
\\
\htszero{\synapp{\synterm_0}{\synterm_1}}
& = &
\synlam{\synvarcont}
  {\synlam{\synvarjumpcont}
    {\synapp
      {\synapp
        {\htszero{\synterm_1}}
        {\synlamp{\synvarval_1}
          {\synapp
            {\synapp
              {\htszero{\synterm_0}}
              {\synlamp{\synvarval_0}
                {\synapp{\synapp{\synapp{\synvarval_0}
                                        {\synvarval_1}}
                                        {\synvarcont}}
                                        {\synvarjumpcont}}}}
            {\synvarjumpcont}}}}
      {\synvarjumpcont}}}
\\
\htszero{\synlam{\synvar}{\synterm}}
& = &
\synlam{\synvarcont}
  {\synlam{\synvarjumpcont}
    {\synapp
      {\synvarcont}
      {\synlam{\synvar}
        {\synlam{\synvarcont'}
          {\synlam{\synvarjumpcont'}
            {\synapp{\synapp{\htszero{\synterm}}
                            {\synvarcont'}}
                            {\synvarcont'}}}}}}}
\\
\htszero{\rawJI}
& = &
\synlam{\synvarcont}
  {\synlam{\synvarjumpcont}
    {\synapp
      {\synvarcont}
      {\synlam{\synvar}
        {\synlam{\synvarcont'}
          {\synlam{\synvarjumpcont'}
            {\synapp{\synvarjumpcont}
                    {\synvar}}}}}}}
\end{array}
\]

\noindent
The continuation $\synvarcont$ is the continuation of the usual
call-by-value CPS transformation.  The continuation $\synvarjumpcont$
is a return continuation, \ie, a snapshot of
the continuation of the caller of a function abstraction.  It is set
to be the continuation of the caller
in the
body of each function abstraction and
it is captured as a first-class function by the JI
operator.  The extra continuation passed to each abstraction is not
necessary (for the encoding of JI), and can be eliminated from the
translation of abstractions and applications, as we did in
Section~\ref{subsubsec:a-specification-with-caller-save-dump-continuations}.

As noted by Thielecke and earlier Landin, J can be expressed in terms
of JI as:
\[
\rawJ
\;\equiv\;
\synapp
    {\synlamp{\synvarcont}
      {\synlam{\synvarval}
        {\synlam{\synvarval'}
          {\synapp{\synvarcont}
            {\synappp{\synvarval}{\synvarval'}}}}}}
    {(\rawJI)}
\]
The $\beta$-expansion is necessary to move the occurrence of $\rawJI$
outside of the outer abstraction, because $\lambda$-abstractions are
CPS-transformed in a non-standard way.  By CPS-transforming this
definition and eliminating the extra continuation for function abstractions, we
derive the same double-barrelled encoding of Landin's J operator
as in Section~\ref{subsec:two-other-simulations-of-J}:
\[
\begin{array}{r@{\ }c@{\ }l}
\htsone{\synvar}
& = &
\synlam{\synvarcont}
 {\synlam{\synvarjumpcont}
   {\synapp{\synvarcont}{\synvar}}}
\\
\htsone{\synapp{\synterm_0}{\synterm_1}}
& = &
\synlam{\synvarcont}
  {\synlam{\synvarjumpcont}
    {\synapp
      {\synapp
        {\htsone{\synterm_1}}
        {\synlamp{\synvarval_1}
        {\synapp
          {\synapp
            {\htsone{\synterm_0}}
            {\synlamp{\synvarval_0}
              {\synapp
                {\synapp
                  {\synvarval_0}
                  {\synvarval_1}}
                {\synvarcont}}}}
          {\synvarjumpcont}}}}
      {\synvarjumpcont}}}
\\
\htsone{\synlam{\synvar}{\synterm}}
& = &
\synlam{\synvarcont}
  {\synlam{\synvarjumpcont}
    {\synapp
      {\synvarcont}
      {\synlam{\synvar}
        {\synlam{\synvarcont'}
          {\synapp
            {\synapp
              {\htsone{\synterm}}
              {\synvarcont'}}
            {\synvarcont'}}}}}}
\\
\htsone{\rawJ}
& = &
\synlam{\synvarcont}
  {\synlam{\synvarjumpcont}
    {\synapp
      {\synvarcont}
      {\synlam{\synvarval}
        {\synlam{\synvarcont'}
          {\synapp
            {\synvarcont'}
            {\ourbox
              {\synlam{\synvarval'}
                {\synlam{\synvarcont''}
                  {\synapp
                    {\synapp
                      {\synvarval}
                      {\synvarval'}}
                    {\synvarjumpcont}}}}}}}}}}
\end{array}
\]

\paragraph{Analysis:} 
In essence, Thielecke's simulation corresponds to an abstract machine
which is the caller-save counterpart of Landin's machine with respect to
the dump.

%
%
%

% \vspace{-1mm}

\subsection{Felleisen}
\label{subsec:Felleisen-s-embedding}

Felleisen showed how to embed Landin's extension of applicative expressions
with J into the Scheme programming language~\cite{Felleisen:CL87}.  The
embedding is defined using Scheme syntactic extensions (\ie, macros).  J is
treated as a dynamic
identifier that is bound in the body of every abstraction,
similarly to the dynamically bound identifier `self' in an embedding of
Smalltalk into Scheme~\cite{Landin:CACM66}.
The control aspect of J
is handled through Scheme's control operator
call/cc.

Here are
the corresponding simulations:
% using $\rawC$ and reset, using shift and reset, and in CPS:
%
\begin{enumerate}[$\bullet$]

\item
In direct style, using either of call/cc, $\rawC$, or shift
($\rawshift$), and
%
% one global
a control delimiter ($\rawreset$):
\[
  \begin{array}{r@{\ }c@{\ }l}
  \mfszero{\synvar}
  & = &
  \synvar
  \\
  \mfszero{\synapp{\synterm_0}{\synterm_1}}
  & = &
  \synapp{\mfszero{\synterm_0}}
         {\mfszero{\synterm_1}}
  \\
  \mfszero{\synlam{\synvar}{\synterm}}
  & = &
  \synlam{\synvar}
         {\synapp{\syncallcc}
                 {\synlam{\synvardump}
                         {\synhlet{\joperator}
                                  {\synlam{\synvarval}
                                          {\ourbox{\synlam{\synvarval'}
                                                  {\synapp{\synvardump}
                                                          {\synappp{\synvarval}
                                                                   {\synvarval'}}}}}}
                                  {\mfszero{\synterm}}}}}
  \\[0.5mm]
  & = &
  \synlam{\synvar}
         {\hspace{0.35mm}\synC{\synvardump}
               {\synhlet{\joperator}
                        {\synlam{\synvarval}
                                {\ourbox{\synlam{\synvarval'}
                                        {\synapp{\synvardump}
                                                {\synappp{\synvarval}
                                                         {\synvarval'}}}}}}
                        {\synapp{\synvardump}{\mfszero{\synterm}}}}}
  \\[0.5mm]
  & = &
  \synlam{\synvar}
         {\synshift{\synvardump}
                   {\synhlet{\joperator}
                            {\synlam{\synvarval}
                                    {\ourbox{\synlam{\synvarval'}
                                            {\synshift{\synvarcont'}
                                                      {\synapp{\synvardump}
                                                              {\synappp{\synvarval}
                                                                       {\synvarval'}}}}}}}
                            {\synapp{\synvardump}{\mfszero{\synterm}}}}}
  \end{array}
\]

\noindent
A
program $p$ is translated as
$\synhlet{J}
         {\synlam{\synvarval}
                 {\synlam{\synvarval'}
                         {\synreset{\synapp{\synvarval}
                                           {\synvarval'}}}}}
         {\synreset{\mfszero{p}}}$.

\item
In CPS:
\[
  \begin{array}{r@{\ }c@{\ }l}
  \mfsone{\synvar}
  & = &
  \synlam{\synvarcont}
         {\synapp{\synvarcont}
                 {\synvar}}
  \\
  \mfsone{\synapp{\synterm_0}{\synterm_1}}
  & = &
  \synlam{\synvarcont}
         {\synapp{\mfsone{\synterm_1}}
                 {\synlam{\synvarval_1}
                         {\synapp{\mfsone{\synterm_0}}
                                 {\synlam{\synvarval_0}
                                         {\synapp{\synapp{\synvarval_0}
                                                         {\synvarval_1}}
                                                 {\synvarcont}}}}}}
  \\
  \mfsone{\synlam{\synvar}{\synterm}}
  & = &
  \synlam{\synvarcont}
         {\synapp{\synvarcont}
                 {\synlamp{\synvar}
                          {\synlam{\synvardump}
                          {\synhlet{\joperator}
                                   {\synlam{\synvarval}
                                           {\synlam{\synvarcont}
                                                   {\synapp{\synvarcont}

{\ourbox{\synlam{\synvarval'}

{\synlam{\synvarcont'}

     {\synapp{\synapp{\synvarval}

                     {\synvarval'}}

             {\synvardump}}}}}}}}
                                   {\synapp{\mfsone{\synterm}}
                                           {\synvardump}}}}}}
  \end{array}
\]

\noindent
A
program $p$ is translated as
$\synhlet{J}
         {\synlam{\synvarval}
                 {\synlam{\synvarcont}
                         {\synapp{\synvarcont}
                                 {\synlamp{\synvarval'}
                                          {\synlam{\synvarcont'}
                                                  {\synapp{\synapp{\synvarval}
                                                                  {\synvarval'}}
                                                          {\synlam{\synvarval''}
                                                                  {\synvarval''}}}}}}}}
         {\synapp{\mfsone{p}}{\synlam{\synvarval}{\synvarval}}}$.

\end{enumerate}

\paragraph{Analysis:}
The simulation of variables and applications is
standard.
The
continuation of the body of each $\lambda$-abstraction is captured, and
the identifier J is dynamically bound to a function closure
(the state appender)
which holds
the continuation captive.  Applying this function closure to a
value $v$ yields a program closure (boxed in the
simulations
above).  Applying this program closure to a value $v'$ has the effect of
applying $v$ to $v'$ and resuming the captured continuation with the
result, abandoning the current continuation.

The evaluator corresponding to
% Felleisen's simulation
these simulations always has a
binding of J in the environment when evaluating the body of an
abstraction (see Section~\ref{sec:syntactic-theory-environment}).
Under the assumption that J is never shadowed in a program,
passing this value as a separate argument to the evaluator leads one
towards the definition of \stt{evaluate2'\_alt} in
Section~\ref{subsubsec:a-specification-with-caller-save-dump-continuations}
(see Section~\ref{sec:syntactic-theory-explicit-caller-save-dumps}).

\section{Related work}
\label{sec:rw}

%

% \vspace{-1mm}

\subsection{Landin and Burge}

Landin~\cite{Landin:TR65-Generalization} introduced the J operator as a new language
feature motivated by three questions about labels and jumps:

\begin{enumerate}[$\bullet$]

\item
Can a language
have jumps without having assignments?

\item
Is there some component of jumping
that is independent of labels?

\item
Is there some feature that corresponds to
functions with arguments in the same sense that labels correspond to
procedures without arguments?

\end{enumerate}

\noindent
Landin gave the semantics of the J operator
by extending the SECD machine.  In addition to using J to model jumps in
Algol 60~\cite{Landin:CACM65},
he gave examples of programming with the J operator, using it to
represent failure actions as program closures where it is essential that
they
abandon the context of their application.

In his textbook~\cite[Section~2.10]{Burge:75}, Burge
adjusted Landin's original
specification of the J operator.
Indeed, 
in Landin's extension of the SECD machine, J could only occur in the
context of an application.
Burge
adjusted the
original specification
so that J could occur in arbitrary contexts.
To this end, he introduced the notion of
a ``state appender'' as the denotation of J.

Thielecke~\cite{Thielecke:HOSC98} gave a detailed introduction to the J
operator as presented by Landin and Burge.
Burstall~\cite{Burstall:MI69-writing-search-algorithms} illustrated
the use of the J operator by simulating threads for parallel search
algorithms, which in
retrospect is the first simulation of threads in terms of first-class
continuations ever.

\vspace{-1mm}

\subsection{Reynolds}

Reynolds~\cite{Reynolds:72} gave a comparison of J to escape, the binder
form of Scheme's
call/cc~\cite{Clinger-Friedman-Wand:85-one}.\footnote{$\synescape{\synvarkont}{\synterm}
\;\equiv\;
\synapp{\syncallcc}{\synlam{\synvarkont}{\synterm}}$}
He gave encodings
of Landin's J
(\ie, restricted to the context of an application) and escape in terms of
each other.

His encoding of escape in terms of J reads as follows:
\[
  \begin{array}{r@{\hspace{2mm}}c@{\hspace{2mm}}l}
  \jrscallcc{\synescapep{\synvarkont}{\synterm}}
  & = &
  \synhlet{\synvarkont}
          {\synJ{\synvarval}{\synvarval}}
          {\jrscallcc{\synterm}}
  \end{array}
\]

\noindent
As Thielecke notes~\cite{Thielecke:HOSC98}, this encoding is only valid
immediately inside an abstraction.
Indeed, the dump continuation
captured by J only coincides with the continuation captured by escape if
the control continuation is the initial one (\ie, immediately inside a
control delimiter).  Thielecke therefore generalized the encoding by adding a dummy
abstraction:
\[
  \begin{array}{r@{\hspace{2mm}}c@{\hspace{2mm}}l}
  \jrscallcc{\synescapep{\synvarkont}{\synterm}}
  & = & 
  \synapp{\synlamp{()}
            {\synhlet{\synvarkont}
                     {\synJ{\synvar}{\synvar}}
                     {\jrscallcc{\synterm}}}}
         {()}
  \end{array}
\]

\noindent
 From the point of view of the rational deconstruction
of Section~\ref{sec:deconstruction-2}, this dummy
abstraction implicitly inserts a control delimiter.

Reynolds's converse encoding of J in terms of escape reads as follows:
\[
  \begin{array}{r@{\hspace{2mm}}c@{\hspace{2mm}}l}
  \jrsJ{\synhletp{\synvarmcont}
                 {\synJ{\synvar}{\synterm_1}}
                 {\synterm_0}}
  & = &
  \synescape{\synvarkont}
            {\synhletp{\synvarmcont}
                      {\synlam{\synvar}
                        {\synapp{\synvarkont}{\jrsJ{\synterm_1}}}}
                      {\jrsJ{\synterm_0}}}
  \end{array}
\]
where $\synvarkont$
does not occur free in $\synterm_0$ and $\synterm_1$.
For the same reason as above, this
encoding is only valid immediately inside an abstraction and therefore it
can be generalized by adding a dummy abstraction:
\[
  \begin{array}{r@{\hspace{2mm}}c@{\hspace{2mm}}l}
  \jrsJ{\synhletp{\synvarmcont}
                 {\synJ{\synvar}{\synterm_1}}
                 {\synterm_0}}
  & = &
  \synapp{\synlamp{()}{
  \synescape{\synvarkont}
            {\synhletp{\synvarmcont}
                      {\synlam{\synvar}
                        {\synapp{\synvarkont}{\jrsJ{\synterm_1}}}}
                      {\jrsJ{\synterm_0}}}
  }}{()}
  \end{array}
\]

\subsection{Felleisen and Burge}
\label{subsec:Felleisen-and-Burge}

Felleisen's version of the SECD machine with
the J operator differs from Burge's.  In the notation of
Section~\ref{subsec:prerequisites-and-domain-of-discourse-fc},
Burge's clause for
applying program closures reads

% \begin{quote}
\begin{verbatim}
    | run ((PGMCLO (v, (s', e', c') :: d')) :: v' :: s, e, APPLY :: c, d)
      = run (v :: v' :: s', e', APPLY :: c', d')
\end{verbatim}
% \end{quote}

\noindent
instead of

% \begin{quote}
\begin{verbatim}
    | run ((PGMCLO (v, d')) :: v' :: s, e, APPLY :: c, d)
      = run (v :: v' :: nil, e_init, APPLY :: nil, d')
\end{verbatim}
% \end{quote}

%

\noindent
Felleisen's version delays the consumption of the dump until the
function, in the program closure, completes, whereas Burge's version
does not.  The modification is unobservable because a program cannot
capture the control continuation and because applying the argument of a
state appender pushes the data stack, the environment, and the control
stack on the dump.
Felleisen's modification can
be characterized as wrapping a control delimiter around the argument of a dump
continuation, similarly to the simulation of static delimited continuations in
terms of dynamic ones~\cite{Biernacki-Danvy:JFP06}.

Burge's version, however, is not in defunctionalized
form.  In
Section~\ref{sec:alt-deconstruction},
we put it in defunctionalized
form without
resorting to a control delimiter and we
outline the corresponding compositional evaluation functions and
simulations.

\section{Deconstruction of the original SECD machine with the J operator}
\label{sec:alt-deconstruction}

We now outline the deconstruction of Burge's specification of the SECD
machine with the J operator.

\subsection{Our starting point: Burge's specification}
\label{subsec:Burge-SECD-starting}

As pointed out in Section~\ref{subsec:Felleisen-and-Burge},
Fell\-eisen's version of the SECD machine applies the value contained in a
program closure before restoring the
components of the captured dump.  Burge's version differs by restoring the
components of the captured dump before applying the value
contained in the program closure.  In other words,
\begin{enumerate}[$\bullet$]

\item
Felleisen's version applies the value contained in a program closure with
an empty data stack, a dummy environment, an empty control stack, and the
captured dump, whereas

\item
Burge's version applies the value contained in a program closure with
the captured data stack, environment, control stack, and previous dump.

\end{enumerate}

\noindent
The versions induce a minor programming difference because
the first makes it possible to use J in any context whereas the
second restricts J to occur only inside a $\lambda$-abstraction.

Burge's specification of the SECD machine with J follows.  Ellipses mark what
does not change from the specification of
Section~\ref{subsec:prerequisites-and-domain-of-discourse-fc}:

% \begin{quote}
\begin{verbatim}
  (*  run : S * E * C * D -> value                                             *)
  fun run (v :: nil, e, nil, d)
      = ...
    | run (s, e, (TERM t) :: c, d)
      = ...
    | run (SUCC :: (INT n) :: s, e, APPLY :: c, d)
      = ...
    | run ((FUNCLO (e', x, t)) :: v :: s, e, APPLY :: c, d)
      = ...
    | run ((STATE_APPENDER d') :: v :: s, e, APPLY :: c, d)
      = ...
    | run ((PGMCLO (v, (s', e', c') :: d')) :: v' :: s, e, APPLY :: c, d)
      = run (v :: v' :: s', e', APPLY :: c', d')
\end{verbatim}
% \end{quote}

% \begin{quote}
\begin{verbatim}
  fun evaluate0_alt t                    (*  evaluate0_alt : program -> value  *)
      = ...
\end{verbatim}
% \end{quote}

\noindent
Just as in
Section~\ref{subsec:SECD-description-disentangled}, Burge's specification
can be disentangled into four mutually-recursive transition functions.  The
disentangled specification, however, is not in defunctionalized form.
We put it next
in defunctionalized form without
resorting to a control delimiter, and
then
outline the rest of the rational deconstruction.

\subsection{Burge's specification in defunctionalized form}
\label{subsec:Burge-SECD-modified}

The disentangled specification of Burge is not in
defunctionalized form because the dump does not have a single point of
consumption.  It is consumed by \stt{run\_d} for values yielded by the body
of $\lambda$-abstractions and in \stt{run\_a} for values thrown to program
closures.  In order to be in the image of defunctionalization and
have \stt{run\_d} as the apply function, the dump should be solely consumed
by \stt{run\_d}.  We therefore distinguish
values yielded by normal evaluation and values thrown to program
closures, and we make \stt{run\_d} dispatch over these two kinds of
returned values.  For values yielded by normal evaluation (\ie, in the
call from \stt{run\_c} to \stt{run\_d}), \stt{run\_d} proceeds as before.
For values thrown to program closures, \stt{run\_d} calls \stt{run\_a}.
Our modification therefore adds one transition (from \stt{run\_a} to
\stt{run\_d}) for values thrown to program closures.

The change only concerns three clauses and ellipses mark what does not
change from the evaluator of
Section~\ref{subsec:SECD-description-disentangled}:

% \begin{quote}
\begin{verbatim}
  datatype returned_value = YIELD of value
                          | THROW of value * value
\end{verbatim}
% \end{quote}

% \begin{quote}
\begin{verbatim}
  (*  run_c :                 S * E * C * D -> value                           *)
  (*  run_d :            returned_value * D -> value                           *)
  (*  run_t :          term * S * E * C * D -> value                           *)
  (*  run_a : value * value * S * E * C * D -> value                           *)
  fun run_c (v :: nil, e, nil, d)
      = run_d (YIELD v, d)                                                (* 1 *)
    | run_c ...
      = ...
  and run_d (YIELD v, nil)
      = v
    | run_d (YIELD v, (s, e, c) :: d)
      = run_c (v :: s, e, c, d)
    | run_d (THROW (v, v'), (s, e, c) :: d)
      = run_a (v, v', s, e, c, d)                                         (* 2 *)
  and run_t ...
      = ...
  and run_a ...
      = ...
    | run_a (PGMCLO (v, d'), v', s, e, c, d)
      = run_d (THROW (v, v'), d')                                         (* 3 *)
\end{verbatim}
% \end{quote}

% \begin{quote}
\begin{verbatim}
  fun evaluate1_alt t                    (*  evaluate1_alt : program -> value  *)
      = ...
\end{verbatim}
% \end{quote}

%

\noindent
\stt{YIELD} is used to tag values returned
by function closures
(in the clause marked ``\stt{1}'' above), and
\stt{THROW} is used to tag values sent to
program closures (in
the clause marked ``\stt{3}'').  \stt{THROW} tags a pair of values, which
will be applied in \stt{run\_d} (by calling \stt{run\_a} in the clause
marked ``\stt{2}'').

\begin{prop}[full correctness]
Given a program, \stt{evaluate0\_alt} and \stt{evalu\-ate1\_alt} either both
diverge or both yield values that are structurally equal.
\end{prop}

\subsection{A higher-order counterpart}
\label{subsec:Burge-SECD-refunctionalized}

In the modified specification of Section~\ref{subsec:Burge-SECD-modified},
the data types of control stacks and dumps are identical to those of the
disentangled machine of Section~\ref{subsec:SECD-description-disentangled}.
These data types, together with \stt{run\_d} and \stt{run\_c}, are in the
image of defunctionalization (\stt{run\_d} and \stt{run\_c} are their apply
functions).  The corresponding higher-order
evaluator reads as follows:

%

% \begin{quote}
\begin{verbatim}
  datatype value = INT of int
                 | SUCC
                 | FUNCLO of E * string * term
                 | STATE_APPENDER of D
                 | PGMCLO of value * D
       and returned_value = YIELD of value
                          | THROW of value * value
  withtype S = value list                                        (* data stack *)
       and E = value Env.env                                    (* environment *)
       and D = returned_value -> value                    (* dump continuation *)
       and C = S * E * D -> value                      (* control continuation *)
\end{verbatim}
% \end{quote}

% \begin{quote}
\begin{verbatim}
  val e_init = Env.extend ("succ", SUCC, Env.empty)
\end{verbatim}
% \end{quote}

% \begin{quote}
\begin{verbatim}
  (*  run_t :          term * S * E * C * D -> value                           *)
  (*  run_a : value * value * S * E * C * D -> value                           *)
  (*  where S = value list, E = value Env.env, C = S * E * D -> value          *)
  (*    and D = returned_value -> value                                        *)
  fun run_t ...
      = ...
\end{verbatim}
% \end{quote}

% \begin{quote}
\begin{verbatim}
  and run_a (SUCC, INT n, s, e, c, d)
      = c ((INT (n+1)) :: s, e, d)
    | run_a (FUNCLO (e', x, t), v, s, e, c, d)
      = run_t (t, nil, Env.extend (x, v, e'),
               fn (v :: nil, e, d) => d (YIELD v),
               fn (YIELD v)
                  => c (v :: s, e, d)
                | (THROW (f, v))
                  => run_a (f, v, s, e, c, d))
    | run_a (STATE_APPENDER d', v, s, e, c, d)
      = c ((PGMCLO (v, d')) :: s, e, d)
    | run_a (PGMCLO (v, d'), v', s, e, c, d)
      = d' (THROW (v, v'))
\end{verbatim}
% \end{quote}

% \begin{quote}
\begin{verbatim}
  fun evaluate2_alt t                    (*  evaluate2_alt : program -> value  *)
      = run_t (t, nil, e_init, fn (v :: nil, e, d) => d (YIELD v),
               fn (YIELD v) => v)
\end{verbatim}
% \end{quote}

%

\noindent
As before, the resulting evaluator is in continuation-passing style (CPS),
with two layered continuations.  It threads a stack of intermediate results,
a (callee-save) environment, a control continuation, and a dump
continuation.  The values sent to dump continuations are tagged to indicate
whether they represent the result of a function closure or an application of a
program closure.  Defunctionalizing this evaluator yields the definition of
Section~\ref{subsec:Burge-SECD-modified}:

\begin{prop}[full correctness]
Given a program,
\stt{evaluate1\_alt} and \stt{evaluate2\_alt}
either both diverge
or yield expressible values; and if these values have an integer type, they are
the same integer.
\end{prop}

\subsection{The rest of the rational deconstruction}
\label{subsec:the-rest-of-the-rational-deconstruction-Burge}

The evaluator of Section~\ref{subsec:Burge-SECD-refunctionalized} can be
transformed exactly as the higher-order evaluator of
Section~\ref{subsec:SECD-refunctionalized}:

\begin{enumerate}
\item
Eliminating the data stack and the callee-save environment yields a
traditional eval--apply evaluator, with \stt{run\_t} as eval and
\stt{run\_a} as apply.  The evaluator is in CPS with two layers of
continuations.

\item
A first direct-style transformation
with respect to the dump yields an
evaluator that uses shift and reset
(or $\rawC$ and a global reset, or again call/cc and a global reset)
to manipulate the
implicit
dump continuation.

\item 
A second direct-style transformation with respect to the control stack
yields an evaluator in direct style that uses the delimited-control
operators shift$_1$, reset$_1$, shift$_2$, and reset$_2$
(or $\rawC_1$, reset$_1$, $\rawC_2$, and reset$_2$)
to manipulate the
implicit
control and dump continuations.

\item
Refunctionalizing the applicable values yields a compositional,
higher-order, direct-style evaluator corresponding to Burge's specification
of the J operator.  The result is presented as a syntax-directed encoding
next.

\end{enumerate}

\subsection{Three simulations of the J operator}

As in Section~\ref{subsubsec:three-simulations-of-J},
the compositional counterpart of the evaluators of
Section~\ref{subsec:the-rest-of-the-rational-deconstruction-Burge}
can be viewed as syntax-directed encodings
into their meta-language.
Below, we state these encodings as three
simulations of J: one in direct style, one in CPS with one layer of
continuations, and one in CPS with two layers of continuations.
Again, we assume a call-by-value meta-language with right-to-left
evaluation and with a sum (to distinguish values returned by functions and
values sent to program closures), a case expression (for the body of
$\lambda$-abstractions) and a destructuring let expression (at the top
level).

\begin{enumerate}[$\bullet$]

\item
In direct style, using either of shift$_2$, reset$_2$, shift$_1$, and
reset$_1$ or $\rawC_2$, reset$_2$, $\rawC_1$, and reset$_1$, based on
the compositional evaluator in direct style:
\[
  \begin{array}{r@{\ }c@{\ }l@{\hspace{8.5mm}}}
  \dmszero{\synlit}
  & = &
  \synlit
  \\
  \dmszero{\synvar}
  & = &
  \synvar
  \\
  \dmszero{\synapp{\synterm_0}
                  {\synterm_1}}
  & = &
  \synapp{\dmszero{\synterm_0}}{\dmszero{\synterm_1}}
  \\
  \dmszero{\synlam{\synvar}
                  {\synterm}}
  & = &
  \synlam{\synvar}
         {\syncase{\synresetn{1}
                             {\inleft{\dmszero{\synterm}}}}
                  {\synvarvalp}
                  {\synvarval}
                  {\synpair{\synvarval}
                           {\synvarval'}}
                  {\synapp{\synvarval}
                          {\synvarval'}}}
  \\
  \dmszero{\rawJ}
  & = &
  \synshiftn{1}
            {\synvarcont}
            {\synshiftn{2}
                       {\synvarmcont}
                       {\synapp{\synvarmcont}
                               {\synappp{\synvarcont}
                                        {\synlam{\synvarval}
                                                 {\ourbox{\synlam{\synvarval'}
                                                         {\synshiftn{1}
                                                                    {\synvarcont'}
                                                         {\synshiftn{2}
                                                                    {\synvarmcont'}
                                                                    {\synapp{\synvarmcont}
                                                                            {\inrightp{\synpair{\synvarval}
                                                                                               {\synvarval'}}}}}}}}}}}}
  \\[0.5mm]
  & = &
  \hspace{0.35mm}\synCn{1}
        {\synvarcont}
        {\hspace{0.35mm}\synCn{2}
               {\synvarmcont}
               {\synapp{\synvarmcont}
                       {\synappp{\synvarcont}
                                {\synlam{\synvarval}
                                         {\ourbox{\synlam{\synvarval'}
                                                 {\synapp{\synvarmcont}
                                                         {\inrightp{\synpair{\synvarval}
                                                                            {\synvarval'}}}}}}}}}}
  \end{array}
\]

\noindent
A
program $p$ is translated as
$\synresetn{2}
           {\synhlet{\inleft{\synvarvalp}}
                    {\synresetn{1}
                               {\inleft{\dmszerop{p}}}}
                    {\synvarval}}$.

\item
In CPS with one layer of continuations, using either of shift
and reset, $\rawC$ and reset, or call/cc and reset, based on the
compositional evaluator in CPS with one layer of continuations:
\[
  \begin{array}{r@{\ }c@{\ }l@{\hspace{12mm}}}
  \dmsone{\synlit}
  & = &
  \synlam{\synvarcont}
         {\synapp{\synvarcont}
                 {\synlit}}
  \\
  \dmsone{\synvar}
  & = &
  \synlam{\synvarcont}
         {\synapp{\synvarcont}
                 {\synvar}}
  \\
  \dmsone{\synapp{\synterm_0}{\synterm_1}}
  & = &
  \synlam{\synvarcont}
         {\synapp{\dmsone{\synterm_1}}
                 {\synlamp{\synvarval_1}
                          {\synapp{\dmsone{\synterm_0}}
                                  {\synlam{\synvarval_0}
                                          {\synapp{\synapp{\synvarval_0}
                                                          {\synvarval_1}}
                                                  {\synvarcont}}}}}}
  \\
  \dmsone{\synlam{\synvar}
                 {\synterm}}
  & = &
  \synlam{\synvarcont}
         {\synapp{\synvarcont}
                 {(\synlam{\synvar}
                          {\synlam{\synvarcont}
                          {\syncase{\synapp{\dmsone{t}}
                                           {\synlam{\synvarval}
                                                   {\inleft{\synvarvalp}}}}
                                   {\synvarvalp}
                                   {\synapp{\synvarcont}
                                           {\synvarval}}
                                   {\synpair{\synvarval}
                                            {\synvarval'}}
                                   {\synapp{\synapp{\synvarval}
                                                   {\synvarval'}}
                                           {\synvarcont})}}}}}
  \end{array}
\]

\[
  \begin{array}{r@{\ }c@{\ }l@{\hspace{12mm}}}
  \dmsone{\rawJ}
  & = &
  \synlam{\synvarcont}
    {\synshift{\synvarmcont}
      {\synapp
        {\synvarmcont}
        {\synappp{\synvarcont}
                 {\synlam{\synvarval}
                   {\synlam{\synvarcont}
                     {\synapp{\synvarcont}
                             {\ourbox{\synlam{\synvarval'}
                                     {\synlam{\synvarcont'}
                                       {\synshift{\synvarmcont'}
                                         {\synapp{\synvarmcont}
                                                 {\inrightp{\synpair{\synvarval}
                                                                    {\synvarval'}}}}}}}}}}}}}}
  \\[0.5mm]
  & = &
  \synlam{\synvarcont}
    {\hspace{0.35mm}\synC{\synvarmcont}
      {\synapp
        {\synvarmcont}
        {\synappp{\synvarcont}
                 {\synlam{\synvarval}
                   {\synlam{\synvarcont}
                     {\synapp{\synvarcont}
                             {\ourbox{\synlam{\synvarval'}
                                     {\synlam{\synvarcont'}
                                       {\synapp{\synvarmcont}
                                                 {\inrightp{\synpair{\synvarval}
                                                                    {\synvarval'}}}}}}}}}}}}}
  \\[0.5mm]
  & = &
  \synlam{\synvarcont}
    {\synapp{\rawcallcc}{\synlam{\synvarmcont}
        {\synapp{\synvarcont}
                 {\synlam{\synvarval}
                   {\synlam{\synvarcont}
                     {\synapp{\synvarcont}
                             {\ourbox{\synlam{\synvarval'}
                                     {\synlam{\synvarcont'}
                                       {\synapp{\synvarmcont}
                                                 {\inrightp{\synpair{\synvarval}
                                                                    {\synvarval'}}}}}}}}}}}}}
  \end{array}
\]

\noindent
A
program $p$ is translated as
$\synreset{\synhlet{\inleft{\synvarvalp}}
                   {\synapp{\dmsone{p}}
                           {\synlam{\synvarval}
                                   {\inleft{\synvarvalp}}}}
                   {\synvarval}}$.

\item
In CPS with two layers of continuations, based on the compositional
evaluator in CPS with two layers of continuations:
\[
  \begin{array}{@{\hspace{-5.25mm}}r@{\ }c@{\ }l@{\hspace{5.25mm}}}
  \dmstwo{\synlit}
  & = &
  \synlam{\synvarcont}
         {\synlam{\synvarmcont}
                 {\synapp{\synapp{\synvarcont}
                                 {\synlit}}
                         {\synvarmcont}}}
  \\
  \dmstwo{\synvar}
  & = &
  \synlam{\synvarcont}
         {\synlam{\synvarmcont}
                 {\synapp{\synapp{\synvarcont}
                                 {\synvar}}
                         {\synvarmcont}}}
  \\
  \dmstwo{\synapp{\synterm_0}{\synterm_1}}
  & = &
  \synlam{\synvarcont}
    {\synlam{\synvarmcont}
      {\synapp{\synapp{\dmstwo{\synterm_1}}
                      {\synlamp{\synvarval_1}
                        {\synlam{\synvarmcont}
                          {\synapp{\synapp{\dmstwo{\synterm_0}}
                                          {\synlamp{\synvarval_0}
                                            {\synlam{\synvarmcont}
                                              {\synapp{\synapp{\synapp{\synvarval_0}
                                                                      {\synvarval_1}}
                                                              {\synvarcont}}
                                                      {\synvarmcont}}}}}
                                  {\synvarmcont}}}}}
              {\synvarmcont}}}
  \end{array}
\]

\[
  \begin{array}{@{\hspace{5.5mm}}r@{\ }c@{\ }l@{}}
  \dmstwo{\synlam{\synvar}{\synterm}}
  & = &
  \synlam{\synvarcont}
    {\synlam{\synvarmcont}
      {\synapp{\synvarcont}
                      {(\synlam{\synvar}
                        {\synlam{\synvarcont}
                          {\synlam{\synvarmcont}
                            {\synapp{\dmstwo{\synterm}}
                                    {\begin{array}[t]{@{}l}
                                     \synlamp{\synvarval}
                                             {\synlam{\synvarmcont}
                                                     {\synapp{\synvarmcont}
                                                             {\inleftp{\synvarvalp}}}}
                                     \\
                                     \synlam{\synvarval''}
                                      {\syncase{\synvarval''}
                                               {\synvarvalp}
                                               {\synapp{\synapp{\synvarcont}
                                                               {\synvarval}}
                                                       {\synvarmcont}}
                                               {\synpair{\synvarval}
                                                        {\synvarval'}}
                                               {\synapp{\synapp{\synapp{\synvarval}
                                                                       {\synvarval'}}
                                                               {\synvarcont}}
                                                       {\synvarmcont})\,{\synvarmcont}}}
				     \end{array}}}}}}}}
  \\
  \dmstwo{\rawJ}
  & = &
  \synlam{\synvarcont}
    {\synlam{\synvarmcont}
      {\synapp{\synapp{\synvarcont}
                      {\synlamp{\synvarval}
                        {\synlam{\synvarcont}
                          {\synlam{\synvarmcont'''}
                            {\synapp{\synapp{\synvarcont}
                                            {\ourbox{\synlamp{\synvarval'}
                                                    {\synlam{\synvarcont'}
                                                      {\synlam{\synvarmcont'}
                                                        {\synapp{\synvarmcont}
                                                                {\inrightp{\synpair{\synvarval}
                                                                                   {\synvarval'}}}}}}}}}
                                    {\synvarmcont'''}}}}}}
              {\synvarmcont}}}
  \end{array}
\]

\noindent
A
program $p$ is translated as
$\synapp{\synapp{\dmstwo{p}}
                {\synlamp{\synvarval}
                         {\synlam{\synvarmcont}
                                 {\synapp{\synvarmcont}
                                         {\inleftp{\synvarvalp}}}}}}
        {\synlamp{\synvarval}
                 {\synhlet{\inleft{(\synvarval')}}
                          {\synvarval}
                          {\synvarval'}}}$.

\end{enumerate}

\paragraph{Analysis:}
The simulation of literals, variables, and applications is standard.  The
body of each $\lambda$-abstraction is evaluated with a control
continuation injecting the resulting value into the sum
type\footnote{This machine is therefore not properly tail recursive.}
to indicate normal completion and resuming the current dump
continuation, and with a dump continuation
inspecting the resulting sum to determine whether to continue normally
or to apply a program closure.  Continuing normally consists of
invoking the control continuation with the resulting value and the dump
continuation.  Applying a program closure consists of restoring
the components of the dump and then performing the application.  The J
operator abstracts both the control continuation and
the dump continuation and immediately restores them, resuming the
computation with a state appender holding the abstracted dump
continuation captive.  Applying this state appender to a value $v$ yields a
program closure (boxed in the three simulations above).  Applying this
program closure to a value $v'$ has the effect of
discarding both the
current control continuation and the current dump continuation, 
injecting $v$ and $v'$ into the sum type to indicate
exceptional completion, and resuming the captured dump continuation.
It is an error to evaluate J outside of a $\lambda$-abstraction.

\subsection{Related work}

Kiselyov's encoding of dynamic delimited continuations in terms of the
static delimited-continuation operators shift and
reset~\cite{Kiselyov:TR05} is similar to this alternative encoding of
the J operator in that both encodings tag the argument to the
meta-continuation to indicate whether it represents a normal return or
a value thrown to a first-class continuation.
In addition though, Kiselyov uses a recursive meta-continuation
in order to encode dynamic delimited continuations.

\renewcommand{\contractsto}[1]{\ensuremath{\mapsto}}
\newcommand{\plugsto}{\ensuremath{\rightarrow}}
\newcommand{\decomposesto}{\ensuremath{\rightarrow}}
\newcommand{\syndompotred}{\mathrm{PotRed}}
\newcommand{\syndomclos}{\mathrm{Closure}}
\newcommand{\syndomval}{\mathrm{Value}}
\newcommand{\syndomcontcont}{\mathrm{Control}}
\newcommand{\syndomdumpcont}{\mathrm{Dump}}

\newcommand{\rawvaluetoclosure}{\uparrow}
\newcommand{\evaluetoclosure}[1]{{\rawvaluetoclosure} \, {#1}}

\newcommand{\rawplug}{\mathsf{plug}}
\newcommand{\rawcontract}{\mathsf{contract}}
\newcommand{\rawdecompose}{\mathsf{decompose}}

\newcommand{\rawsecddecompose}{\mathsf{decompose}}
\newcommand{\rawsecddecomposep}{\mathsf{decompose}'_{\mathrm{clos}}}
\newcommand{\rawsecddecomposepaux}{\mathsf{decompose}'_{\mathrm{cont}}}
\newcommand{\rawsecddecomposepauxaux}{\mathsf{decompose}'_{\mathrm{dump}}}

\newcommand{\varpotred}{r}

\newcommand{\eplugone}[1]{\rawplug\,({#1})}
\newcommand{\eplugtwo}[2]{\rawplug\,({#1},\,{#2})}
\newcommand{\eplug}[3]{\rawplug\,({#1},\,{#2},\,{#3})}
\newcommand{\econtract}[3]{\rawcontract\,({#1},\,{#2},\,{#3})}
\newcommand{\econtractone}[1]{\rawcontract\,({#1})}
\newcommand{\econtracttwo}[2]{\rawcontract\,({#1},\,{#2})}
\newcommand{\edecompose}[1]{\rawdecompose\,({#1})}
\newcommand{\secddecomposep}[3]{\rawsecddecomposep\,({#1},\,{#2},\,{#3})}
\newcommand{\secddecomposepaux}[3]{\rawsecddecomposepaux\,({#1},\,{#2},\,{#3})}
\newcommand{\secddecomposepauxaux}[2]{\rawsecddecomposepauxaux\,({#1},\,{#2})}

\newcommand{\rawreduce}{\mathsf{reduce}}
\newcommand{\ereduce}[1]{\rawreduce\,({#1})}

\newcommand{\rawiterate}{\mathsf{iterate}}
\newcommand{\eiterate}[1]{\rawiterate\,({#1})}

\newcommand{\rawevaluate}{\mathsf{evaluate}}
\newcommand{\eevaluate}[1]{\rawevaluate\,({#1})}

\newcommand{\rawrefocus}{\mathsf{refocus}}
\newcommand{\erefocus}[3]{\rawrefocus\,({#1},\,{#2},\,{#3})}
\newcommand{\erefocusone}[1]{\rawrefocus\,({#1})}

\newcommand{\synboundary}[1]{\ensuremath{\langlethick{#1}\ranglethick}}

\newcommand{\decthreep}[3]{\decthree{#1}{#2}{#3}'}
\newcommand{\decthreepaux}[3]{\decthree{#1}{#2}{#3}'_{\mathrm{aux}}}

\newcommand{\dectwopaux}[2]{\dectwo{#1}{#2}'_{\mathrm{aux}}}

\newcommand{\secddecompositionSOME}[3]{\mathrm{DEC}\:({#1},\,{#2},\,{#3})}
\newcommand{\secddecompositionNONE}[1]{\mathrm{VAL}\:({#1})}

\newcommand{\secddecompositionSOMEtwo}[2]{\mathrm{DEC}\:({#1},\,{#2})}

\newcommand{\vcase}[5]{\begin{array}[t]{@{}r@{\ }l@{\ }c@{\ }l}
                       \mathbf{case} & \multicolumn{3}{@{}l}{#1}
                       \\
                       \mathbf{of}   & {#2} & \Rightarrow & {#3}
                       \\
                       \mathbf{\mid} & {#4} & \Rightarrow & {#5}
                       \end{array}}

\section{A syntactic theory of applicative expressions with the J operator:
         explicit, callee-save dumps}
\label{sec:syntactic-theory-callee-save-dumps}

Symmetrically to the functional correspondence between evaluation
functions and abstract machines that was sparked by the first rational
deconstruction of the SECD machine ~\cite{Ager:PhD, Ager-al:PPDP03,
Ager-al:IPL04, Ager-al:TCS05, Biernacka-al:LMCS05, Biernacki:PhD,
Danvy:IFL04, Danvy:DSc}, a syntactic correspondence exists between
calculi and abstract machines, as investigated by Biernacka, Danvy, and
Nielsen~\cite{Biernacka:PhD, Biernacka-Danvy:TCS07,
Biernacka-Danvy:TOCL07, Danvy:WRS04, Danvy:DSc, Danvy-Nielsen:RS-04-26}.
This syntactic correspondence is also derivational, and hinges not on
defunctionalization but on a `refocusing' transformation that
mechanically connects an evaluation function defined as the
iteration of one-step reduction, and an abstract machine.

The goal of this section is to present the reduction semantics
and the reduction-based evaluation function that correspond to the
modernized SECD machine of
Section~\ref{subsec:a-specification-with-no-data-stack-and-caller-save-environments}.
We successively present this machine
(Section~\ref{subsec:transmogrified-SECD-machine}), the syntactic
correspondence (Section~\ref{subsec:from-redsem-to-absmac}), 
a reduction semantics for applicative expressions with the J operator
(Section~\ref{subsec:syntactic-theory-J}), and the derivation from this
reduction semantics to this SECD machine
(Section~\ref{subsec:from-red-sem-to-secd-mach}).  We consider a
calculus of explicit substitutions because the explicit substitutions
directly correspond to the environments of the modernized SECD machine.
In turn, this calculus of explicit substitutions directly corresponds to
a calculus with actual substitutions.

\newcommand{\secdconfeval}[4]{{\confquadruple{#1}{#2}{#3}{#4}}_{\mathrm{eval}}}
\newcommand{\secdconfapply}[4]{{\confquadruple{#1}{#2}{#3}{#4}}_{\mathrm{apply}}}
\newcommand{\secdconfcont}[3]{{\conftriple{#1}{#2}{#3}}_{\mathrm{cont}}}
\newcommand{\secdconfdump}[2]{{\confpair{#1}{#2}}_{\mathrm{dump}}}

\subsection{The SECD machine with no data stack and caller-save environments,
            revisited}
\label{subsec:transmogrified-SECD-machine}

The terms, values, environments, and contexts are defined as in
Section~\ref{subsec:prerequisites-and-domain-of-discourse-fc}:

\[
\begin{array}{@{}rc@{\ }c@{\ }l@{\hspace{2.4cm}}}
\textrm{(programs)} & p & ::= &
\esub{\tm}{\envextend{\synsucc}{\semsucc}{\envempty}}
\\[0.5mm]
\textrm{(terms)} & \tm & ::= &
\represent{n} \Mid x \Mid \synlam{x}{\tm} \Mid \app{\tm}{\tm} \Mid J
\\[0.5mm]
\textrm{(values)} & \val & ::= &
\represent{n}
\Mid
\semsucc
\Mid
\synpair{\synlam{x}{\tm}}{\sub}
\Mid
\stateappend{\represent{\ctxtwo}}{\val}
\Mid
\represent{\ctxtwo}
\\[0.5mm]
\textrm{(environments)}
& \sub & ::= &
\envempty \Mid \envextend{x}{\val}{\sub}
\\[0.5mm]
\textrm{(control contexts)} & \ctxone & ::= &
\mtctx \Mid
\apctx{\ctxone}{\synpair{\tm}{\sub}} \Mid
\argctx{\ctxone}{\val}
\\[0.5mm]
\textrm{(dump contexts)} & \ctxtwo & ::= &
\mtmetactx \Mid \metactx{\ctxtwo}{\ctxone}
\end{array}
\]

\noindent
The following four transition functions are the stackless, caller-save
respective counterparts of \stt{run\_t}, \stt{run\_a}, \stt{run\_c}, and
\stt{run\_d} in Section~\ref{subsec:SECD-description-disentangled}.
This abstract machine
is implemented by the modernized and disentangled evaluator
\stt{evaluate1'} in the diagram at the end of
Section~\ref{subsec:a-specification-with-no-data-stack-and-caller-save-environments}:

\[
\begin{array}{@{}r@{\hspace{1.75mm}}l@{\hspace{1.75mm}}l@{}l}
\namedtransition{}
{\secdconfeval{\represent{n}}{\sub}{\ctxone}{\ctxtwo}}
{\secdconfcont{\ctxone}{\represent{n}}{\ctxtwo}}
\\[1mm]
\namedtransition{}
{\secdconfeval{x}{\sub}{\ctxone}{\ctxtwo}}
{\secdconfcont{\ctxone}{\val}{\ctxtwo}}
&
\textrm{if }lookup (x, \sub) = \val
\\[1mm]
\namedtransition{}
{\secdconfeval{\synlam{x}{\tm}}{\sub}{\ctxone}{\ctxtwo}}
{\secdconfcont{\ctxone}{\synpair{\synlam{x}{\tm}}{\sub}}{\ctxtwo}}
\\[1mm]
\namedtransition{}
{\secdconfeval{\app{\tmone}{\tmtwo}}{\sub}{\ctxone}{\ctxtwo}}
{\secdconfeval{\tmtwo}{\sub}{\apctx{\ctxone}{\synpair{\tmone}{\sub}}}{\ctxtwo}}
\\[1mm]
\namedtransition{}
{\secdconfeval{J}{\sub}{\ctxone}{\ctxtwo}}
{\secdconfcont{\ctxone}{\represent{\ctxtwo}}{\ctxtwo}}
\\[4mm]
\namedtransition{}
{\secdconfapply{\semsucc}{\represent{n}}{\ctxone}{\ctxtwo}}
{\secdconfcont{\ctxone}{\represent{n+1}}{\ctxtwo}}
\\[1mm]
\namedtransition{}
{\secdconfapply{\synpair{\synlam{x}{\tm}}{\sub}}{\val}{\ctxone}{\ctxtwo}}
{\secdconfeval{\tm}{\sub'}{\mtctx}{\metactx{\ctxtwo}{\ctxone}}}
&
\textrm{where }\sub' = extend (x, \val, \sub)
\\[1mm]
\namedtransition{}
{\secdconfapply{\stateappend{\represent{\ctxtwo'}}{\val'}}{\val}{\ctxone}{\ctxtwo}}
{\secdconfapply{\val}{\val'}{\mtctx}{\ctxtwo'}}
\\[1mm]
\namedtransition{}
{\secdconfapply{\represent{\ctxtwo'}}{\val}{\ctxone}{\ctxtwo}}
{\secdconfcont{\ctxone}{\stateappend{\represent{\ctxtwo'}}{\val}}{\ctxtwo}}
\\[4mm]
\namedtransition{}
{\secdconfcont{\mtctx}{\val}{\ctxtwo}}
{\secdconfdump{\ctxtwo}{\val}}
\\[1mm]
\namedtransition{}
{\secdconfcont{\apctx{\ctxone}{\synpair{\tm}{\sub}}}{\val}{\ctxtwo}}
{\secdconfeval{\tm}{\sub}{\argctx{\ctxone}{\val}}{\ctxtwo}}
\\[1mm]
\namedtransition{}
{\secdconfcont{\argctx{\ctxone}{\val'}}{\val}{\ctxtwo}}
{\secdconfapply{\val}{\val'}{\ctxone}{\ctxtwo}}
\\[4mm]
\namedtransition{}
{\secdconfdump{\mtmetactx}{\val}}
\val\\[1mm]
\namedtransition{}
{\secdconfdump{\metactx{\ctxtwo}{\ctxone}}{\val}}
{\secdconfcont{\ctxone}{\val}{\ctxtwo}}
\\[4mm]
\end{array}
\]

\noindent
% This machine evaluates a
% %
% program $\tm$
A program $\tm$ is evaluated
by starting in the
configuration
$\secdconfeval{\tm}
              {\envextend{\synsucc}{\semsucc}{\envempty}}
              {\mtctx}
              {\mtmetactx}$.
% It
The machine halts with a value $\val$ if it reaches a configuration
$\secdconfdump{\mtmetactx}{\val}$.

\subsection{From reduction semantics to abstract machine}
\label{subsec:from-redsem-to-absmac}

Consider a calculus together with a reduction strategy expressed as a
Felleisen-style
reduction semantics satisfying the unique-decomposition
property~\cite{Felleisen:PhD}.  In such a reduction semantics,
a one-step reduction function is defined
as the composition of three functions:

\begin{description}

\item[decomposition]
a total function mapping a value term to itself and decomposing
a non-value term into a potential redex and a reduction context
(decomposition is a function because of the unique-decomposition property);

\item[contraction]
a partial function mapping an actual redex to its contractum; and

\item[plugging]
a total function mapping a term and a reduction context to a new term by
filling the hole in the context with the term.

\end{description}

\noindent
The one-step reduction function is partial because it is the composition
of two total functions and a partial function.

An 
evaluation function is traditionally defined 
as
the
iteration of the one-step reduction function:
{\let\labelstyle=\textstyle
\spreaddiagramrows{-0.1cm}
\spreaddiagramcolumns{0.33cm}
 \newcommand{\atadless}{\hspace{-2mm}}
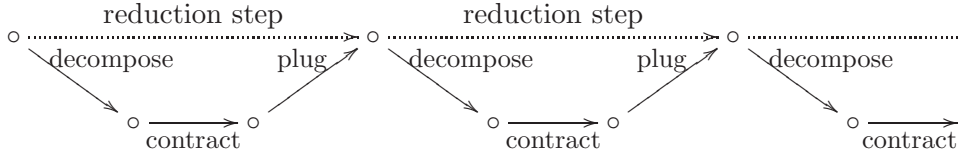
 $$
 \diagram
 \circ
 \drto^{\atadless\text{{\small decompose}}}
 \ar@{..>}[rrr]^{\text{reduction step}}
 &
 &
 &
 \circ
 \drto^{\atadless\text{{\small decompose}}}
 \ar@{..>}[rrr]^{\text{reduction step}}
 &
 &
 &
 \circ
 \drto^{\atadless\text{{\small decompose}}}
 \ar@{..}[rr]
 &
 &
 \\
 &
 \circ
 \rto_{\text{{\small contract}}}
 &
 \circ
 \urto^{\text{{\small plug}}\atadless}
 &
 &
 \circ
 \rto_{\text{{\small contract}}}
 &
 \circ
 \urto^{\text{{\small plug}}\atadless}
 &
 &
 \circ
 \rto_{\text{{\small contract}}}
 &
 \enddiagram
 $$
}

\noindent
Danvy and Nielsen have observed that composing the two total functions
plug and decompose into a `refocus' function could avoid the construction
of intermediate terms:
{\let\labelstyle=\textstyle
\spreaddiagramrows{-0.1cm}
\spreaddiagramcolumns{0.33cm}
 \newcommand{\atadless}{\hspace{-2mm}}
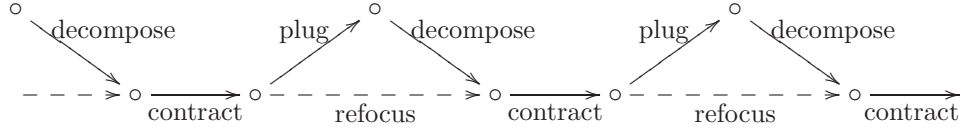
 $$
 \diagram
 \circ
 \drto^{\atadless\text{{\small decompose}}}
 &
 &
 &
 \circ
 \drto^{\atadless\text{{\small decompose}}}
 &
 &
 &
 \circ
 \drto^{\atadless\text{{\small decompose}}}
 &
 &
 \\
 \ar@{-->}[r]
 &
 \circ
 \rto_{\text{{\small contract}}}
 &
 \circ
 \urto^{\text{{\small plug}}\atadless}
 \ar@{-->}[rr]_{\text{{\small refocus}}}
 &
 &
 \circ
 \rto_{\text{{\small contract}}}
 &
 \circ
 \urto^{\text{{\small plug}}\atadless}
 \ar@{-->}[rr]_{\text{{\small refocus}}}
 &
 &
 \circ
 \rto_{\text{{\small contract}}}
 &
 \enddiagram
 $$
}

\noindent
The resulting `refocused' evaluation function is defined as the
iteration of refocusing and contraction.  CPS transformation
and defunctionalization make it take the form of a state-transition
function, \ie, an abstract machine.  Short-circuiting its intermediate
transitions yields abstract machines that are often independently
known~\cite{Danvy-Nielsen:RS-04-26}.

Biernacka and Danvy then showed that the refocusing technique could be
applied to the very first calculus of explicit substitutions, Curien's
simple calculus of closures~\cite{Curien:TCS91}, and that depending on
the reduction order, it gave rise to a collection of both known and new
environment-based abstract machines such as Felleisen et al.'s CEK
machine (for left-to-right applicative order), the Krivine machine (for
normal order), Krivine's machine (for normal order with generalized
reduction), and Leroy's ZINC machine (for right-to-left applicative order
with generalized reduction)~\cite{Biernacka-Danvy:TOCL07}.  They then
turned to context-sensitive contraction functions, as first proposed by
Felleisen~\cite{Felleisen:PhD}, and showed that refocusing mechanically
gives rise to an even larger
collection of both known and new environment-based abstract machines for
languages with computational effects such as Krivine's machine with
call/cc, the $\lambda\mu$-calculus, static and dynamic delimited
continuations, input/output, stack inspection, proper tail-recur\-sion, and lazy
evaluation~\cite{Biernacka-Danvy:TCS07}.

The next section presents the calculus of closures corresponding to the
abstract machine of Section~\ref{subsec:transmogrified-SECD-machine}.

\subsection{A reduction semantics for applicative expressions with the J operator}
\label{subsec:syntactic-theory-J}

The $\lrhJ$-calculus is an extension of Biernacka and Danvy's
$\lrh$-calculus~\cite{Biernacka-Danvy:TOCL07}, which is itself a minimal
extension of Curien's original calculus of closures
$\lrho$~\cite{Curien:TCS91} to make it closed under one-step reduction.
We use it here to formalize Landin's applicative expressions with the J
operator as a reduction semantics.
To this end, we present its syntactic categories
(Section~\ref{subsubsec:syntactic-categories}); a plug function mapping a
closure and a two-layered reduction context into a closure by filling the
given context with the given closure (Section~\ref{subsubsec:plugging});
a contraction function implementing a context-sensitive notion of
reduction (Section~\ref{subsubsec:contraction}) and therefore mapping a
potential redex and its reduction context into a contractum and a
reduction context (possibly another one); and a decomposition function
mapping a non-value term into a potential redex and a reduction context
(Section~\ref{subsubsec:decomposition}).  We are then in position to
define a one-step reduction function
(Section~\ref{subsubsec:one-step-reduction}), and a reduction-based
evaluation function (Section~\ref{subsubsec:reduction-based-evaluation}).

Before delving into this section, the reader might want to first
browse through Section~\ref{app:from-RS-to-AM}, in the appendix.  This
section has the same structure as the present one but instead of the SECD
machine, it addresses the CEK machine, which is simpler.

\subsubsection{Syntactic categories}
\label{subsubsec:syntactic-categories}

We consider a variant of the $\lrhJ$-calculus with names instead of de
Bruijn indices, and with two layers of contexts $\ctxone$ and $\ctxtwo$
that embody the right-to-left applicative-order reduction strategy
favored by Landin:
$\ctxone$ is
the control context and
$\ctxtwo$ is the dump context.  In the syntactic category of closures,
$\represent{\ctxtwo}$ and $\stateappend{\represent{\ctxtwo}}{\val}$
respectively denote a state appender and a program closure, and
$\synboundary{\clo}$ (which is shaded below)
marks the boundary between the context of a
$\beta$-redex that has been contracted, \ie, a function closure that has
been applied, and the body of the $\lambda$-abstraction in this function
closure:

\[
\begin{array}{@{}rc@{\ }c@{\ }l@{}}
\textrm{(programs)} & p & ::= &
\esub{\tm}{\envextend{\synsucc}{\semsucc}{\envempty}}
\\[0.5mm]
\textrm{(terms)} & \tm & ::= &
\represent{n} \Mid x \Mid \synlam{x}{\tm} \Mid \app{\tm}{\tm} \Mid J
\\[0.5mm]
\textrm{(closures)} & \clo & ::= &
\represent{n}
\Mid
\semsucc
\Mid
\esub{\tm}{\sub}
\Mid
\comp{\clo}{\clo}
\Mid
\represent{\ctxtwo}
\Mid
\stateappend{\represent{\ctxtwo}}{\val}
\Mid
\shadethis{$\synboundary{\clo}$}
\\[0.5mm]
\textrm{(values)} & \val & ::= &
\represent{n}
\Mid
\semsucc
\Mid
\esub{\synlamp{x}{\tm}}{\sub}
\Mid
\represent{\ctxtwo}
\Mid
\stateappend{\represent{\ctxtwo}}{\val}
\\[0.5mm]
\textrm{(potential redexes)} & \varpotred & ::= &
\esub{x}{\sub} \Mid
\comp{\val}{\val} \Mid
J
\\[0.5mm]
\textrm{(substitutions)} & \sub & ::= &
\envempty \Mid \envextend{x}{\val}{\sub}
\\[0.5mm]
\textrm{(control contexts)} & \ctxone & ::= &
\mtctx \Mid
\apctx{\ctxone}{\clo} \Mid
\argctx{\ctxone}{\val}
\\[0.5mm]
\textrm{(dump contexts)} & \ctxtwo & ::= &
\mtmetactx \Mid \metactx{\ctxtwo}{\ctxone}
\end{array}
\]

\noindent
Values are therefore a syntactic subcategory of closures, and in this
section, we make use of the syntactic coercion $\rawvaluetoclosure$
mapping a value into a closure.

\newcommand{\secdplugconf}[3]{\langle {#1},\,{#2},\,{#3} \rangle_{\mathrm{plug/cont}}}
\newcommand{\secdplugconfp}[2]{\langle {#1},\,{#2} \rangle_{\mathrm{plug/dump}}}

\subsubsection{Plugging}
\label{subsubsec:plugging}

Plugging a closure in the two layered contexts is defined by induction
over these two contexts.  We express this definition as a
state-transition system with two intermediate states,
$\secdplugconf{\ctxone}{\clo}{\ctxtwo}$ and
$\secdplugconfp{\ctxtwo}{\clo}$, an initial state
$\secdplugconf{\ctxone}{\clo}{\ctxtwo}$, and a final state $\clo$.  The
transition function from the state
$\secdplugconf{\ctxone}{\clo}{\ctxtwo}$ incrementally peels off the given
control context and the transition function from the state
$\secdplugconfp{\ctxtwo}{\clo}$ dispatches over the given dump context:
  
\[
  \begin{array}{rcl@{\hspace{0.5cm}}l}
  \secdplugconf{\mtctx}{\clo}{\ctxtwo}
  & \plugsto &
  \secdplugconfp{\ctxtwo}{\clo}
  \\[1mm]
  \secdplugconf{\apctx{\ctxone}{\clo_0}}
               {\clo_1}
               {\ctxtwo}
  & \plugsto &
  \secdplugconf{\ctxone}
               {\comp{\clo_0}{\clo_1}}
               {\ctxtwo}
  \\[1mm]
  \secdplugconf{\argctx{\ctxone}{\val_1}}
               {\clo_0}
               {\ctxtwo}
  & \plugsto &
  \secdplugconf{\ctxone}
               {\comp{\clo_0}{\clo_1}}
               {\ctxtwo}
  &
  \mathrm{where} \; \clo_1 = \evaluetoclosure{\val_1}
  \\[4mm]
  \secdplugconfp{\mtmetactx}
                {\clo}
  & \plugsto &
  \clo
  \\[1mm]
  \secdplugconfp{\metactx{\ctxtwo}{\ctxone}}
                {\clo}
  & \plugsto &
  \secdplugconf{\shadethis{$\synboundary{\clo}$}}
               {\ctxone}
               {\ctxtwo}
  \end{array}
\]

We can now define a total function $\rawplug$ over closures, control
contexts, and dump contexts that fills the given closure into the given
control context, and further fills the result into the given dump
context:

\[
  \begin{array}{r@{\ }c@{\ }l}
  \rawplug :
  \syndomclos \times \syndomcontcont \times \syndomdumpcont
  \rightarrow
  \syndomclos
  \end{array}
\]

\begin{defi}
For any closure $\clo$, control context $\ctxone$, and dump
context $\ctxtwo$,
$\eplug{\ctxone}{\clo}{\ctxtwo}
 =
 \clo'
$
if and only if
$\secdplugconf{\ctxone}{\clo}{\ctxtwo}
 \plugsto^*
 \clo'
$.
\end{defi}

\subsubsection{Notion of contraction}
\label{subsubsec:contraction}

The notion of reduction over applicative expressions with the J operator
is specified by the following context-sensitive contraction rules over
actual redexes:

\[
\begin{array}{@{}r@{\ }r@{\hspace{1mm}}c@{\hspace{1mm}}l@{\hspace{3mm}}l@{}}
  \labl{Var}
  &
  \decthree{\esub{x}{\sub}}{\ctxone}{\ctxtwo}
  &
  \contractsto{\shift}
  &
  \decthree{\val}{\ctxone}{\ctxtwo}
  &
  \textrm{if }lookup (x, \sub) = \val
\end{array}
\]

\clearpage

\ 

\vspace{-1.0cm}

\[
\begin{array}{@{}r@{\ }r@{\hspace{1mm}}c@{\hspace{1mm}}l@{\hspace{-3mm}}l@{}}
%   \labl{Var}
%   &
%   \decthree{\esub{x}{\sub}}{\ctxone}{\ctxtwo}
%   &
%   \contractsto{\shift}
%   &
%   \decthree{\val}{\ctxone}{\ctxtwo}
%   &
% %
%   \textrm{if }lookup (x, \sub) = \val
%   \\[2mm]
  \labl{Beta$_{succ}$}
  &
  \decthree{\comp{\semsucc}{\represent{n}}}{\ctxone}{\ctxtwo}
  &
  \contractsto{\shift}
  &
  \decthree{\represent{n+1}}{\ctxone}{\ctxtwo}
  \\[2mm]
  \labl{Beta$_{FC}$}
  &
  \decthree{\comp{\esubp{\synlamp{x}{\tm}}{\sub}}{\val}}{\ctxone}{\ctxtwo}
  &
  \contractsto{\shift}
  &
  \decthree{\esub{\tm}{\sub'}}{\mtctx}{\metactx{\ctxtwo}{\ctxone}}
  &
  \textrm{where }\sub' = extend (x, \val, \sub) = \envextend{x}{\val}{\sub}
  \\[2mm]
  \labl{Beta$_{SA}$}
  &
  \decthree{\comp{\represent{\ctxtwo'}}{\val}}{\ctxone}{\ctxtwo}
  &
  \contractsto{\shift}
  &
  \decthree{\stateappend{\represent{\ctxtwo'}}{\val}}{\ctxone}{\ctxtwo}
  \\[2mm]
  \labl{Beta$_{PC}$}
  &
  \decthree{\comp{\stateappendp{\represent{\ctxtwo'}}{\val'}}{\val}}{\ctxone}{\ctxtwo}
  &
  \contractsto{\shift}
  &
  \decthree{\comp{\val'}{\val}}{\mtctx}{\ctxtwo'}
  \\[2mm]
  \labl{Prop}
  &
  \decthree{\esub{\appp{\tmone}{\tmtwo}}{\sub}}{\ctxone}{\ctxtwo}
  &
  \contractsto{\shift}
  &
  \decthree{\comp{\esubp{\tmone}{\sub}}{\esubp{\tmtwo}{\sub}}}{\ctxone}{\ctxtwo}
  \\[2mm]
  \labl{${\shift}$}
  &
  \decthree{J}{\ctxone}{\ctxtwo}
  &
  \contractsto{\shift}
  &
  \decthree{\represent{\ctxtwo}}{\ctxone}{\ctxtwo}
\end{array}
\]

\noindent
Three of these contraction rules depend on the contexts: the $J$ rule
captures a copy of the dump context and yields a state appender; the
$\beta$-rule for function closures resets the control context and pushes
it on the dump context; and the $\beta$-rule for program closures resets
the control context and reinstates a previously captured copy of the dump
context.

Among the potential redexes, only the ones listed above are actual ones.
The other applications of one value to another are stuck.

We now can define by cases a partial function
$\rawcontract$ over potential redexes that contracts an actual redex and
its two layers of contexts into the corresponding contractum and
contexts:
\[
  \begin{array}{r@{\ }c@{\ }l}
  \rawcontract :
  \syndompotred \times \syndomcontcont \times \syndomdumpcont
  \rightharpoonup
  \syndomclos \times \syndomcontcont \times \syndomdumpcont
  \end{array}
\]

\begin{defi}
For any potential redex $\varpotred$, control context $\ctxone$, and dump
context $\ctxtwo$,
$\econtract{\varpotred}{\ctxone}{\ctxtwo}
 =
 \decthree{\clo}{\ctxone'}{\ctxtwo'}
$
if and only if
$\decthree{\varpotred}{\ctxone}{\ctxtwo}
 \contractsto{\shift}
 \decthree{\clo}{\ctxone'}{\ctxtwo'}
$.
\end{defi}

\newcommand{\secddecomposeconf}[3]{\langle{#1},\,{#2},\,{#3}\rangle_{\mathrm{dec/clos}}}
\newcommand{\secddecomposeauxconf}[3]{\langle{#1},\,{#2},\,{#3}\rangle_{\mathrm{dec/cont}}}
\newcommand{\secddecomposeauxauxconf}[2]{\langle{#1},\,{#2}\rangle_{\mathrm{dec/dump}}}

\subsubsection{Decomposition}
\label{subsubsec:decomposition}

There are many ways to define a total function mapping a value closure
to itself and a non-value closure to a potential redex and a reduction
context.  In our experience, the following definition is a convenient
one.  It is a state-transition system with three intermediate states,
$\secddecomposeconf{\clo}{\ctxone}{\ctxtwo}$,
$\secddecomposeauxconf{\ctxone}{\val}{\ctxtwo}$, and
$\secddecomposeauxauxconf{\ctxtwo}{\val}$, an initial state
$\secddecomposeconf{\clo}{\mtctx}{\mtmetactx}$ and two final states
$\secddecompositionNONE{\val}$ and
$\secddecompositionSOME{\varpotred}{\ctxone}{\ctxtwo}$.  If possible, the
transition function from the state
$\secddecomposeconf{\clo}{\ctxone}{\ctxtwo}$
decomposes the given closure $\clo$ and accumulates the
corresponding two layers of reduction context, $\ctxone$ and $\ctxtwo$.
The transition function from the state
$\secddecomposeauxconf{\ctxone}{\val}{\ctxtwo}$ dispatches over the given
control context, and the transition function from the state
$\secddecomposeauxauxconf{\ctxtwo}{\val}$ dispatches over the given dump
context.

\[
  \begin{array}{rcl@{\hspace{0.5cm}}l}
  \secddecomposeconf{\represent{n}}
            {\ctxone}
            {\ctxtwo}
  & \decomposesto &
  \secddecomposeauxconf{\ctxone}
               {\represent{n}}
               {\ctxtwo}
  \\[1mm]
  \secddecomposeconf{\semsucc}
            {\ctxone}
            {\ctxtwo}
  & \decomposesto &
  \secddecomposeauxconf{\ctxone}
               {\semsucc}
               {\ctxtwo}
  \\[1mm]
  \secddecomposeconf{\esub{\represent{n}}{\sub}}
            {\ctxone}
            {\ctxtwo}
  & \decomposesto &
  \secddecomposeauxconf{\ctxone}
               {\represent{n}}
               {\ctxtwo}
  \\[1mm]
  \secddecomposeconf{\esub{x}{\sub}}
            {\ctxone}
            {\ctxtwo}
  & \decomposesto &
  \secddecompositionSOME{\esub{x}{\sub}}
                    {\ctxone}
                    {\ctxtwo}
  \\[1mm]
  \secddecomposeconf{\esub{\synlamp{x}{\tm}}{\sub}}
            {\ctxone}
            {\ctxtwo}
  & \decomposesto &
  \secddecomposeauxconf{\ctxone}
               {\esub{\synlamp{x}{\tm}}{\sub}}
               {\ctxtwo}
  \\[1mm]
  \secddecomposeconf{\esub{\appp{\tm_0}{\tm_1}}{\sub}}
            {\ctxone}
            {\ctxtwo}
  & \decomposesto &
  \secddecompositionSOME{\esub{\appp{\tm_0}{\tm_1}}{\sub}}
                    {\ctxone}
                    {\ctxtwo}
  \\[1mm]
  \secddecomposeconf{\esub{J}{\sub}}
            {\ctxone}
            {\ctxtwo}
  & \decomposesto &
  \secddecompositionSOME{J}
                    {\ctxone}
                    {\ctxtwo}
  \\[1mm]
  \secddecomposeconf{\comp{\clo_0}{\clo_1}}
            {\ctxone}
            {\ctxtwo}
  & \decomposesto &
  \secddecomposeconf{\clo_1}
            {\apctx{\ctxone}{\clo_0}}
            {\ctxtwo}
  \\[1mm]
  \secddecomposeconf{\represent{\ctxtwo'}}
            {\ctxone}
            {\ctxtwo}
  & \decomposesto &
  \secddecomposeauxconf{\ctxone}
               {\represent{\ctxtwo'}}
               {\ctxtwo}
  \\[1mm]
  \secddecomposeconf{\stateappend{\represent{\ctxtwo}}{\val}}
            {\ctxone}
            {\ctxtwo}
  & \decomposesto &
  \secddecomposeauxconf{\ctxone}
               {\stateappend{\represent{\ctxtwo}}{\val}}
               {\ctxtwo}
  \\[1mm]
  \secddecomposeconf{\shadethis{$\synboundary{\clo}$}}
            {\ctxone}
            {\ctxtwo}
  & \decomposesto &
  \secddecomposeconf{\clo}
            {\mtctx}
            {\metactx{\ctxtwo}{\ctxone}}
  \end{array}
\]

\clearpage

\ 

\vspace{-0.8cm}

\[
  \begin{array}{rcl@{\hspace{0.5cm}}l}
%  \\[4mm]
%
%
%
%
%
  \secddecomposeauxconf{\mtctx}
               {\val}
               {\ctxtwo}
  & \decomposesto &
  \secddecomposeauxauxconf{\ctxtwo}
             {\val}
  \\[1mm]
  \secddecomposeauxconf{\apctx{\ctxone}{\clo_0}}
               {\val_1}
               {\ctxtwo}
  & \decomposesto &
  \secddecomposeconf{\clo_0}
            {\argctx{\ctxone}{\val_1}}
            {\ctxtwo}
  \\[1mm]
  \secddecomposeauxconf{\argctx{\ctxone}{\val_1}}
               {\val_0}
               {\ctxtwo}
  & \decomposesto &
  \secddecompositionSOME{\comp{\val_0}{\val_1}}
                    {\ctxone}
                    {\ctxtwo}
  \\[4mm]
  \secddecomposeauxauxconf{\mtmetactx}
             {\val}
  & \decomposesto &
  \secddecompositionNONE{\val}
  \\[1mm]
  \secddecomposeauxauxconf{\metactx{\ctxtwo}{\ctxone}}
             {\val}
  & \decomposesto &
  \secddecomposeauxconf{\ctxone}
               {\val}
               {\ctxtwo}
  \end{array}
\]

We now can define a total function $\rawsecddecompose$ over
closures that maps a value closure to itself and a non-value closure to a
decomposition into a potential redex, a control context, and a dump
context.  This total function uses three auxiliary functions
$\rawsecddecomposep$, $\rawsecddecomposepaux$, and
$\rawsecddecomposepauxaux$:

\[
  \begin{array}{l@{\ }c@{\ \ }l@{\ }c@{\ }l}
  \rawsecddecompose
  & : &
  \syndomclos
  &
  \rightarrow
  \syndomval + (\syndompotred \times \syndomcontcont \times \syndomdumpcont)
  \\
  \rawsecddecomposep
  & : &
  \syndomclos \times \syndomcontcont \times \syndomdumpcont
  &
  \rightarrow
  \syndomval + (\syndompotred \times \syndomcontcont \times \syndomdumpcont)
  \\
  \rawsecddecomposepaux
  & : &
  \syndomcontcont \times \syndomval \times \syndomdumpcont
  &
  \rightarrow
  \syndomval + (\syndompotred \times \syndomcontcont \times \syndomdumpcont)
  \\
  \rawsecddecomposepauxaux
  & : &
  \syndomdumpcont \times \syndomval
  &
  \rightarrow
  \syndomval + (\syndompotred \times \syndomcontcont \times \syndomdumpcont)
  \end{array}
\]

\begin{defi}
For any closure $\clo$, control context $\ctxone$, and dump
context $\ctxtwo$,
\[\enspace
  \begin{array}{@{}rcl@{}}
  \secddecomposep{\clo}{\ctxone}{\ctxtwo}\!\!
  & = &
  \!\!\left\{
    \begin{array}{l@{\hspace{0.5cm}}l@{}}
    \secddecompositionNONE{\val'}
    &
    \mbox{if }
    \secddecomposeconf{\clo}{\ctxone}{\ctxtwo}
    \decomposesto^*
    \secddecompositionNONE{\val'}
    \\
    \secddecompositionSOME{\varpotred}{\ctxone'}{\ctxtwo'}
    &
    \mbox{if }
    \secddecomposeconf{\clo}{\ctxone}{\ctxtwo}
    \decomposesto^*
    \secddecompositionSOME{\varpotred}{\ctxone'}{\ctxtwo'}
    \end{array}
  \right.
  \\[4mm]
  \secddecomposepaux{\ctxone}{\val}{\ctxtwo}\!\!
  & = &
  \!\!\left\{
    \begin{array}{l@{\hspace{0.5cm}}l@{}}
    \secddecompositionNONE{\val'}
    &
    \mbox{if }
    \secddecomposeauxconf{\ctxone}{\val}{\ctxtwo}
    \decomposesto^*
    \secddecompositionNONE{\val'}
    \\
    \secddecompositionSOME{\varpotred}{\ctxone'}{\ctxtwo'}
    &
    \mbox{if }
    \secddecomposeauxconf{\ctxone}{\val}{\ctxtwo}
    \decomposesto^*
    \secddecompositionSOME{\varpotred}{\ctxone'}{\ctxtwo'}
    \end{array}
  \right.
  \\[4mm]
  \secddecomposepauxaux{\ctxtwo}{\val}\!\!
  & = &
  \!\!\left\{
    \begin{array}{l@{\hspace{0.5cm}}l@{}}
    \secddecompositionNONE{\val'}
    &
    \mbox{if }
    \secddecomposeauxauxconf{\ctxtwo}{\val}
    \decomposesto^*
    \secddecompositionNONE{\val'}
    \\
    \secddecompositionSOME{\varpotred}{\ctxone'}{\ctxtwo'}
    &
    \mbox{if }
    \secddecomposeauxauxconf{\ctxtwo}{\val}
    \decomposesto^*
    \secddecompositionSOME{\varpotred}{\ctxone'}{\ctxtwo'}
    \end{array}
  \right.
  \end{array}
\]
and
$\edecompose{\clo}
 =
 \secddecomposep{\clo}{\mtctx}{\mtmetactx}$.
\end{defi}

\subsubsection{One-step reduction}
\label{subsubsec:one-step-reduction}

We are now in position to define a partial function $\rawreduce$ over
closed closures that
maps a value closure to itself and a non-value closure to the next
closure in the reduction sequence.  This function is defined by
composing the three functions above:

\[
  \ereduce{\clo}
  =
  \begin{array}[t]{@{}l}
  \vcase{\edecompose{\clo}}
        {\secddecompositionNONE{\val}}
        {\evaluetoclosure{\val}}
        {\secddecompositionSOME{\varpotred}{\ctxone}{\ctxtwo}}
        {\eplugone{\econtract{\varpotred}{\ctxone}{\ctxtwo}}}
  \end{array}
\]

\noindent
The function $\rawreduce$ is partial because of $\rawcontract$, which is
undefined for stuck closures.

\begin{defi}[One-step reduction]
For any
closure $\clo$,
$ \clo \rightarrow \clo' $
if and only if
$ \ereduce{\clo} = \clo' $.
\end{defi}

\subsubsection{Reduction-based evaluation}
\label{subsubsec:reduction-based-evaluation}

Iterating
$\rawreduce$
defines a reduction-based evaluation function.  The definition below uses
$\rawdecompose$ to distinguish between values and non-values,
and implements iteration (tail-) recursively with the partial function
$\rawiterate$:

\[
  \begin{array}{rcl@{\hspace{1.9cm}}}
  \eevaluate{\clo}
  & = &
  \eiterate{\edecompose{\clo}}
  \end{array}
\]
\noindent
where
\[
  \left\{
  \begin{array}{lcl}
  \eiterate{\secddecompositionNONE{\val}}
  & = &
  \val
  \\
  \eiterate{\secddecompositionSOME{\varpotred}{\ctxone}{\ctxtwo}}
  & = &
  \eiterate{\edecompose{\eplugone{\econtract{\varpotred}{\ctxone}{\ctxtwo}}}}
  \end{array}
  \right.
\]

\clearpage

\noindent
The function $\rawevaluate$ is partial because a given closure might be
stuck or reducing it might not converge.

\begin{defi}[Reduction-based evaluation]
For any
closure $\clo$,
$ \clo \rightarrow^* \val $
if and only if
$ \eevaluate{\clo}$
$= \val $.
\end{defi}

To close, let us adjust the definition of $\rawevaluate$ by exploiting
the fact that for any closure $\clo$, $\eplug{\clo}{\mtctx}{\mtmetactx} =
\clo$:

\[
   \eevaluate{\clo}
   =
   \eiterate{\edecompose{\eplug{\clo}{\mtctx}{\mtmetactx}}}
\]

\noindent
In this adjusted definition, $\rawdecompose$ is always applied to the
result of $\rawplug$.

\subsection{From the reduction semantics for applicative expressions
            to the SECD machine}
\label{subsec:from-red-sem-to-secd-mach}

Deforesting the intermediate terms in the reduction-based evaluation
function of Section~\ref{subsubsec:reduction-based-evaluation} yields a
reduction-free evaluation function in the form of a small-step abstract
machine (Section~\ref{subsubsec:refocusing}).
We simplify this small-step abstract machine by fusing a part of its
driver loop with the contraction function
(Section~\ref{subsubsec:lightweight-fusion}) and compressing its
`corridor' transitions
(Section~\ref{subsubsec:inlining-and-transition-compression}).  Unfolding
the recursive data type of closures precisely yields the caller-save,
stackless SECD abstract machine of
Section~\ref{subsec:transmogrified-SECD-machine}
(Section~\ref{subsubsec:opening-closures}).

\subsubsection{Refocusing: from reduction-based to reduction-free evaluation}
\label{subsubsec:refocusing}

Following Danvy and Nielsen~\cite{Danvy-Nielsen:RS-04-26}, we deforest
the intermediate closure in the reduction sequence by replacing the
composition of $\rawplug$ and $\rawdecompose$ by a call to a composite
function $\rawrefocus$:

\[
  \begin{array}{@{\hspace{0.5cm}}rcl}
   \eevaluate{\clo}
   & = &
   \eiterate{\erefocus{\clo}{\mtctx}{\mtmetactx}}
  \end{array}
\]
where

\[
  \left\{
  \begin{array}{lcl}
  \eiterate{\secddecompositionNONE{\val}}
  & = &
  \val
  \\
  \eiterate{\secddecompositionSOME{\varpotred}{\ctxone}{\ctxtwo}}
  & = &
  \eiterate{\erefocusone{\econtract{\varpotred}{\ctxone}{\ctxtwo}}}
  \end{array}
  \right.
\]

\noindent
and $\rawrefocus$ is optimally defined as continuing the decomposition in
the current reduction context~\cite{Danvy-Nielsen:RS-04-26}:

\[
  \begin{array}{rcl@{\hspace{1.2cm}}}
   \erefocus{\clo}{\ctxone}{\ctxtwo}
   & = &
   \secddecomposep{\clo}{\ctxone}{\ctxtwo}
  \end{array}
\]

\begin{defi}[Reduction-free evaluation]
For any
closure $\clo$,
$ \clo \mapsto_{\shift}^* \val $
if and only if
$ \eevaluate{\clo} = \val $.
\end{defi}

\newcommand{\secditerate}[3]{\langle {#1},\,{#2},\,{#3} \rangle_{\mathrm{iter}}}
\newcommand{\secdrefocusesto}{\Rightarrow}

\newcommand{\secdrefocusconf}[3]{\langle {#1},\,{#2},\,{#3} \rangle_{\mathrm{eval}}}
\newcommand{\secdrefocusauxconf}[3]{\langle {#1},\,{#2},\,{#3} \rangle_{\mathrm{cont}}}
\newcommand{\secdrefocusauxauxconf}[2]{\langle {#1},\,{#2} \rangle_{\mathrm{dump}}}

\subsubsection{Lightweight fusion: making do without driver loop}
\label{subsubsec:lightweight-fusion}

In effect, $\rawiterate$ is as the `driver loop' of a small-step abstract
machine that refocuses and contracts.  Instead, let us fuse
$\rawcontract$ and $\rawiterate$ and express the result with rewriting
rules over a configuration $\secditerate{\varpotred}{\ctxone}{\ctxtwo}$.  We clone
the rewriting rules for $\rawsecddecomposep$, $\rawsecddecomposepaux$, and
$\rawsecddecomposepauxaux$ into refocusing rules,
respectively indexing the
configuration $\secddecomposeconf{\clo}{\ctxone}{\ctxtwo}$
as $\secdrefocusconf{\clo}{\ctxone}{\ctxtwo}$, the
configuration $\secddecomposeauxconf{\ctxone}{\val}{\ctxtwo}$
as $\secdrefocusauxconf{\ctxone}{\val}{\ctxtwo}$, and the
configuration $\secddecomposeauxauxconf{\ctxtwo}{\val}$
as $\secdrefocusauxauxconf{\ctxtwo}{\val}$:

\begin{enumerate}[$\bullet$]

\item
instead of rewriting to $\secddecompositionNONE{\val}$, the cloned
rules rewrite to $\val$;

\item
instead of rewriting to $\secddecompositionSOME{\varpotred}{\ctxone}{\ctxtwo}$,
the cloned
rules rewrite to $\secditerate{\varpotred}{\ctxone}{\ctxtwo}$.

\end{enumerate}

\noindent
The result reads as follows:
\[
  \begin{array}{rcl@{\hspace{0.5cm}}l}
  \secdrefocusconf{\represent{n}}
            {\ctxone}
            {\ctxtwo}
  & \secdrefocusesto &
  \secdrefocusauxconf{\ctxone}
               {\represent{n}}
               {\ctxtwo}
  \\[1mm]
  \secdrefocusconf{\semsucc}
            {\ctxone}
            {\ctxtwo}
  & \secdrefocusesto &
  \secdrefocusauxconf{\ctxone}
               {\semsucc}
               {\ctxtwo}
  \\[1mm]
  \secdrefocusconf{\esub{\represent{n}}{\sub}}
            {\ctxone}
            {\ctxtwo}
  & \secdrefocusesto &
  \secdrefocusauxconf{\ctxone}
               {\represent{n}}
               {\ctxtwo}
  \\[1mm]
  \secdrefocusconf{\esub{x}{\sub}}
            {\ctxone}
            {\ctxtwo}
  & \secdrefocusesto &
  \secditerate{\esub{x}{\sub}}
                    {\ctxone}
                    {\ctxtwo}
  \\[1mm]
  \secdrefocusconf{\esub{\synlamp{x}{\tm}}{\sub}}
            {\ctxone}
            {\ctxtwo}
  & \secdrefocusesto &
  \secdrefocusauxconf{\ctxone}
               {\esub{\synlamp{x}{\tm}}{\sub}}
               {\ctxtwo}
  \\[1mm]
  \secdrefocusconf{\esub{\appp{\tm_0}{\tm_1}}{\sub}}
            {\ctxone}
            {\ctxtwo}
  & \secdrefocusesto &
  \secditerate{\esub{\appp{\tm_0}{\tm_1}}{\sub}}
                    {\ctxone}
                    {\ctxtwo}
  \\[1mm]
  \secdrefocusconf{\esub{J}{\sub}}
            {\ctxone}
            {\ctxtwo}
  & \secdrefocusesto &
  \secditerate{J}
                    {\ctxone}
                    {\ctxtwo}
  \\[1mm]
  \secdrefocusconf{\comp{\clo_0}{\clo_1}}
            {\ctxone}
            {\ctxtwo}
  & \secdrefocusesto &
  \secdrefocusconf{\clo_1}
            {\apctx{\ctxone}{\clo_0}}
            {\ctxtwo}
  \\[1mm]
  \secdrefocusconf{\represent{\ctxtwo'}}
            {\ctxone}
            {\ctxtwo}
  & \secdrefocusesto &
  \secdrefocusauxconf{\ctxone}
               {\represent{\ctxtwo'}}
               {\ctxtwo}
  \\[1mm]
  \secdrefocusconf{\stateappend{\represent{\ctxtwo'}}{\val}}
            {\ctxone}
            {\ctxtwo}
  & \secdrefocusesto &
  \secdrefocusauxconf{\ctxone}
               {\stateappend{\represent{\ctxtwo'}}{\val}}
               {\ctxtwo}
  \\[1mm]
  \secdrefocusconf{\shadethis{$\synboundary{\clo}$}}
            {\ctxone}
            {\ctxtwo}
  & \secdrefocusesto &
  \secdrefocusconf{\clo}
            {\mtctx}
            {\metactx{\ctxtwo}{\ctxone}}
  \\[4mm]
  \secdrefocusauxconf{\mtctx}
               {\val}
               {\ctxtwo}
  & \secdrefocusesto &
  \secdrefocusauxauxconf{\ctxtwo}
             {\val}
  \\[1mm]
  \secdrefocusauxconf{\apctx{\ctxone}{\clo_0}}
               {\val_1}
               {\ctxtwo}
  & \secdrefocusesto &
  \secdrefocusconf{\clo_0}
            {\argctx{\ctxone}{\val_1}}
            {\ctxtwo}
  &
  \phantom{\textrm{where }\sub' = extend (x, \val, \sub) = \envextend{x}{\val}{\sub}}
  \\[1mm]
  \secdrefocusauxconf{\argctx{\ctxone}{\val_1}}
               {\val_0}
               {\ctxtwo}
  & \secdrefocusesto &
  \secditerate{\comp{\val_0}{\val_1}}
                    {\ctxone}
                    {\ctxtwo}
  \\[4mm]
  \secdrefocusauxauxconf{\mtmetactx}
             {\val}
  & \secdrefocusesto &
  \val
  \\[1mm]
  \secdrefocusauxauxconf{\metactx{\ctxtwo}{\ctxone}}
             {\val}
  & \secdrefocusesto &
  \secdrefocusauxconf{\ctxone}
               {\val}
               {\ctxtwo}
  \\[4mm]
  \secditerate{\esub{x}{\sub}}
                    {\ctxone}
                    {\ctxtwo}
  & \secdrefocusesto &
  \secdrefocusconf{\val}
            {\ctxone}
            {\ctxtwo}
  &
  \textrm{if }lookup (x, \sub) = \val
  \\[1mm]
  \secditerate{\comp{\semsucc}{\represent{n}}}
                    {\ctxone}
                    {\ctxtwo}
  & \secdrefocusesto &
  \secdrefocusconf{\represent{n+1}}
            {\ctxone}
            {\ctxtwo}
  \\[1mm]
  \secditerate{\comp{\esubp{\synlamp{x}{\tm}}{\sub}}{\val}}
                    {\ctxone}
                    {\ctxtwo}
  & \secdrefocusesto &
  \secdrefocusconf{\esub{\tm}{\sub'}}
            {\mtctx}
            {\metactx{\ctxtwo}{\ctxone}}
  &
  \textrm{where }\sub' = extend (x, \val, \sub) = \envextend{x}{\val}{\sub}
  \\[1mm]
  \secditerate{\comp{\represent{\ctxtwo'}}{\val}}
                    {\ctxone}
                    {\ctxtwo}
  & \secdrefocusesto &
  \secdrefocusconf{\stateappend{\represent{\ctxtwo'}}{\val}}
            {\ctxone}
            {\ctxtwo}
  \\[1mm]
  \secditerate{\comp{\stateappendp{\represent{\ctxtwo'}}{\val'}}{\val}}
                    {\ctxone}
                    {\ctxtwo}
  & \secdrefocusesto &
  \secdrefocusconf{\comp{\val'}{\val}}
            {\mtctx}
            {\ctxtwo'}
  \\[1mm]
  \secditerate{\esub{\appp{\tm_0}{\tm_1}}{\sub}}
                    {\ctxone}
                    {\ctxtwo}
  & \secdrefocusesto &
  \rlap{$
  \secdrefocusconf{\comp{\esubp{\tmone}{\sub}}{\esubp{\tmtwo}{\sub}}}
            {\ctxone}
            {\ctxtwo}$}
  \\[1mm]
  \secditerate{J}
                    {\ctxone}
                    {\ctxtwo}
  & \secdrefocusesto &
  \secdrefocusconf{\represent{\ctxtwo}}
            {\ctxone}
            {\ctxtwo}
  \end{array}
\]

\noindent
The following proposition summarizes the situation:

\begin{prop}
For any closure $\clo$,
$ \eevaluate{\clo} = \val $
if and only if
$ \secdrefocusconf{\clo}{\mtctx}{\mtmetactx} \Rightarrow^* \val $.
\end{prop}

\noindent
Proof: straightforward.  The two machines operate in lockstep.
\hfill $\Box$

\subsubsection{Inlining and transition compression}
\label{subsubsec:inlining-and-transition-compression}

The abstract machine of Section~\ref{subsubsec:lightweight-fusion},
while interesting in its own right (it is `staged' in that the
contraction rules are implemented separately from the congruence
rules~\cite{Biernacka-Danvy:TOCL07, Hardin-al:JFP98}), is not minimal: a
number of transitions yield a configuration whose transition is uniquely
determined.  Let us carry out these hereditary, ``corridor'' transitions
once and for all:

\begin{enumerate}[$\bullet$]

\item
$\secdrefocusconf{\esub{x}{\sub}}{\ctxone}{\ctxtwo}
 \secdrefocusesto
 \secditerate{\esub{x}{\sub}}{\ctxone}{\ctxtwo}
 \secdrefocusesto
 \secdrefocusconf{\val}{\ctxone}{\ctxtwo}
 \secdrefocusesto
 \secdrefocusauxconf{\ctxone}{\val}{\ctxtwo}$
 \hfill
 $\textrm{if }lookup (x, \sub) = \val$

\item
$\secdrefocusconf{\esub{\appp{\tm_0}{\tm_1}}{\sub}}{\ctxone}{\ctxtwo}
 \secdrefocusesto
 \secditerate{\esub{\appp{\tm_0}{\tm_1}}{\sub}}{\ctxone}{\ctxtwo}
 \secdrefocusesto$

$\secdrefocusconf{\comp{\esubp{\tmone}{\sub}}{\esubp{\tmtwo}{\sub}}}{\ctxone}{\ctxtwo}
 \secdrefocusesto
 \secdrefocusconf{\esub{\tm_1}{\sub}}{\apctx{\ctxone}{\esubp{\tm_0}{\sub}}}{\ctxtwo}$

\item
$\secdrefocusconf{\esub{J}{\sub}}{\ctxone}{\ctxtwo}
 \secdrefocusesto
 \secditerate{J}{\ctxone}{\ctxtwo}
 \secdrefocusesto
 \secdrefocusconf{\represent{\ctxtwo}}{\ctxone}{\ctxtwo}
 \secdrefocusesto
 \secdrefocusauxconf{\ctxone}{\represent{\ctxtwo}}{\ctxtwo}$

\item
$\secditerate{\comp{\semsucc}{\represent{n}}}{\ctxone}{\ctxtwo}
 \secdrefocusesto
 \secdrefocusconf{\represent{n+1}}{\ctxone}{\ctxtwo}
 \secdrefocusesto
 \secdrefocusauxconf{\ctxone}{\represent{n+1}}{\ctxtwo}$

\item
$\secditerate{\comp{\represent{\ctxtwo'}}{\val}}{\ctxone}{\ctxtwo}
 \secdrefocusesto
 \secdrefocusconf{\stateappend{\represent{\ctxtwo'}}{\val}}{\ctxone}{\ctxtwo}
 \secdrefocusesto
 \secdrefocusauxconf{\ctxone}{\stateappend{\represent{\ctxtwo'}}{\val}}{\ctxtwo}$

\end{enumerate}

\noindent
The result reads as follows:
\[
\begin{array}{@{}r@{\hspace{1.75mm}}l@{\hspace{1.75mm}}l@{}l}
\namedtransition{}
{\secdrefocusconf{\esub{\represent{n}}{\sub}}{\ctxone}{\ctxtwo}}
{\secdrefocusauxconf{\ctxone}{\represent{n}}{\ctxtwo}}
\\[1mm]
\namedtransition{}
{\secdrefocusconf{\esub{x}{\sub}}{\ctxone}{\ctxtwo}}
{\secdrefocusauxconf{\ctxone}{\val}{\ctxtwo}}
&
\textrm{if }lookup (x, \sub) = \val
\\[1mm]
\namedtransition{}
{\secdrefocusconf{\esub{\synlamp{x}{\tm}}{\sub}}{\ctxone}{\ctxtwo}}
{\secdrefocusauxconf{\ctxone}{\esub{\synlamp{x}{\tm}}{\sub}}{\ctxtwo}}
\\[1mm]
\namedtransition{}
{\secdrefocusconf{\esub{\appp{\tmone}{\tmtwo}}{\sub}}{\ctxone}{\ctxtwo}}
{\secdrefocusconf{\esub{\tmtwo}{\sub}}{\apctx{\ctxone}{\esubp{\tmone}{\sub}}}{\ctxtwo}}
\\[1mm]
\namedtransition{}
{\secdrefocusconf{\esub{J}{\sub}}{\ctxone}{\ctxtwo}}
{\secdrefocusauxconf{\ctxone}{\represent{\ctxtwo}}{\ctxtwo}}
\\[4mm]
\namedtransition{}
{\secditerate{\app{\semsucc}{\represent{n}}}{\ctxone}{\ctxtwo}}
{\secdrefocusauxconf{\ctxone}{\represent{n+1}}{\ctxtwo}}
\\[1mm]
\namedtransition{}
{\secditerate{\app{\esubp{\synlamp{x}{\tm}}{\sub}}{\val}}{\ctxone}{\ctxtwo}}
{\secdrefocusconf{\esub{\tm}{\sub'}}{\mtctx}{\metactx{\ctxtwo}{\ctxone}}}
&
\textrm{where }\sub' = extend (x, \val, \sub)
\\[1mm]
\namedtransition{}
{\secditerate{\app{\stateappendp{\represent{\ctxtwo'}}{\val'}}{\val}}{\ctxone}{\ctxtwo}}
{\secditerate{\app{\val}{\val'}}{\mtctx}{\ctxtwo'}}
\\[1mm]
\namedtransition{}
{\secditerate{\app{\represent{\ctxtwo'}}{\val}}{\ctxone}{\ctxtwo}}
{\secdrefocusauxconf{\ctxone}{\stateappend{\represent{\ctxtwo'}}{\val}}{\ctxtwo}}
\\[4mm]
\namedtransition{}
{\secdrefocusauxconf{\mtctx}{\val}{\ctxtwo}}
{\secdrefocusauxauxconf{\ctxtwo}{\val}}
\\[1mm]
\namedtransition{}
{\secdrefocusauxconf{\apctx{\ctxone}{\esubp{\tm}{\sub}}}{\val}{\ctxtwo}}
{\secdrefocusconf{\esub{\tm}{\sub}}{\argctx{\ctxone}{\val}}{\ctxtwo}}
\\[1mm]
\namedtransition{}
{\secdrefocusauxconf{\argctx{\ctxone}{\val'}}{\val}{\ctxtwo}}
{\secditerate{\app{\val}{\val'}}{\ctxone}{\ctxtwo}}
\\[4mm]
\namedtransition{}
{\secdrefocusauxauxconf{\mtmetactx}{\val}}
\val\\[1mm]
\namedtransition{}
{\secdrefocusauxauxconf{\metactx{\ctxtwo}{\ctxone}}{\val}}
{\secdrefocusauxconf{\ctxone}{\val}{\ctxtwo}}
\end{array}
\]

\noindent
The eval-clauses for $\represent{n}$, $\semsucc$ (which only occurs in
the initial environment), $\comp{\clo_0}{\clo_1}$, $\represent{\ctxtwo}$,
and $\stateappend{\represent{\ctxtwo}}{\val}$ and the iter-clauses for
$\esub{x}{\sub}$, $\esub{\appp{\tmone}{\tmtwo}}{\sub}$, and $J$ all have
disappeared: they were only transitory.  The eval-clause for
$\synboundary{\clo}$ has also disappeared: it is a dead clause here since
$\rawplug$ has been refocused away.

\begin{prop}
For any closure $\clo$,
$ \eevaluate{\clo} = \val $
if and only if
$ \secdrefocusconf{\clo}{\mtctx}{\mtmetactx} \Rightarrow^* \val $.
\end{prop}

\noindent
Proof: immediate.  We have merely compressed corridor transitions and
removed one dead clause.
\hfill $\Box$

\subsubsection{Opening closures: from explicit substitutions
               to terms and environments}
\label{subsubsec:opening-closures}

The abstract machine above solely operates on ground closures and the
iter-clauses solely dispatch on applications of one value to another.  If
we (1) open the closures $\esub{\tm}{\sub}$ into pairs
$\synpair{\tm}{\sub}$ and flatten the configuration
$\secdrefocusconf{\synpair{\tm}{\sub}}{\ctxone}{\ctxtwo}$ into a
quadruple $\secdconfeval{\tm}{\sub}{\ctxone}{\ctxtwo}$ and (2) flatten
the configuration $\secditerate{\app{\val}{\val'}}{\ctxone}{\ctxtwo}$
into a quadruple $\secdconfapply{\val}{\val'}{\ctxone}{\ctxtwo}$, we
obtain an abstract machine that coincides with the caller-save, stackless
SECD machine of Section~\ref{subsec:transmogrified-SECD-machine}.

The following proposition captures
that the SECD machine implements the reduction semantics of
Section~\ref{subsec:syntactic-theory-J}.

\begin{prop}[syntactic correspondence]
For any program $\tm$ in the $\lrhJ$-calculus,
$$\esub{\tm}{\envextend{\synsucc}{\semsucc}{\envempty}}
  \rightarrow^*
  \val
  \quad
  \textrm{if and only if}
  \quad
  \secdrefocusconf{\esub{\tm}
                        {\envextend{\synsucc}{\semsucc}{\envempty}}}
                  {\mtctx}
                  {\mtmetactx}
  \Rightarrow^*
  \val.$$
\end{prop}

\noindent
Proof: this proposition is a simple corollary of the above series of
propositions and of the observation just above.
\hfill $\Box$

\subsection{Summary and conclusion}

All in all, the syntactic and the functional correspondences provide a
method to mechanically build compatible small-step semantics in the form
of calculi
(reduction semantics) and abstract machines, and big-step
semantics in the form of evaluation functions.  We have illustrated this
method here for applicative expressions with the J operator,
providing their first big-step semantics and their first reduction
semantics.

\section{A syntactic theory of applicative expressions with the J operator:
         implicit, caller-save dumps}
\label{sec:syntactic-theory-implicit-caller-save-dumps}

The J operator capture the continuation of the caller and accordingly,
the SECD machine is structured as the expression continuation of the
current function up to its point of call (the C component) and as a list
of the delimited expression continuations of the previously called
functions (the D component).  This architecture stands both for the
original SECD machine (Section~\ref{sec:deconstruction-1}) and for its
modernized instances, whether the dump is managed in a callee-save fashion
(Section~\ref{sec:deconstruction-2}) or in a caller-save fashion
(Section~\ref{sec:deconstruction-3}).  In this section, we study a single
representation of the context that is dynamically scanned in search for
the context of the caller, as in Felleisen et al.'s initial take on
delimited continuations~\cite{Felleisen-al:LFP88} and in John Clements's
PhD thesis work on continuation marks~\cite{Clements:PhD}.  We start from
a reduction semantics (Section~\ref{subsec:RS-implicit}) and refocus it
into an abstract machine (Section~\ref{subsec:AM-implicit}).

\newcommand{\plugctxq}[3]{\ensuremath{#1[#2]#3}}
\newcommand{\apctxq}[2]{\ensuremath{\plugctx{#1}{\app{#2}{\ctxhole}}}}
\newcommand{\delctx}[1]{\ensuremath{\langlethick{#1}\ranglethick}}

\subsection{Reduction semantics}
\label{subsec:RS-implicit}

We specify the reduction semantics as in
Sections~\ref{subsec:syntactic-theory-J} and~\ref{subsec:cek-RS}, \ie,
with its syntactic categories, a plugging function, a notion of
contraction, a decomposition function, a one-step reduction function, and
a reduction-based evaluation function.

\subsubsection{Syntactic categories}
\label{subsubsec:syntactic-categories-implicit}

We consider a variant of the $\lrhJ$-calculus
with one layer of context $\ctx$
and with delimiters
$\synboundary{\clo}$ and $\delctx{\ctx}$ (shaded below)
to mark the boundary between the context of a
$\beta$-redex that has been contracted, \ie, a function closure that has
been applied, and the body of the $\lambda$-abstraction in this function
closure which is undergoing reduction:
\[
\begin{array}{@{}rc@{\ }c@{\ }l@{}}
\textrm{(programs)} & p & ::= &
\esub{\tm}{\envextend{\synsucc}{\semsucc}{\envempty}}
\\[0.5mm]
\textrm{(terms)} & \tm & ::= &
\represent{n} \Mid x \Mid \synlam{x}{\tm} \Mid \app{\tm}{\tm} \Mid J
\\[0.5mm]
\textrm{(closures)} & \clo & ::= &
\represent{n}
\Mid
\semsucc
\Mid
\esub{\tm}{\sub}
\Mid
\comp{\clo}{\clo}
\Mid
\represent{\ctx}
\Mid
\stateappend{\represent{\ctx}}{\val}
\Mid
\shadethis{$\synboundary{\clo}$}
\\[0.5mm]
\textrm{(values)} & \val & ::= &
\represent{n}
\Mid
\semsucc
\Mid
\esub{\synlamp{x}{\tm}}{\sub}
\Mid
\represent{\ctx}
\Mid
\stateappend{\represent{\ctx}}{\val}
\\[0.5mm]
\textrm{(potential redexes)} & \varpotred & ::= &
\esub{x}{\sub} \Mid
\comp{\val}{\val} \Mid
J
\\[0.5mm]
\textrm{(substitutions)} & \sub & ::= &
\envempty \Mid \envextend{x}{\val}{\sub}
\\[0.5mm]
\textrm{(contexts)} & \ctx & ::= &
\mtctx \Mid
\apctx{\ctx}{\clo} \Mid
\argctx{\ctx}{\val} \Mid
\shadethis{$\delctx{\ctx}$}
\end{array}
\]

\noindent
Again, in the syntactic category of closures,
$\represent{\ctx}$ and $\stateappend{\represent{\ctx}}{\val}$
respectively denote a state appender and a program closure.  Also again,
values are therefore a syntactic subcategory of closures, and
we make use of the syntactic coercion $\rawvaluetoclosure$
mapping a value into a closure.

\newcommand{\secdplugconfq}[2]{\langle {#1},\,{#2} \rangle_{\mathrm{plug/cont}}}

\subsubsection{Plugging}
\label{subsubsec:plugging-implicit}

Plugging a closure in a context is defined by induction
over this context:
\[
  \begin{array}{rcl@{\hspace{0.5cm}}l}
  \secdplugconfq{\mtctx}{\clo}
  & \plugsto &
  \clo
  \\[1mm]
  \secdplugconfq{\apctx{\ctxone}{\clo_0}}{\clo_1}
  & \plugsto &
  \secdplugconfq{\ctx}{\comp{\clo_0}{\clo_1}}
  \\[1mm]
  \secdplugconfq{\argctx{\ctx}{\val_1}}
                {\clo_0}
  & \plugsto &
  \secdplugconfq{\ctx}
                {\comp{\clo_0}{\clo_1}}
  &
  \mathrm{where} \; \clo_1 = \evaluetoclosure{\val_1}
  \\[1mm]
  \secdplugconfq{\shadethis{$\delctx{\ctx}$}}{\clo}
  & \plugsto &
  \secdplugconfq{\ctx}{\shadethis{$\delctx{\clo}$}}
  \end{array}
\]

\begin{defi}
For any closure $\clo$ and context $\ctx$,
$\eplugtwo{\ctx}{\clo}
 =
 \clo'
$
if and only if \\
$\secdplugconfq{\ctxone}{\clo}
 \plugsto^*
 \clo'
$.
\end{defi}

\subsubsection{Notion of contraction}
\label{subsubsec:notion-of-contraction-implicit}

The notion of reduction
is specified by the following context-sensitive contraction rules over
actual redexes:
\[
\begin{array}{@{}r@{\ }r@{\hspace{1mm}}c@{\hspace{1mm}}l@{\hspace{-3mm}}l@{}}
  \labl{Var}
  &
  \dectwo{\esub{x}{\sub}}{\ctx}
  &
  \contractsto{\shift}
  &
  \dectwo{\val}{\ctx}
  &
  \textrm{if }lookup (x, \sub) = \val
  \\[2mm]
  \labl{Beta$_{succ}$}
  &
  \dectwo{\comp{\semsucc}{\represent{n}}}{\ctx}
  &
  \contractsto{\shift}
  &
  \dectwo{\represent{n+1}}{\ctx}
  \\[2mm]
  \labl{Beta$_{FC}$}
  &
  \dectwo{\comp{\esubp{\synlamp{x}{\tm}}{\sub}}{\val}}{\ctx}
  &
  \contractsto{\shift}
  &
  \dectwo{\shadethis{$\synboundary{\esub{\tm}{\sub'}}$}}{\ctx}
  &
  \textrm{where }\sub' = extend (x, \val, \sub) = \envextend{x}{\val}{\sub}
  \\[2mm]
  \labl{Beta$_{SA}$}
  &
  \dectwo{\comp{\represent{\ctx'}}{\val}}{\ctx}
  &
  \contractsto{\shift}
  &
  \dectwo{\stateappend{\represent{\ctx'}}{\val}}{\ctx}
  \\[2mm]
  \labl{Beta$_{PC}$}
  &
  \dectwo{\comp{\stateappendp{\represent{\ctx'}}{\val'}}{\val}}{\ctx}
  &
  \contractsto{\shift}
  &
  \dectwo{\comp{\val'}{\val}}{\ctx'}
  \\[2mm]
  \labl{Prop}
  &
  \dectwo{\esub{\appp{\tmone}{\tmtwo}}{\sub}}{\ctx}
  &
  \contractsto{\shift}
  &
  \dectwo{\comp{\esubp{\tmone}{\sub}}{\esubp{\tmtwo}{\sub}}}{\ctx}
  \\[2mm]
  \labl{${\shift}$}
  &
  \dectwo{J}{\ctx}
  &
  \contractsto{\shift}
  &
  \dectwo{\represent{\ctx'}}{\ctx}
  &
  \textrm{where }\ctx' = previous(\ctx)
\end{array}
\]

\noindent
where
% the following partial function
$previous$ maps a context to its most recent
delimited context, if any:
\[
  \begin{array}{rcl@{\hspace{0.5cm}}l}
  previous({\apctx{\ctxone}{\clo}})
  & = &
  previous(\ctxone)
  \\[1mm]
  previous({\argctx{\ctx}{\val}})
  & = &
  previous({\ctx})
  \\[1mm]
  previous({\shadethis{$\delctx{\ctx}$}})
  & = &
  \ctx
  \end{array}
\]

Two of the contraction rules depend on the context: the $J$ rule
captures a copy of the context of the most recent caller and yields a
state appender,
and the $\beta$-rule for program closures
reinstates a previously captured copy of the context.
As for the $\beta$-rule for function closures, it introduces a delimiter.

\begin{defi}
For any potential redex $\varpotred$ and context $\ctx$,
$\econtracttwo{\varpotred}{\ctx}
 =
 \dectwo{\clo}{\ctx'}
$
if and only if
$\dectwo{\varpotred}{\ctx'}
 \contractsto{\shift}
 \dectwo{\clo}{\ctx'}
$.
\end{defi}

\newcommand{\secddecomposeconfq}[2]{\langle{#1},\,{#2}\rangle_{\mathrm{dec/clos}}}
\newcommand{\secddecomposeauxconfq}[2]{\langle{#1},\,{#2}\rangle_{\mathrm{dec/cont}}}

\subsubsection{Decomposition}
\label{subsubsec:decomposition-implicit}

Decomposition is essentially as in Section~\ref{subsubsec:decomposition},
except that there is no explicit dump component:
\[
  \begin{array}{rcl@{\hspace{0.5cm}}l}
  \secddecomposeconfq{\represent{n}}
            {\ctxone}
  & \decomposesto &
  \secddecomposeauxconfq{\ctxone}
               {\represent{n}}
  \\[1mm]
  \secddecomposeconfq{\semsucc}
            {\ctxone}
  & \decomposesto &
  \secddecomposeauxconfq{\ctxone}
               {\semsucc}
  \\[1mm]
  \secddecomposeconfq{\esub{\represent{n}}{\sub}}
            {\ctxone}
  & \decomposesto &
  \secddecomposeauxconfq{\ctxone}
               {\represent{n}}
  \\[1mm]
  \secddecomposeconfq{\esub{x}{\sub}}
            {\ctxone}
  & \decomposesto &
  \secddecompositionSOMEtwo{\esub{x}{\sub}}
                    {\ctxone}
  \\[1mm]
  \secddecomposeconfq{\esub{\synlamp{x}{\tm}}{\sub}}
            {\ctxone}
  & \decomposesto &
  \secddecomposeauxconfq{\ctxone}
               {\esub{\synlamp{x}{\tm}}{\sub}}
  \\[1mm]
  \secddecomposeconfq{\esub{\appp{\tm_0}{\tm_1}}{\sub}}
            {\ctxone}
  & \decomposesto &
  \secddecompositionSOMEtwo{\esub{\appp{\tm_0}{\tm_1}}{\sub}}
                    {\ctxone}
  \\[1mm]
  \secddecomposeconfq{\esub{J}{\sub}}
            {\ctxone}
  & \decomposesto &
  \secddecompositionSOMEtwo{J}
                    {\ctxone}
  \\[1mm]
  \secddecomposeconfq{\comp{\clo_0}{\clo_1}}
            {\ctxone}
  & \decomposesto &
  \secddecomposeconfq{\clo_1}
            {\apctx{\ctxone}{\clo_0}}
  \end{array}
\]

%  \\[1mm]

\clearpage

\ 

\vspace{-0.8cm}

\[
  \begin{array}{rcl@{\hspace{0.5cm}}l}
  \secddecomposeconfq{\represent{\ctxone'}}
            {\ctxone}
  & \decomposesto &
  \secddecomposeauxconfq{\ctxone}
               {\represent{\ctxone'}}
  \\[1mm]
  \secddecomposeconfq{\stateappend{\represent{\ctxone'}}{\val}}
            {\ctxone}
  & \decomposesto &
  \secddecomposeauxconfq{\ctxone}
               {\stateappend{\represent{\ctxone'}}{\val}}
  \\[1mm]
  \secddecomposeconfq{\shadethis{$\synboundary{\clo}$}}
            {\ctxone}
  & \decomposesto &
  \secddecomposeconfq{\clo}
            {\shadethis{$\delctx{\ctx}$}}
  \\[4mm]
  \secddecomposeauxconfq{\mtctx}
               {\val}
  & \decomposesto &
  \secddecompositionNONE{\val}
  \\[1mm]
  \secddecomposeauxconfq{\apctxq{\ctxone}{\clo_0}}
               {\val_1}
  & \decomposesto &
  \secddecomposeconfq{\clo_0}
            {\argctx{\ctxone}{\val_1}}
  \\[1mm]
  \secddecomposeauxconfq{\argctx{\ctxone}{\val_1}}
               {\val_0}
  & \decomposesto &
  \secddecompositionSOMEtwo{\comp{\val_0}{\val_1}}
                    {\ctxone}
  \\[1mm]
  \secddecomposeauxconfq{\shadethis{$\delctx{\ctxone}$}}
               {\val}
  & \decomposesto &
  \secddecomposeauxconfq{\ctxone}
               {\val}
  \end{array}
\]

\begin{defi}
For any closure $\clo$,
\[
  \begin{array}{@{}rcl}
  \edecompose{\clo}
  & = &
  \left\{
    \begin{array}{l@{\hspace{0.5cm}}l}
    \secddecompositionNONE{\val}
    &
    \mbox{if }
    \secddecomposeconf{\clo}{\mtctx}{\mtctx}
    \decomposesto^*
    \secddecompositionNONE{\val}
    \\
    \secddecompositionSOMEtwo{\varpotred}{\ctxone}
    &
    \mbox{if }
    \secddecomposeconf{\clo}{\mtctx}{\mtctx}
    \decomposesto^*
    \secddecompositionSOMEtwo{\varpotred}{\ctxone}
    \end{array}
  \right.
  \end{array}
\]
\end{defi}

\vspace{-3mm}

\subsubsection{One-step reduction and reduction-based evaluation}

We are now in position to define a one-step reduction function (as in
Sections~\ref{subsubsec:one-step-reduction}
and~\ref{subsubsec:cek-one-step-reduction}) and an evaluation function
iterating this reduction function (as in
Section~\ref{subsubsec:reduction-based-evaluation}
and~\ref{subsubsec:cek-reduction-based-evaluation}).

\vspace{-1mm}

\subsection{From reduction semantics to abstract machine}
\label{subsec:AM-implicit}

Repeating mutatis mutandis the derivation illustrated in
Sections~\ref{subsec:from-red-sem-to-secd-mach}
and~\ref{subsec:from-red-sem-to-cek-mach} leads one to the following
variant of the SECD machine:
\[
\begin{array}{@{}rc@{\ }c@{\ }l@{\hspace{2.4cm}}}
\textrm{(programs)} & p & ::= &
\esub{\tm}{\envextend{\synsucc}{\semsucc}{\envempty}}
\\[0.5mm]
\textrm{(terms)} & \tm & ::= &
\represent{n} \Mid x \Mid \synlam{x}{\tm} \Mid \app{\tm}{\tm} \Mid J
\\[0.5mm]
\textrm{(values)} & \val & ::= &
\represent{n}
\Mid
\semsucc
\Mid
\synpair{\synlam{x}{\tm}}{\sub}
\Mid
\stateappend{\represent{\ctx}}{\val}
\Mid
\represent{\ctx}
\\[0.5mm]
\textrm{(environments)}
& \sub & ::= &
\envempty \Mid \envextend{x}{\val}{\sub}
\\[0.5mm]
\textrm{(contexts)} & \ctx & ::= &
\mtctx \Mid
\apctxq{\ctx}{\synpair{\tm}{\sub}} \Mid
\argctx{\ctx}{\val} \Mid
\shadethis{$\delctx{\ctx}$}
\end{array}
\]

\newcommand{\secdkonfeval}[3]{{\conftriple{#1}{#2}{#3}}_{\mathrm{eval}}}
\newcommand{\secdkonfapply}[3]{{\conftriple{#1}{#2}{#3}}_{\mathrm{apply}}}
\newcommand{\secdkonfcont}[2]{{\confpair{#1}{#2}}_{\mathrm{cont}}}

\[
\begin{array}{@{}r@{\hspace{1.75mm}}l@{\hspace{1.75mm}}l@{}l}
\namedtransition{}
{\secdkonfeval{\represent{n}}{\sub}{\ctxone}}
{\secdkonfcont{\ctxone}{\represent{n}}}
\\[1mm]
\namedtransition{}
{\secdkonfeval{x}{\sub}{\ctxone}}
{\secdkonfcont{\ctxone}{\val}}
&
\textrm{if }lookup (x, \sub) = \val
\\[1mm]
\namedtransition{}
{\secdkonfeval{\synlam{x}{\tm}}{\sub}{\ctxone}}
{\secdkonfcont{\ctxone}{\synpair{\synlam{x}{\tm}}{\sub}}}
\\[1mm]
\namedtransition{}
{\secdkonfeval{\app{\tmone}{\tmtwo}}{\sub}{\ctxone}}
{\secdkonfeval{\tmtwo}{\sub}{\apctxq{\ctxone}{\synpair{\tmone}{\sub}}}}
\\[1mm]
\namedtransition{}
{\secdkonfeval{J}{\sub}{\ctxone}}
{\secdkonfcont{\ctxone}{\represent{\ctx'}}}
&
\textrm{if }\ctx' = previous(\ctx)
\\[1mm]
\namedtransition{}
{\secdkonfapply{\semsucc}{\represent{n}}{\ctxone}}
{\secdkonfcont{\ctxone}{\represent{n+1}}}
\\[1mm]
\namedtransition{}
{\secdkonfapply{\synpair{\synlam{x}{\tm}}{\sub}}{\val}{\ctx}}
{\secdkonfeval{\tm}{\sub'}{\shadethis{$\delctx{\ctx}$}}}
&
\textrm{where }\sub' = extend (x, \val, \sub)
\\[1mm]
\namedtransition{}
{\secdkonfapply{\stateappend{\represent{\ctxone'}}{\val'}}{\val}{\ctxone}}
{\secdkonfapply{\val}{\val'}{\ctxone'}}
\\[1mm]
\namedtransition{}
{\secdkonfapply{\represent{\ctxone'}}{\val}{\ctxone}}
{\secdkonfcont{\ctxone}{\stateappend{\represent{\ctxone'}}{\val}}}
\\[4mm]
\namedtransition{}
{\secdkonfcont{\mtctx}{\val}}
{\val}
\\[1mm]
\namedtransition{}
{\secdkonfcont{\apctxq{\ctxone}{\synpair{\tm}{\sub}}}{\val}}
{\secdkonfeval{\tm}{\sub}{\argctx{\ctxone}{\val}}}
\\[1mm]
\namedtransition{}
{\secdkonfcont{\argctx{\ctxone}{\val'}}{\val}}
{\secdkonfapply{\val}{\val'}{\ctxone}}
\\[1mm]
\namedtransition{}
{\secdkonfcont{\shadethis{$\delctx{\ctx}$}}{\val}}
{\secdkonfcont{\ctx}{\val}}
\end{array}
\]

% \noindent
% This machine evaluates a
% %
% program $\tm$ by starting in the
% configuration
% $$\secdkonfeval{\tm}
%               {\envextend{\synsucc}{\semsucc}{\envempty}}
%               {\mtctx}.$$
% It halts with a value $\val$ if it reaches a configuration
% $\secdkonfcont{\mtctx}{\val}$.

\noindent
Starting in the
configuration
$\secdkonfeval{\tm}
              {\envextend{\synsucc}{\semsucc}{\envempty}}
              {\mtctx}$
makes this machine evaluate the program $\tm$.
The machine halts with a value $\val$ if it reaches a configuration
$\secdkonfcont{\mtctx}{\val}$.

\clearpage

Alternatively (if we allow J to be used outside the body of a
lambda-term and we let it denote the empty context), this machine
evaluates a
program $\tm$ by starting in the
configuration
$\secdkonfeval{\tm}
              {\envextend{\synsucc}{\semsucc}{\envempty}}
              {\shadethis{$\delctx{\mtctx}$}}$.
It halts with a value $\val$ if it reaches a configuration
$\secdkonfcont{\mtctx}{\val}$.

In either case, the machine is not in defunctionalized
form~\cite{Danvy-Nielsen:PPDP01, Danvy-Millikin:SCP0?}.  Therefore, one
cannot immediately map it into an evaluation function in CPS, as in
Sections~\ref{sec:deconstruction-1}, \ref{sec:deconstruction-2},
and~\ref{sec:deconstruction-3}.  The next two sections present two
alternatives, each of which is in defunctionalized form and operates in
lockstep with the present abstract machine.

\section{A syntactic theory of applicative expressions with the J operator:
         explicit, caller-save dumps}
\label{sec:syntactic-theory-explicit-caller-save-dumps}

Instead of marking the context and the intermediate closures, as in
Section~\ref{sec:syntactic-theory-implicit-caller-save-dumps}, one can
cache the context of the caller in a separate register, which leads one
towards \stt{evaluate1'\_alt} in
Section~\ref{subsec:the-rest-of-the-rational-deconstruction}.  For an
analogy, in some formal specifications of
Prolog~\cite{Biernacki-Danvy:LOPSTR03, de-Bruin-de-Vink:TAPSOFT89}, the
cut continuation denotes the previous failure continuation and is cached
in a separate register.

\section{A syntactic theory of applicative expressions with the J operator:
         inheriting the dump through the environment}
\label{sec:syntactic-theory-environment}

Instead of marking the context and the intermediate closures, as in
Section~\ref{sec:syntactic-theory-implicit-caller-save-dumps}, or of
caching the context of the caller in a separate register, as in
Section~\ref{sec:syntactic-theory-explicit-caller-save-dumps}, one can
cache the context of the caller in the environment, which leads one
towards Felleisen's simulation
(Section~\ref{subsec:Felleisen-s-embedding}) and a lightweight extension
of the CEK machine.  Let us briefly outline this reduction semantics and
this abstract machine.

\subsection{Reduction semantics}
\label{subsec:RS-environment}

We specify the reduction semantics as in
Section~\ref{subsec:RS-implicit}.

\subsubsection{Syntactic categories}

We consider a variant of the $\lrhJ$-calculus which is essentially that of
Section~\ref{subsubsec:syntactic-categories-implicit}, except that J is
now an identifier and there are no delimiters:

\[
\begin{array}{@{}rc@{\ }c@{\ }l@{}}
\textrm{(programs)} & p & ::= &
\esub{\tm}{\envextend{\synsucc}{\semsucc}{\envempty}}
\\[0.5mm]
\textrm{(terms)} & \tm & ::= &
\represent{n} \Mid x \Mid \synlam{x}{\tm} \Mid \app{\tm}{\tm}
\\[0.5mm]
\textrm{(closures)} & \clo & ::= &
\represent{n}
\Mid
\semsucc
\Mid
\esub{\tm}{\sub}
\Mid
\comp{\clo}{\clo}
\Mid
\represent{\ctx}
\Mid
\stateappend{\represent{\ctx}}{\val}
\\[0.5mm]
\textrm{(values)} & \val & ::= &
\represent{n}
\Mid
\semsucc
\Mid
\esub{\synlamp{x}{\tm}}{\sub}
\Mid
\represent{\ctx}
\Mid
\stateappend{\represent{\ctx}}{\val}
\\[0.5mm]
\textrm{(potential redexes)} & \varpotred & ::= &
\esub{x}{\sub} \Mid
\comp{\val}{\val}
\\[0.5mm]
\textrm{(substitutions)} & \sub & ::= &
\envempty \Mid \envextend{x}{\val}{\sub}
\\[0.5mm]
\textrm{(contexts)} & \ctx & ::= &
\mtctx \Mid
\apctx{\ctx}{\clo} \Mid
\argctx{\ctx}{\val}
\end{array}
\]

\subsubsection{Plugging}

The notion of reduction is essentially as that of
Section~\ref{subsubsec:plugging-implicit}, except that
there is no control delimiter:

\[
  \begin{array}{rcl@{\hspace{0.5cm}}l}
  \secdplugconfq{\mtctx}{\clo}
  & \plugsto &
  \clo
  \\[1mm]
  \secdplugconfq{\apctx{\ctxone}{\clo_0}}{\clo_1}
  & \plugsto &
  \secdplugconfq{\ctx}{\comp{\clo_0}{\clo_1}}
  \\[1mm]
  \secdplugconfq{\argctx{\ctx}{\val_1}}
                {\clo_0}
  & \plugsto &
  \secdplugconfq{\ctx}
                {\comp{\clo_0}{\clo_1}}
  &
  \mathrm{where} \; \clo_1 = \evaluetoclosure{\val_1}
  \end{array}
\]

\clearpage

\subsubsection{Notion of contraction}

The notion of reduction is essentially as that of
Section~\ref{subsubsec:notion-of-contraction-implicit}, except that
there is no rule for J and there are no delimiters:

\[
\begin{array}{@{}r@{\ }r@{\hspace{1mm}}c@{\hspace{1mm}}l@{\hspace{-3mm}}l@{}}
  \labl{Var}
  &
  \dectwo{\esub{x}{\sub}}{\ctx}
  &
  \contractsto{\shift}
  &
  \dectwo{\val}{\ctx}
  &
  \textrm{if }lookup (x, \sub) = \val
  \\[2mm]
  \labl{Beta$_{succ}$}
  &
  \dectwo{\comp{\semsucc}{\represent{n}}}{\ctx}
  &
  \contractsto{\shift}
  &
  \dectwo{\represent{n+1}}{\ctx}
  \\[2mm]
  \labl{Beta$_{FC}$}
  &
  \dectwo{\comp{\esubp{\synlamp{x}{\tm}}{\sub}}{\val}}{\ctx}
  &
  \contractsto{\shift}
  &
  \dectwo{\esub{\tm}{\sub'}}{\ctx}
  &
  \textrm{where }\sub' = \envextend{J}{\represent{\ctx}}{\envextend{x}{\val}{\sub}}
  \\[2mm]
  \labl{Beta$_{SA}$}
  &
  \dectwo{\comp{\represent{\ctx'}}{\val}}{\ctx}
  &
  \contractsto{\shift}
  &
  \dectwo{\stateappend{\represent{\ctx'}}{\val}}{\ctx}
  \\[2mm]
  \labl{Beta$_{PC}$}
  &
  \dectwo{\comp{\stateappendp{\represent{\ctx'}}{\val'}}{\val}}{\ctx}
  &
  \contractsto{\shift}
  &
  \dectwo{\comp{\val'}{\val}}{\ctx'}
  \\[2mm]
  \labl{Prop}
  &
  \dectwo{\esub{\appp{\tmone}{\tmtwo}}{\sub}}{\ctx}
  &
  \contractsto{\shift}
  &
  \dectwo{\comp{\esubp{\tmone}{\sub}}{\esubp{\tmtwo}{\sub}}}{\ctx}
\end{array}
\]

\noindent
In the $\beta$-rule for function closures,
% the identifier J
$J$ is
dynamically bound to the current context.

\subsubsection{Decomposition}

Decomposition is essentially as in
Section~\ref{subsubsec:decomposition-implicit}, except that there is no
rule for J and there are no delimiters:

\[
  \begin{array}{rcl@{\hspace{0.5cm}}l}
  \secddecomposeconfq{\represent{n}}
            {\ctxone}
  & \decomposesto &
  \secddecomposeauxconfq{\ctxone}
               {\represent{n}}
  \\[1mm]
  \secddecomposeconfq{\semsucc}
            {\ctxone}
  & \decomposesto &
  \secddecomposeauxconfq{\ctxone}
               {\semsucc}
  \\[1mm]
  \secddecomposeconfq{\esub{\represent{n}}{\sub}}
            {\ctxone}
  & \decomposesto &
  \secddecomposeauxconfq{\ctxone}
               {\represent{n}}
  \\[1mm]
  \secddecomposeconfq{\esub{x}{\sub}}
            {\ctxone}
  & \decomposesto &
  \secddecompositionSOMEtwo{\esub{x}{\sub}}
                    {\ctxone}
  \\[1mm]
  \secddecomposeconfq{\esub{\synlamp{x}{\tm}}{\sub}}
            {\ctxone}
  & \decomposesto &
  \secddecomposeauxconfq{\ctxone}
               {\esub{\synlamp{x}{\tm}}{\sub}}
  \\[1mm]
  \secddecomposeconfq{\esub{\appp{\tm_0}{\tm_1}}{\sub}}
            {\ctxone}
  & \decomposesto &
  \secddecompositionSOMEtwo{\esub{\appp{\tm_0}{\tm_1}}{\sub}}
                    {\ctxone}
  \\[1mm]
  \secddecomposeconfq{\comp{\clo_0}{\clo_1}}
            {\ctxone}
  & \decomposesto &
  \secddecomposeconfq{\clo_1}
            {\apctx{\ctxone}{\clo_0}}
  \\[1mm]
  \secddecomposeconfq{\represent{\ctxone'}}
            {\ctxone}
  & \decomposesto &
  \secddecomposeauxconfq{\ctxone}
               {\represent{\ctxone'}}
  \\[1mm]
  \secddecomposeconfq{\stateappend{\represent{\ctxone'}}{\val}}
            {\ctxone}
  & \decomposesto &
  \secddecomposeauxconfq{\ctxone}
               {\stateappend{\represent{\ctxone'}}{\val}}
   \end{array}
 \]
 
 \[
   \begin{array}{rcl@{\hspace{0.5cm}}l}
  \secddecomposeauxconfq{\mtctx}
               {\val}
  & \decomposesto &
  \secddecompositionNONE{\val}
  \\[1mm]
  \secddecomposeauxconfq{\apctxq{\ctxone}{\clo_0}}
               {\val_1}
  & \decomposesto &
  \secddecomposeconfq{\clo_0}
            {\argctx{\ctxone}{\val_1}}
  \\[1mm]
  \secddecomposeauxconfq{\argctx{\ctxone}{\val_1}}
               {\val_0}
  & \decomposesto &
  \secddecompositionSOMEtwo{\comp{\val_0}{\val_1}}
                    {\ctxone}
  \end{array}
\]

\subsection{From reduction semantics to abstract machine}
\label{subsec:AM-environment}

Repeating mutatis mutandis the derivation illustrated in
Sections~\ref{subsec:from-red-sem-to-secd-mach}
and~\ref{subsec:from-red-sem-to-cek-mach} leads one to the following
variant of the CEK machine:

\[
\begin{array}{@{}rc@{\ }c@{\ }l@{\hspace{2.4cm}}}
\textrm{(programs)} & p & ::= &
\esub{\tm}{\envextend{\synsucc}{\semsucc}{\envempty}}
\\[0.5mm]
\textrm{(terms)} & \tm & ::= &
\represent{n} \Mid x \Mid \synlam{x}{\tm} \Mid \app{\tm}{\tm}
\\[0.5mm]
\textrm{(values)} & \val & ::= &
\represent{n}
\Mid
\semsucc
\Mid
\synpair{\synlam{x}{\tm}}{\sub}
\Mid
\stateappend{\represent{\ctx}}{\val}
\Mid
\represent{\ctx}
\\[0.5mm]
\textrm{(environments)}
& \sub & ::= &
\envempty \Mid \envextend{x}{\val}{\sub}
\\[0.5mm]
\textrm{(contexts)} & \ctx & ::= &
\mtctx \Mid
\apctxq{\ctx}{\synpair{\tm}{\sub}} \Mid
\argctx{\ctx}{\val}
\end{array}
\]

\[
\begin{array}{@{}r@{\hspace{1.75mm}}l@{\hspace{1.75mm}}l@{}l}
\namedtransition{}
{\secdkonfeval{\represent{n}}{\sub}{\ctxone}}
{\secdkonfcont{\ctxone}{\represent{n}}}
\\[1mm]
\namedtransition{}
{\secdkonfeval{x}{\sub}{\ctxone}}
{\secdkonfcont{\ctxone}{\val}}
&
\textrm{if }lookup (x, \sub) = \val
\\[1mm]
\namedtransition{}
{\secdkonfeval{\synlam{x}{\tm}}{\sub}{\ctxone}}
{\secdkonfcont{\ctxone}{\synpair{\synlam{x}{\tm}}{\sub}}}
\\[1mm]
\namedtransition{}
{\secdkonfeval{\app{\tmone}{\tmtwo}}{\sub}{\ctxone}}
{\secdkonfeval{\tmtwo}{\sub}{\apctxq{\ctxone}{\synpair{\tmone}{\sub}}}}
\\[4mm]
\namedtransition{}
{\secdkonfapply{\semsucc}{\represent{n}}{\ctxone}}
{\secdkonfcont{\ctxone}{\represent{n+1}}}
\\[1mm]
\namedtransition{}
{\secdkonfapply{\synpair{\synlam{x}{\tm}}{\sub}}{\val}{\ctx}}
{\secdkonfeval{\tm}{\sub'}{\ctx}}
&
\textrm{where }\sub' = extend(J, \represent{\ctx}, extend (x, \val, \sub))
\\[1mm]
\namedtransition{}
{\secdkonfapply{\stateappend{\represent{\ctxone'}}{\val'}}{\val}{\ctxone}}
{\secdkonfapply{\val}{\val'}{\ctxone'}}
\\[1mm]
\namedtransition{}
{\secdkonfapply{\represent{\ctxone'}}{\val}{\ctxone}}
{\secdkonfcont{\ctxone}{\stateappend{\represent{\ctxone'}}{\val}}}
\\[4mm]
\namedtransition{}
{\secdkonfcont{\mtctx}{\val}}
{\val}
\\[1mm]
\namedtransition{}
{\secdkonfcont{\apctxq{\ctxone}{\synpair{\tm}{\sub}}}{\val}}
{\secdkonfeval{\tm}{\sub}{\argctx{\ctxone}{\val}}}
\\[1mm]
\namedtransition{}
{\secdkonfcont{\argctx{\ctxone}{\val'}}{\val}}
{\secdkonfapply{\val}{\val'}{\ctxone}}
\end{array}
\]

\noindent
This machine evaluates a
program $\tm$ by starting in the
configuration
$$\secdkonfeval{\tm}
              {\envextend{\synsucc}{\semsucc}{\envempty}}
              {\mtctx}.$$
It halts with a value $\val$ if it reaches a configuration
$\secdkonfcont{\mtctx}{\val}$.

Alternatively (if we allow J to be used outside the body of a
lambda-term and we let it denote the empty context), this machine
evaluates a
program $\tm$ by starting in the
configuration
$$\secdkonfeval{\tm}
              {\envextend{J}{\mtctx}{\envextend{\synsucc}{\semsucc}{\envempty}}}
              {\mtctx}.$$
It halts with a value $\val$ if it reaches a configuration
$\secdkonfcont{\mtctx}{\val}$.

In either case, the machine is in defunctionalized form.
Refunctionalizing it yields a continuation-passing evaluation function.
Refunctionalizing its closures and mapping the result back to direct
style yields the compositional evaluation functions displayed in
Section~\ref{subsec:Felleisen-s-embedding}, \ie, Felleisen's embedding of
the J operator in Scheme~\cite{Felleisen:CL87}.

\section{Summary and conclusion}
\label{sec:concl}

We have presented a rational deconstruction of the SECD machine with the J
operator, through a series of alternative implementations,
in the form of abstract machines and
compositional evaluation functions, all of which are new.
We have also presented the first
syntactic theories of applicative expressions with the J operator.
In passing, we have shown new
applications of refocusing and defunctionalization and new examples of
control delimiters and of both pushy and jumpy delimited continuations in
programming practice.

Even though they were the first of their kind, the SECD machine and the J
operator remain computationally relevant today:

\begin{enumerate}[$\bullet$]

\item
Architecturally, and until the advent of
JavaScript run-time systems~\cite{Flanagan:06},
the SECD machine has been superseded by
abstract machines with a single control component instead of two (namely
C and D).  In some JavaScript run-time systems, however, methods have a
local stack similar to C to implement and manage their expression
continuation, and a global stack similar to D to implement and manage
command continuations, \ie, the continuation of their caller.

\item
Programmatically, and until the advent of first-class continuations in
JavaScript~\cite{Clements-al:Scheme08},
the J operator has been superseded by control operators
that capture the current continuation (\ie, both C and D) instead of the
continuation of the caller (\ie, D).
In the Rhino implementation of JavaScript, however, the control operator
captures the continuation of the caller of the current method, \ie, the
command continuation instead of both the expression continuation and the
command continuation.

\end{enumerate}

\noindent
At any rate, as we have shown here, both the SECD
machine and the J operator fit the functional
correspondence~\cite{Ager:PhD, Ager-al:PPDP03, Ager-al:IPL04, Ager-al:TCS05,
Biernacka-al:LMCS05, Biernacki:PhD, Danvy:IFL04, Danvy:DSc} as well as
the syntactic correspondence~\cite{Biernacka:PhD, Biernacka-Danvy:TCS07,
Biernacka-Danvy:TOCL07, Danvy:WRS04, Danvy:DSc, Danvy-Nielsen:RS-04-26},
which made it possible for us to mechanically characterize them in new and
precise ways.

All of the points above make us conclude that new abstract machines should be
defined in defunctionalized form today, or at least be made to work in
lockstep with an abstract machine in defunctionalized form.

\section{On the origin of first-class continuations}
\label{sec:on-the-origin-of-continuations}

\begin{flushright}
\begin{minipage}{13.5cm}
\begin{em}
We have shown that jumping and
labels are not essentially connected with strings of imperatives
and in particular, with assignment. Second, that jumping is not
essentially connected with labels.  In performing this piece of
logical analysis we have provided a precisely limited sense in
which the ``value of a label'' has meaning.  Also, we have discovered
a new language feature, not present in current programming languages,
that promises to clarify and simplify a notoriously untidy area of
programming---that concerned with success/failure situations, and
the actions needed on failure.
\end{em}
\hfill-- Peter J. Landin, 1965~\cite[page~133]{Landin:TR65-Generalization}
\end{minipage}
\end{flushright}

\noindent
It was Strachey who coined the term ``first-class
functions''~\cite[Section~3.5.1]{Strachey:67}.%
\footnote{\emph{``Out of Quine's dictum: To be is to be the value of a
  variable, grew Strachey's `first-class citizens'.''}~Peter J.~Landin,
  2000\cite[page~75]{Landin:HOSC00}}
In turn it was Landin who, through the J operator, invented what we know
today as first-class continuations~\cite{Friedman-Haynes:POPL85}:
like Reynolds for escape~\cite{Reynolds:72},
Landin defined J in an unconstrained
way, \ie, with no regard for it to be compatible with the last-in,
first-out allocation discipline prevalent for control stacks since Algol
60.%
\footnote{\emph{``Dumps and program-closures are data-items, with all the
  implied latency for unruly multiple use and other privileges of
  first-class-citizenship.''}~Peter J. Landin,
  1997~\cite[Section~1]{Landin:CW97}}

Today, `continuation' is an overloaded term, that may refer

\begin{enumerate}[$\bullet$]

\item
to the original semantic description technique for representing `the
meaning of the rest of the program' as a function, the continuation,
as multiply co-discovered
in the early
1970's~\cite{Reynolds:LaSC93-one}; or

\item
to the programming-language feature of first-class
continuations as typically provided by a control operator such as J,
escape, or call/cc, as invented by Landin.

\end{enumerate}

\noindent
Whether a semantic description technique or a programming-language
feature, the goal of continuations was the same: to formalize Algol's
labels and jumps.  But where Wadsworth and Abdali gave a continuation
semantics to Algol, and as illustrated in the beginning of
Section~\ref{sec:intro},
Landin translated Algol programs into applicative
expressions in direct style.
In turn, he specified the semantics of applicative expressions with the
SECD machine, \ie,
using first-order means.
The meaning of an Algol label was an ISWIM `program closure' as obtained
by the J operator.  Program closures were defined by extending the SECD
machine, \ie,
still using first-order means.

Landin did not use an explicit representation of the rest of the
computation in his direct semantics of Algol 60, and
for that reason he is not listed
among the co-discoverers of continuations~\cite{Reynolds:LaSC93-one}.
Such an explicit representation, however, exists in the SECD machine, in
first-order form---the dump---which represents the rest of the computation
after returning from the current function call.
In an earlier work~\cite{Danvy:IFL04},
Danvy has shown that the SECD machine, even though it is first-order,
directly corresponds to a compositional evaluation function in
CPS---the tool of choice for specifying control operators since
Rey\-nolds's work~\cite{Reynolds:72}.
In
particular, the dump
directly corresponds to
a functional representation of control, since it
is a defunctionalized continuation.  In the light of
defunctionalization, Landin therefore did use an explicit representation
of the rest of the computation that corresponds to a function,
and for that reason
we wish to see his name added to the list of
co-discoverers of continuations.

\section*{Acknowledgments}
Thanks are due to
Ma{\l}gorzata Biernacka,
Dariusz Biernacki,
Julia L. Lawall,
Johan Munk,
Kristian St{\o}vring,
and
the anonymous reviewers of IFL'05 and LMCS
for comments.
We are also grateful to
Andrzej Filinski,
Dan Friedman,
Lockwood Morris,
John Reynolds,
Guy Steele,
Carolyn Talcott,
Bob Tennent,
Hayo Thielecke,
and
Chris Wadsworth
for their feedback on Section~\ref{sec:on-the-origin-of-continuations}
in November 2005.

This work was
partly carried out while the two authors visited the TOPPS group at DIKU
({\small\url{http://www.diku.dk/topps}}).
It is partly supported by the Danish Natural
Science Research Council, Grant no.~21-03-0545.

\appendix
\appendixpage
\addappheadtotoc

Appendix~\ref{app:fib} demonstrates how two programs, before and after
defunctionalization, do not just yield the same result but also operate
in lockstep.
The three following
appendices illustrate the callee-save, stack-threading features of
the evaluator corresponding to the SECD machine by contrasting them with
a caller-save, stackless evaluator for the pure $\lambda$-calculus.
We successively consider a caller-save, stackless evaluator and the
corresponding abstract machine
(Appendix~\ref{app:caller-save-stackless-evaluator}),
a callee-save, stackless evaluator and the
corresponding abstract machine
(Appendix~\ref{app:callee-save-stackless-evaluator}),
and a caller-save, stack-threading evaluator and the
corresponding abstract machine
(Appendix~\ref{app:caller-save-stack-threading-evaluator}).
Finally, Appendix~\ref{app:from-RS-to-AM} demonstrates how to go from a
reduction semantics of the $\lambda\widehat{\rho}$-calculus to the CEK
machine.

\section{Defunctionalizing a continuation-passing version 
  \texorpdfstring{\\}{}
  of the Fibonacci function}
\label{app:fib}

We start with the traditional Fibonacci function in direct style
(Section~\ref{subsec:fib-orig}), and then present its
continuation-passing counterpart before (Section~\ref{subsec:fib-cps})
and after (Section~\ref{subsec:fib-cps-defunct}) defunctionalization.  To
pinpoint that these two functions operate in lockstep, we equip them with
a trace recording their calling sequence, and we show that they yield the
same result and the same trace.  (One can use the same tracing technique to
prove Proposition~\ref{prop:evaluate1-and-evaluate2} in
Section~\ref{subsec:SECD-refunctionalized}.)

\subsection{The traditional Fibonacci function}
\label{subsec:fib-orig}

We start from the traditional definition of the Fibonacci function in ML:

% \begin{quote}
\begin{verbatim}
  fun fib n
      = if n <= 1
        then n
        else (fib (n - 1)) + (fib (n - 2))
  
  fun main0 n
      = fib n
\end{verbatim}
% \end{quote}

\noindent
So for example, evaluating \stt{main0 5} yields \stt{5}.

\subsection{The Fibonacci function in CPS}
\label{subsec:fib-cps}

To CPS-transform, we first name all intermediate results and
sequentialize their computation, assuming a left-to-right order of
evaluation~\cite{Danvy:3Steps}:

% \begin{quote}
\begin{verbatim}
  fun fib n
      = if n <= 1
        then n
        else let val v1 = fib (n - 1)
                 val v2 = fib (n - 2)
             in v1 + v2
             end
  
  fun main0' n
      = let val v = fib n
        in v
        end
\end{verbatim}
% \end{quote}

\noindent
We then give \stt{fib} an extra argument, the continuation:

% \begin{quote}
\begin{verbatim}
  fun fib_c (n, k)
      = if n <= 1
        then k n
        else fib_c (n - 1,
                    fn v1 => fib_c (n - 2,
                                    fn v2 => k (v1 + v2)))
  
  fun main1 n
      = fib_c (n, fn v => v)
\end{verbatim}
% \end{quote}

\noindent
So for example, evaluating \stt{main1 5} yields \stt{5}.

\subsection{The Fibonacci function in CPS, defunctionalized}
\label{subsec:fib-cps-defunct}

To defunctionalize the Fibonacci function in CPS, we consider its
continuation, which has type \stt{int~->~int}.  Each inhabitant of this
function space arises as an instance of the initial continuation
in \stt{main1} or of the two continuations in \stt{fib\_c}.  We therefore
represent the function space as a sum with three summands, one for each
$\lambda$-abstraction, and we interpret each summand with the body of
each of these $\lambda$-abstractions, using \stt{apply\_cont}:

% \begin{quote}
\begin{verbatim}
  type res = int
  
  datatype cont = C0
                | C1 of res * cont
                | C2 of int * cont
  
  fun apply_cont (C0, v)
      = v
    | apply_cont (C1 (v1, c), v2)
      = apply_cont (c, v1 + v2)
    | apply_cont (C2 (n, c), v1)
      = fib_c_def (n - 2, C1 (v1, c))
  and fib_c_def (n, c)
      = if n <= 1
        then apply_cont (c, n)
        else fib_c_def (n - 1, C2 (n, c))
  
  fun main2 n
      = fib_c_def (n, C0)
\end{verbatim}
% \end{quote}

\noindent
Defunctionalization is summarized with the following two tables, the
first one for the function abstractions and the corresponding sum
injections into the data type \stt{cont},\footnote{Which the cognoscenti
will recognize as Daniel P. Friedman's ``data-structure
continuations''~\cite{Friedman-Wand:08, Wand:JACM80}.}
and the second one for the
function applications and the corresponding calls to the apply function
dispatching over summands:

\begin{enumerate}[$\bullet$]

\item
introduction

\begin{tabular}{|l|l@{\hspace{2.12cm}}|}
\hline
function abstraction
&
sum injection
\\
\hline
\hline
\stt{fn v => v}
&
\stt{C0}
\\
\hline
\stt{fn v2 => k (v1 + v2)}
&
\stt{C1 (v1, c)}
\\
\hline
\stt{fn v1 => fib\_c (n - 2, fn v2 => k (v1 + v2))}
&
\stt{C2 (n, c)}
\\
\hline
\end{tabular}

\item
elimination

\begin{tabular}{|l@{\hspace{4.88cm}}|l|}
\hline
function application
&
case dispatch
\\
\hline
\hline
\stt{k n}
&
\stt{apply\_cont (c, n)}
\\
\hline
\stt{k (v1 + v2)}
&
\stt{apply\_cont (c, v1 + v2)}
\\
\hline
\end{tabular}

\end{enumerate}

\noindent
So for example, evaluating \stt{main2 5} yields \stt{5}.

\subsection{The Fibonacci function in CPS with a trace}
\label{subsec:fib-cps-trace}

We can easily show that applying \stt{main1} and \stt{main2} as defined
above to the same integer yields the same
result, but we want to show a stronger property, namely that they operate
in lockstep.  To this end, we equip \stt{fib\_c}
with a trace recording its calls with the value of its first
argument.  (It would be simple to trace its returns as well, \ie, the
calls to the continuation.)

Representing the trace as a list, the Fibonacci function in CPS reads as
follows:

% \begin{quote}
\begin{verbatim}
  type res = int
  
  (*  fib_c : int * (res * int list -> 'a) -> 'a  *)
  fun fib_c (n, k, T)
      = if n <= 1
        then k (n, T)
        else fib_c (n - 1,
                    fn (v1, T) => fib_c (n - 2,
                                         fn (v2, T) => k (v1 + v2, T),
                                         (n - 2) :: T),
                    (n - 1) :: T)
  
  (*  main3 : int -> res * int list  *)
  fun main3 n
      = fib_c (n, fn (v, T) => (v, T), n :: nil)
\end{verbatim}
% \end{quote}

\noindent
So for example, evaluating \stt{main3 5} yields
\stt{(5,[1,0,1,2,3,0,1,2,1,0,1,2,3,4,5])}.

\subsection{The Fibonacci function in CPS with a trace, defunctionalized}
\label{subsec:fib-cps-defunct-trace}

Proceeding as in Section~\ref{subsec:fib-cps-defunct}, the corresponding
defunctionalized version reads as follows; \stt{fib\_c\_def} is equipped
with a trace recording its calls with the value of its first argument.
(Its returns, \ie, the calls to \stt{apply\_cont}, 
could be traced as well.)

% \begin{quote}
\begin{verbatim}
  type res = int
  
  datatype cont = C0
                | C1 of res * cont
                | C2 of int * cont
  
  (*  apply_cont : cont * res * int list -> res * int list  *)
  fun apply_cont (C0, v, T)
      = (v, T)
    | apply_cont (C1 (v1, c), v2, T)
      = apply_cont (c, v1 + v2, T)
    | apply_cont (C2 (n, c), v1, T)
      = fib_c_def (n - 2, C1 (v1, c), (n - 2) :: T)
  (*  fib_c_def : int * cont * int list -> res * int list  *)
  and fib_c_def (n, c, T)
      = if n <= 1
        then apply_cont (c, n, T)
        else fib_c_def (n - 1, C2 (n, c), (n - 1) :: T)
  
  (*  main4 : int -> res * int list  *)
  fun main4 n
      = fib_c_def (n, C0, n :: nil)
\end{verbatim}
% \end{quote}

\noindent
So for example, evaluating \stt{main4 5} yields
\stt{(5,[1,0,1,2,3,0,1,2,1,0,1,2,3,4,5])}.

\subsection{Lockstep correspondence}
\label{subsec:fib-lockstep}

\begin{defi}
We define $\isrelated{\funcontide}{\defcontide}$ as
\[
  \forall \valide . \forall \traceide .
  \mbox{\stt{k (v, T) = a}}
  \Leftrightarrow
  \mbox{\stt{apply\_cont (c, v, T) = a}}
\]
where ``\stt{e = a}'' means ``there exists an ML value \stt{a} such that
evaluating the ML expression \stt{e} yields \stt{a}.''
\end{defi}

\begin{lem}
\label{lem:one}
$\isrelated{\mbox{\stt{fn (v, T) => (v, T)}}}{\mbox{\stt{C0}}}$
\end{lem}

\noindent
Proof: immediate.
\hfill $\Box$

\begin{lem}
$\forall \valideone . \forall \funcontide \wedge \defcontide \hbox{ such
that } \isrelated{\funcontide}{\defcontide} . 
\isrelated{\mbox{\stt{fn (v2, T) => k (v1 + v2, T)}}}
           {\mbox{\stt{C1 (v1, c)}}}$.
\end{lem}

\noindent
Proof:

By $\beta_v$ reduction, 
\stt{(fn (v2, T) => k (v1 + v2, T)) (v2, T)}
yields the same value as
\stt{k (v1 + v2, T)}.

By definition, 
\stt{apply\_cont (C1 (v1, c), v2, T)}
yields the same value as
\stt{apply\_cont (c, v1 + v2, T)}.

Suppose that \stt{k (v1 + v2, T) = a} holds.
Then since $\isrelated{\funcontide}{\defcontide}$,
\stt{apply\_cont (c, v1 + v2, T) = a} also holds,
and vice-versa.
\hfill $\Box$

\begin{lem}
\label{lem:three}
$\forall \intide .
 \forall \funcontide \wedge \defcontide
 \hbox{ such that } \isrelated{\funcontide}{\defcontide} .
  $

% \begin{itemize}
% 
% \item[a.]
$1.$
\stt{fib\_c (n, k, T) = a}
$\Leftrightarrow$
\stt{fib\_c\_def (n, c, T) = a}

% \item[b.]
$2.$
$\isrelated{\hbox{\stt{fn (v1, T) => fib\_c (n, fn (v2, T) => k (v1 + v2, T), n~::~T)}}}
           {\hbox{\stt{C2 (n+2, c)}}}$
% \end{itemize}
\end{lem}

\noindent
Proof: by simultaneous course-of-value induction. \hfill $\Box$

\begin{thm}
$\forall \intide .
 \hbox{\stt{main3 n = a}}
 \Leftrightarrow
 \hbox{\stt{main4 n = a}}$
\end{thm}

\Proof a consequence of Lemmas~\ref{lem:one} and \ref{lem:three}.\qed

The two versions, before and after defunctionalization, therefore operate
in lockstep, since they yield the same trace and the same result.

\section{A caller-save, stackless evaluator 
  \texorpdfstring{\\}{}
  and the corresponding abstract machine}
\label{app:caller-save-stackless-evaluator}

\subsection{The evaluator}

The following evaluator for the pure call-by-value $\lambda$-calculus
(\ie, the language of
Section~\ref{subsec:prerequisites-and-domain-of-discourse-fc} without
constants and the J operator) is standard.  As pointed out by
Reynolds~\cite{Reynolds:72}, it depends on the evaluation order of its
metalanguage (here, call by value):

% \begin{quote}
\begin{verbatim}
  datatype value = FUN of value -> value
\end{verbatim}
% \end{quote}

% \begin{quote}
\begin{verbatim}
  (*  eval : term * value Env.env -> value  *)
  fun eval (VAR x, e)
      = Env.lookup (x, e)
    | eval (LAM (x, t), e)
      = FUN (fn v => eval (t, Env.extend (x, v, e)))
    | eval (APP (t0, t1), e)
      = let val (FUN f) = eval (t0, e)
        in f (eval (t1, e))
        end
\end{verbatim}
% \end{quote}

% \begin{quote}
\begin{verbatim}
  fun evaluate t
      = eval (t, Env.mt)
\end{verbatim}
% \end{quote}

The evaluator is stackless because it does not thread any data stack.  It
is also caller-save because in the clause for applications, when \stt{t0}
is evaluated, the environment is implicitly saved in the context in order
to evaluate \stt{t1} later on.  In other words, the environment is solely an
inherited attribute.

\newcommand{\confquadrupleeval}[4]{\langle{#1},\,{#2},\,{#3},\,{#4}\rangle_{\mathrm{eval}}}
\newcommand{\conftripleeval}[3]{\langle{#1},\,{#2},\,{#3}\rangle_{\mathrm{eval}}}
\newcommand{\conftriplecont}[3]{\langle{#1},\,{#2},\,{#3}\rangle_{\mathrm{cont}}}
\newcommand{\confpaircont}[2]{\langle{#1},\,{#2}\rangle_{\mathrm{cont}}}

\subsection{The abstract machine}
\label{app:subsec:CEK-AM}

As initiated by Reynolds~\cite{Ager-al:PPDP03, Reynolds:72},
closure-converting the data values of an evaluator, CPS transforming
its control flow, and defunctionalizing its continuations yields an
abstract machine.  For the evaluator above, this machine is the CEK
machine~\cite{Felleisen-Friedman:FDPC3}, \ie, an eval-continue abstract
machine where the evaluation contexts and the continue transition function
are the defunctionalized counterparts of the continuations of the
evaluator just above:

\[
\begin{array}{@{}rc@{\ }c@{\ }l@{\hspace{2.2cm}}}
\textrm{(terms)} & \tm & ::= &
x \Mid \synlam{x}{\tm} \Mid \app{\tm}{\tm}
\\[0.5mm]
\textrm{(values)} & \valmeta & ::= &
\closure{\varmeta}{\termmeta}{\envmeta}
\\[0.5mm]
\textrm{(environments)}
& \sub & ::= &
\envempty \Mid \envextend{x}{\val}{\sub}
\\[0.5mm]
\textrm{(contexts)} & \ctxtmeta & ::= &
\cekendctxt
\Mid
\cekargctxt{\termmeta}{\envmeta}{\ctxtmeta}
\Mid
\cekfunctxt{\valmeta}{\ctxtmeta}
\end{array}
\]

\[
\begin{array}{@{}r@{\hspace{1.75mm}}l@{\hspace{1.75mm}}l@{}l@{\hspace{-3mm}}l@{}}
\namedtransition{}
{\conftripleeval{\varmeta}{\envmeta}{\ctxtmeta}}
{\confpaircont{\ctxtmeta}{\valmeta}}
&
\textrm{if }lookup (\varmeta, \envmeta) = \valmeta
\\[1mm]
\namedtransition{}
{\conftripleeval{\synlam{\varmeta}{\termmeta}}{\envmeta}{\ctxtmeta}}
{\confpaircont{\ctxtmeta}{\closure{\varmeta}{\termmeta}{\envmeta}}}
\\[1mm]
\namedtransition{}
{\conftripleeval{\synapp{\termmeta_0}{\termmeta_1}}{\envmeta}{\ctxtmeta}}
{\conftripleeval{\termmeta_0}
            {\envmeta}
            {\cekargctxt{\termmeta_1}{\envmeta}{\ctxtmeta}}}
\\[4mm]
\namedtransition{}
{\confpaircont{\cekeendctxt}{\valmeta}}
{\valmeta}
\\[1mm]
\namedtransition{}
{\confpaircont{\cekargctxt{\termmeta}{\envmeta}{\ctxtmeta}}{\valmeta}}
{\conftripleeval{\termmeta}{\envmeta}{\cekfunctxt{\valmeta}{\ctxtmeta}}}
\\[1mm]
\namedtransition{}
{\confpaircont{\cekfunctxt{\closure{\varmeta}{\termmeta}{\envmeta}}{\ctxtmeta}}
          {\valmeta}}
{\conftripleeval{\termmeta}{\envmeta'}{\ctxtmeta}}
&
\textrm{where }\envmeta' = extend (\varmeta, \valmeta, \envmeta)
\end{array}
\]

\noindent
This machine evaluates a closed term $\termmeta$ by starting in the
configuration $\conftripleeval{\termmeta}{\envempty}{\cekendctxt}$.
It halts with a value $\valmeta$ if it reaches a configuration
$\confpaircont{\cekendctxt}{\valmeta}$.

\section{A callee-save, stackless evaluator 
  \texorpdfstring{\\}{}
  and the corresponding abstract machine}
\label{app:callee-save-stackless-evaluator}

\subsection{The evaluator}

The following evaluator is a callee-save version of the evaluator of
Appendix~\ref{app:caller-save-stackless-evaluator}.  Whereas the
evaluator of Appendix~\ref{app:caller-save-stackless-evaluator} maps a
term and an environment to the corresponding value, this evaluator maps a
term and an environment to the corresponding value \emph{and the
environment}.  This way, in the clause for applications, the environment
does not need to be implicitly saved since it is explicitly returned
together with the value of \stt{t0}.  In other words, the environment is
not solely an inherited attribute as in the evaluator of
Appendix~\ref{app:caller-save-stackless-evaluator}: it is a synthesized
attribute as well.

Functional values are passed the environment of their caller, and
eventually they return it.  The body of function abstractions is still
evaluated in an extended lexical environment, which is returned but then
discarded.  Otherwise, environments are threaded through the evaluator as
inherited attributes:

% \begin{quote}
\begin{verbatim}
  datatype value = FUN of value * value Env.env -> value * value Env.env
\end{verbatim}
% \end{quote}

% \begin{quote}
\begin{verbatim}
  (*  eval : term * value Env.env -> value * value Env.env  *)
  fun eval (VAR x, e)
      = (Env.lookup (x, e), e)
    | eval (LAM (x, t), e)
      = (FUN (fn (v0, e0) => let val (v1, e1) = eval (t, Env.extend (x, v0, e))
                             in (v1, e0) end),
         e)
\end{verbatim}
% \end{quote}

% \begin{quote}
\begin{verbatim}
    | eval (APP (t0, t1), e)
      = let val (FUN f, e0) = eval (t0, e)
            val     (v, e1) = eval (t1, e0)
        in f (v, e1) end
\end{verbatim}
% \end{quote}

% \begin{quote}
\begin{verbatim}
  fun evaluate t
      = let val (v, e) = eval (t, Env.mt)
        in v end
\end{verbatim}
% \end{quote}

Operationally, one may wish to note that unlike the evaluator of
Appendix~\ref{app:caller-save-stackless-evaluator}, this evaluator is not
properly tail recursive since the evaluation of the body of a function
abstraction no longer occurs in tail position~\cite{Clinger:PLDI98,
Ramsdell:JAR99}.

\subsection{The abstract machine}

As in Appendix~\ref{app:caller-save-stackless-evaluator},
closure-converting the data values of this evaluator, CPS-transforming
its control flow, and defunctionalizing its continuations yields an
abstract machine.  This machine is a variant of the CEK machine with
callee-save environments; its terms, values, and environments remain the
same:

\[
\begin{array}{@{\hspace{1.1cm}}rc@{\ }c@{\ }l@{}}
\textrm{(contexts)} & \ctxtmeta & ::= &
\cekeendctxt
\Mid
\cekeargctxt{\termmeta}{\ctxtmeta}
\Mid
\cekefunctxt{\valmeta}{\ctxtmeta}
\Mid
\cekeretctxt{\envmeta}{\ctxtmeta}
\end{array}
\]

\[
\begin{array}{@{}r@{\hspace{1.75mm}}l@{\hspace{1.75mm}}l@{}l@{\hspace{-3mm}}l@{}}
\namedtransition{\mathit{E}}
{\conftripleeval{\varmeta}{\envmeta}{\ctxtmeta}}
{\conftriplecont{\ctxtmeta}{\valmeta}{\envmeta}}
&
\textrm{if }lookup (\varmeta, \envmeta) = \valmeta
\\[1mm]
\namedtransition{\mathit{E}}
{\conftripleeval{\synlam{\varmeta}{\termmeta}}{\envmeta}{\ctxtmeta}}
{\conftriplecont{\ctxtmeta}{\closure{\varmeta}{\termmeta}{\envmeta}}{\envmeta}}
\\[1mm]
\namedtransition{\mathit{E}}
{\conftripleeval{\synapp{\termmeta_0}{\termmeta_1}}{\envmeta}{\ctxtmeta}}
{\conftripleeval{\termmeta_0}{\envmeta}{\cekeargctxt{\termmeta_1}{\ctxtmeta}}}
\\[4mm]
\namedtransition{\mathit{E}}
{\conftriplecont{\cekeendctxt}{\valmeta}{\envmeta}}
{\valmeta}
\\[1mm]
\namedtransition{\mathit{E}}
{\conftriplecont{\cekeargctxt{\termmeta}{\ctxtmeta}}{\valmeta}{\envmeta}}
{\conftripleeval{\termmeta}{\envmeta}{\cekefunctxt{\valmeta}{\ctxtmeta}}}
\\[1mm]
\namedtransition{\mathit{E}}
{\conftriplecont{\cekefunctxt{\closure{\varmeta}{\termmeta}{\envmeta'}}{\ctxtmeta}}
            {\valmeta}
            {\envmeta}}
{\conftripleeval{\termmeta}
            {\envmeta''}
            {\cekeretctxt{\envmeta}{\ctxtmeta}}}
&
\textrm{where }\envmeta'' = extend (\varmeta, \valmeta, \envmeta')
\\[1mm]
\namedtransition{\mathit{E}}
{\conftriplecont{\cekeretctxt{\envmeta'}{\ctxtmeta}}
            {\valmeta}
            {\envmeta}}
{\conftripleeval{\ctxtmeta}
            {\valmeta}
            {\envmeta'}}
\end{array}
\]

\noindent
This
machine evaluates a closed term $\termmeta$ by starting in the
configuration $\conftripleeval{\termmeta}{\envempty}{\cekeendctxt}$.  It halts
with a value $\valmeta$ if it reaches a configuration
$\conftriplecont{\cekeendctxt}{\valmeta}{\envmeta}$.

\subsection{Analysis}

Compared to the CEK machine in Section~\ref{app:subsec:CEK-AM},
there are two differences in the datatype of
contexts and one new transition rule.  The first difference is that
environments are no longer saved by the caller in $\rawarg$ contexts.
The second difference is that there is an extra context constructor,
$\rawret$, to represent the continuation of the non-tail call to the
evaluator over the body of function abstractions.  The new transition
interprets a $\rawret$ constructor by restoring the environment of the
caller before returning.

It is simple to construct a bisimulation between this callee-save machine
and the CEK machine.

\section{A caller-save, stack-threading evaluator 
  \texorpdfstring{\\}{}
  and the corresponding abstract machine}
\label{app:caller-save-stack-threading-evaluator}

\subsection{The evaluator}

In a stack-threading evaluator, a data stack
stores intermediate values after they have been
computed but before they are used.  Evaluating an expression leaves its
value on top of the data stack.  Applications therefore expect to find their
argument and function on top of the data stack.\footnote{If evaluation is
left-to-right, the argument will be evaluated after the function and thus
will be on top of the data stack.  Some shuffling of the stack can be
avoided if the evaluation order is right-to-left, as in the SECD machine or
the ZINC abstract machine.}

Several design possibilities arise.  First, one can choose between a
single global data stack used for all intermediate values (\ie, as in Forth)
or one can use a local data stack for each function application (\ie, as in
the SECD machine and in the JVM).  For the purpose of illustration, we adopt
the latter since it matches the design of the SECD machine.

Since there is one local data stack per function application, then this
data stack can be chosen to be saved by the caller or by the callee.
Though the former design might be more natural, we again adopt the latter
in this illustration since it matches the design of the SECD machine.

If there is a local, callee-save data stack,
then functional values
are passed their argument and a data stack, and return a value and
a data stack.  One can choose instead to pass the argument to the function
on top of the stack and leave the return value on top of the stack (\ie, as
in Forth).  We adopt this design here, for a local callee-save data stack:

% \begin{quote}
\begin{verbatim}
  datatype value = FUN of value list -> value list
\end{verbatim}
% \end{quote}

% \begin{quote}
\begin{verbatim}
  (*  eval : term * value list * value Env.env -> value  *)
  fun eval (VAR x, s, e)
      = Env.lookup (x, e) :: s
    | eval (LAM (x, t), s, e)
      = FUN (fn (v0 :: s0)
                => let val (v1 :: s1) = eval (t, nil, Env.extend (x, v0, e))
                   in (v1 :: s0) end) :: s
    | eval (APP (t0, t1), s, e)
      = let val                 s0 = eval (t0, s, e)
            val (v :: FUN f :: s1) = eval (t1, s0, e)
        in f (v :: s1) end
\end{verbatim}
% \end{quote}

% \begin{quote}
\begin{verbatim}
  fun evaluate t
      = let val (v :: s) = eval (t, nil, Env.mt)
        in v end
\end{verbatim}
% \end{quote}

Functional values are now passed the data stack of their caller and they
find their argument on top of it.  The body of a function abstraction is
evaluated with an empty data stack, and yields a stack with the value of
the body on top.  This value is returned to the caller on top of its
stack.

\subsection{The abstract machine}

As in Appendix~\ref{app:callee-save-stackless-evaluator},
one may wish to note that functions using
local callee-save data stacks are not properly tail-recursive, though
functions using global or local caller-save data stacks can be made to be.

As in Appendix~\ref{app:caller-save-stackless-evaluator}
and \ref{app:callee-save-stackless-evaluator},
closure converting the data values of this evaluator,
CPS transforming its control flow, and
defunctionalizing its continuations yields an
abstract machine.  This machine is another variant of the CEK
machine with a data stack; its terms, values, and environments remain the
same:

\[
\begin{array}{@{\hspace{0.9cm}}rc@{\ }c@{\ }l@{}}
\textrm{(contexts)} & \ctxtmeta & ::= &
\ceksendctxt
\Mid
\ceksargctxt{\termmeta}{\envmeta}{\ctxtmeta}
\Mid
\ceksfunctxt{\ctxtmeta}
\Mid
\ceksretctxt{\stackmeta}{\ctxtmeta}
\end{array}
\]

\[
\begin{array}{@{}r@{\hspace{1.75mm}}l@{\hspace{1.75mm}}l@{}l@{\hspace{-3mm}}l@{}}
\namedtransition{\mathit{S}}
{\confquadrupleeval{\varmeta}{\stackmeta}{\envmeta}{\ctxtmeta}}
{\confpaircont{\ctxtmeta}
          {\stackpush{\valmeta}{\stackmeta}}}
&
\textrm{if }lookup (\varmeta, \envmeta) = \valmeta
\\[1mm]
\namedtransition{\mathit{S}}
{\confquadrupleeval{\synlam{\varmeta}{\termmeta}}{\stackmeta}{\envmeta}{\ctxtmeta}}
{\confpaircont{\ctxtmeta}
          {\stackpush{\closure{\varmeta}{\termmeta}{\envmeta}}{\stackmeta}}}
\\[1mm]
\namedtransition{\mathit{S}}
{\confquadrupleeval{\synapp{\termmeta_0}{\termmeta_1}}
               {\stackmeta}
               {\envmeta}
               {\ctxtmeta}}
{\confquadrupleeval{\termmeta_0}
               {\stackmeta}
               {\envmeta}
               {\ceksargctxt{\termmeta_1}{\envmeta}{\ctxtmeta}}}
\\[4mm]
\namedtransition{\mathit{S}}
{\confpaircont{\ceksendctxt}{\stackpush{\valmeta}{\stackmeta}}}
{\valmeta}
\\[1mm]
\namedtransition{\mathit{S}}
{\confpaircont{\ceksargctxt{\termmeta}{\envmeta}{\ctxtmeta}}{\stackmeta}}
{\confquadrupleeval{\termmeta}{\stackmeta}{\envmeta}{\ceksfunctxt{\ctxtmeta}}}
\\[1mm]
\namedtransition{\mathit{S}}
{\confpaircont{\ceksfunctxt{\ctxtmeta}}
          {\stackpush{\valmeta}
                     {\stackpush{\closure{\varmeta}{\termmeta}{\envmeta}}
                                {\stackmeta}}}}
{\confquadrupleeval{\termmeta}
               {\stackempty}
               {\envmeta'}
               {\ceksretctxt{\stackmeta}{\ctxtmeta}}}
&
\textrm{where }\envmeta' = extend (\varmeta, \valmeta, \envmeta)
\\[1mm]
\namedtransition{\mathit{S}}
{\confpaircont{\ceksretctxt{\stackmeta'}{\ctxtmeta}}
          {\stackpush{\valmeta}{\stackmeta}}}
{\confpaircont{\ctxtmeta}{\stackpush{\valmeta}{\stackmeta'}}}
\end{array}
\]

\noindent
This
machine evaluates a closed term $\termmeta$ by starting in the
configuration
$\confquadrupleeval{\termmeta}{\stackempty}{\envempty}{\ceksendctxt}$.  It halts
with a value $\valmeta$ if it reaches a configuration
$\confpaircont{\ceksendctxt}{\stackpush{\valmeta}{\stackmeta}}$.

\subsection{Analysis}

Compared to the CEK machine in Section~\ref{app:subsec:CEK-AM},
there are two differences in the datatype of
contexts and one new transition rule.  The first difference is that
intermediate values are no longer saved in $\rawfun$ contexts, since they
are stored on the data stack instead.  The second difference is that
there is an extra context constructor, $\rawret$, to represent the
continuation of the non-tail call to the evaluator over the body of function
abstractions (\ie, a continuation that restores the caller's data stack
and pushes the function return value on top).  The new transition
interprets a $\rawret$ constructor by restoring the data stack of the
caller and pushing the returned value on top of it before returning.

It is simple to construct a bisimulation between this stack-threading machine
and the CEK machine.

\section{From reduction semantics to abstract machine}
\label{app:from-RS-to-AM}

As a warmup to Sections~\ref{subsec:syntactic-theory-J}
and~\ref{subsec:from-red-sem-to-secd-mach}, we present a reduction
semantics for applicative expressions (Section~\ref{subsec:cek-RS}) and
we derive the CEK machine from this reduction semantics
(Section~\ref{subsec:from-red-sem-to-cek-mach}).

\newcommand{\functx}[2]{\plugctx{#1}{\app{#2}{\ctxhole}}}

\newcommand{\cekplugconf}[2]{\langle {#1},\,{#2} \rangle_{\mathrm{plug}}}
\newcommand{\cekdecomposeconf}[2]{\langle {#1},\,{#2} \rangle_{\mathrm{dec/clos}}}
\newcommand{\cekdecomposeconfq}[2]{\langle {#1},\,{#2} \rangle_{\mathrm{dec/clos}}}
\newcommand{\cekdecomposeauxconf}[2]{\langle {#1},\,{#2} \rangle_{\mathrm{dec/cont}}}

\newcommand{\cekdecompositionSOME}[2]{\mathrm{DEC}\:({#1},\,{#2})}
\newcommand{\cekdecompositionNONE}[1]{\mathrm{VAL}\:({#1})}

\newcommand{\cekdecomposesto}{\ensuremath{\rightarrow}}

\newcommand{\syndomcont}{\mathrm{Context}}

\newcommand{\rawcekdecompose}{\mathsf{decompose}}
\newcommand{\ecekdecompose}[1]{\rawcekdecompose\,({#1})}
\newcommand{\rawcekdecomposep}{\mathsf{decompose}'_{\mathrm{clos}}}
\newcommand{\ecekdecomposep}[2]{\rawcekdecomposep\,({#1},\,{#2})}
\newcommand{\rawcekdecomposepaux}{\mathsf{decompose}'_{\mathrm{cont}}}
\newcommand{\ecekdecomposepaux}[2]{\rawcekdecomposepaux\,({#1},\,{#2})}

\newcommand{\rawcekrefocus}{\mathsf{refocus}}
\newcommand{\ecekrefocus}[2]{\rawcekrefocus\,({#1},\,{#2})}

\subsection{A reduction semantics for applicative expressions}
\label{subsec:cek-RS}

The $\lrh$-calculus is a minimal extension of Curien's original calculus
of closures $\lrho$~\cite{Curien:TCS91} to make it closed under one-step
reduction~\cite{Biernacka-Danvy:TOCL07}.
We use it here to illustrate how to go from a reduction semantics to an
abstract machine.
To this end, we present its syntactic categories
(Section~\ref{subsubsec:cek-syntactic-categories}); a plug function mapping a
closure and a reduction context into a closure by filling the given
context with the given closure (Section~\ref{subsubsec:cek-plugging}); a
contraction function implementing a context-insensitive notion of reduction
(Section~\ref{subsubsec:cek-contraction}) and therefore mapping a potential
redex into a contractum; and a decomposition function mapping a non-value
term into a potential redex and a reduction context
(Section~\ref{subsubsec:cek-decomposition}).  We are then in position to
define a one-step reduction function
(Section~\ref{subsubsec:cek-one-step-reduction}) and a reduction-based
evaluation function (Section~\ref{subsubsec:cek-reduction-based-evaluation}).

\subsubsection{Syntactic categories}
\label{subsubsec:cek-syntactic-categories}

We consider a variant of the $\lrh$-calculus with names instead of de
Bruijn indices, and with the usual
reduction context $\ctxone$
embodying a left-to-right applicative-order reduction strategy.
\[
\begin{array}{@{}rc@{\ }c@{\ }l@{}}
\textrm{(terms)} & \tm & ::= &
x \Mid \synlam{x}{\tm} \Mid \app{\tm}{\tm}
\\[0.5mm]
\textrm{(closures)} & \clo & ::= &
\esub{\tm}{\sub}
\Mid
\comp{\clo}{\clo}
\\[0.5mm]
\textrm{(values)} & \val & ::= &
\esub{\synlamp{x}{\tm}}{\sub}
\\[0.5mm]
\textrm{(potential redexes)} & \varpotred & ::= &
\esub{x}{\sub} \Mid
\comp{\val}{\val}
\\[0.5mm]
\textrm{(substitutions)} & \sub & ::= &
\envempty \Mid \envextend{x}{\val}{\sub}
\\[0.5mm]
\textrm{(contexts)} & \ctxone & ::= &
\mtctx \Mid
\argctx{\ctxone}{\clo} \Mid
\functx{\ctxone}{\val}
\end{array}
\]

\noindent
Values are therefore a syntactic subcategory of closures, and in this
section, we make use of the syntactic coercion $\rawvaluetoclosure$
mapping a value into a closure.

\subsubsection{Plugging}
\label{subsubsec:cek-plugging}

Plugging a closure in a context is defined by induction
over this context.  We express this definition as a
state-transition system with one intermediate state,
$\cekplugconf{\clo}{\ctxone}$, an initial state
$\cekplugconf{\clo}{\ctxone}$, and a final state $\clo$.  The
transition function
incrementally peels off the given control context:
  
\[
  \begin{array}{rcl@{\hspace{0.5cm}}l}
  \cekplugconf{\mtctx}
              {\clo}
  & \plugsto &
  \clo
  \\[1mm]
  \cekplugconf{\argctx{\ctxone}{\clo_1}}
              {\clo_0}
  & \plugsto &
  \cekplugconf{\ctxone}
              {\comp{\clo_0}{\clo_1}}
  \\[1mm]
  \cekplugconf{\functx{\ctxone}{\val_0}}
              {\clo_1}
  & \plugsto &
  \cekplugconf{\ctxone}
             {\comp{\clo_0}{\clo_1}}
  &
  \mathrm{where} \; \clo_0 = \evaluetoclosure{\val_0}
  \end{array}
\]

We now can define a total function $\rawplug$ over closures and contexts
that fills the given closure into the given context:

\[
  \begin{array}{r@{\ }c@{\ }l}
  \rawplug :
  \syndomclos \times \syndomcontcont
  \rightarrow
  \syndomclos
  \end{array}
\]

\begin{defi}
For any closure $\clo$ and context $\ctxone$,
$\eplugtwo{\ctxone}{\clo}
 =
 \clo'
$
if and only if
$\cekplugconf{\clo}{\ctxone}
 \plugsto^*
 \clo'
$.
\end{defi}

\subsubsection{Notion of contraction}
\label{subsubsec:cek-contraction}

The notion of reduction over applicative expressions
is specified by the following context-insensitive contraction rules over
actual redexes:

\[
\begin{array}{@{}r@{\ }r@{\hspace{1mm}}c@{\hspace{1mm}}l@{\hspace{1cm}}l@{}}
  \labl{Var}
  &
  {\esub{x}{\sub}}
  &
  \contractsto{\shift}
  &
  {\val}
  &
  \textrm{if }lookup (x, \sub) = \val
  \\[2mm]
  \labl{Beta}
  &
  {\comp{\esubp{\synlamp{x}{\tm}}{\sub}}{\val}}
  &
  \contractsto{\shift}
  &
  {\esub{\tm}{s'}}
  &
  \textrm{where }s' = extend (x, \val, \sub) = \envextend{x}{\val}{\sub}
  \\[2mm]
  \labl{Prop}
  &
  {\esub{\appp{\tmone}{\tmtwo}}{\sub}}
  &
  \contractsto{\shift}
  &
  {\comp{\esubp{\tmone}{\sub}}{\esubp{\tmtwo}{\sub}}}
\end{array}
\]

\noindent
For closed closures (\ie, closures with no free variables), all potential
redexes are actual ones.

We now can define by cases a total function $\rawcontract$
that maps a redex to the corresponding contractum:

\[
  \begin{array}{r@{\ }c@{\ }l}
  \rawcontract :
  \syndompotred
  \rightarrow
  \syndomclos
  \end{array}
\]

\begin{defi}
For any potential redex $\varpotred$,
$\econtractone{\varpotred}
 =
 \clo
$
if and only if
$\varpotred
 \contractsto{\shift}
 \clo
$.
\end{defi}

\subsubsection{Decomposition}
\label{subsubsec:cek-decomposition}

There are many ways to define a total function mapping a value closure
to itself and a non-value closure to a potential redex and a reduction
context.  In our experience, the following definition is a convenient
one.  It is a state-transition system with two intermediate states,
$\cekdecomposeconf{\clo}{\ctxone}$ and
$\cekdecomposeauxconf{\ctxone}{\val}$, an initial state
$\cekdecomposeconf{\clo}{\mtctx}$ and two final states
$\cekdecompositionNONE{\val}$ and
$\cekdecompositionSOME{\varpotred}{\ctxone}$.  If possible, the
transition function from the state $\cekdecomposeconf{\clo}{\ctxone}$
decomposes the given closure $\clo$ and accumulates the corresponding
reduction context $\ctxone$.  The transition function from the state
$\cekdecomposeauxconf{\ctxone}{\val}$ dispatches over the given context.

\[
  \begin{array}{rcl@{\hspace{0.5cm}}l}
  \cekdecomposeconf{\esub{x}{\sub}}
                   {\ctxone}
  & \cekdecomposesto &
  \cekdecompositionSOME{\esub{x}{\sub}}
                       {\ctxone}
  \\[1mm]
  \cekdecomposeconf{\esub{\synlamp{x}{\tm}}{\sub}}
                   {\ctxone}
  & \cekdecomposesto &
  \cekdecomposeauxconf{\ctxone}
                      {\esub{\synlamp{x}{\tm}}{\sub}}
  \\[1mm]
  \cekdecomposeconf{\esub{\appp{\tm_0}{\tm_1}}{\sub}}
                   {\ctxone}
  & \cekdecomposesto &
  \cekdecompositionSOME{\esub{\appp{\tm_0}{\tm_1}}{\sub}}
                       {\ctxone}
  \\[1mm]
  \cekdecomposeconf{\comp{\clo_0}{\clo_1}}
                   {\ctxone}
  & \cekdecomposesto &
  \cekdecomposeconf{\clo_0}
                   {\argctx{\ctxone}{\clo_1}}
  \\[4mm]
  \cekdecomposeauxconf{\mtctx}
                      {\val}
  & \cekdecomposesto &
  \cekdecompositionNONE{\val}
  \\[1mm]
  \cekdecomposeauxconf{\argctx{\ctxone}{\clo_1}}
                      {\val_0}
  & \cekdecomposesto &
  \cekdecomposeconf{\clo_1}
                   {\functx{\ctxone}{\val_0}}
  \\[1mm]
  \cekdecomposeauxconf{\functx{\ctxone}{\val_0}}
                      {\val_1}
  & \cekdecomposesto &
  \cekdecompositionSOME{\comp{\val_0}{\val_1}}
                       {\ctxone}
  \end{array}
\]

We now can define a total function $\rawcekdecompose$ over
closures that maps a value closure to itself and a non-value closure to a
decomposition into a potential redex, a control context, and a dump
context.  This total function uses two auxiliary functions
$\rawcekdecomposep$ and $\rawcekdecomposepaux$:

\[
  \begin{array}{l@{\ }c@{\ \ }l@{\ }c@{\ }l}
  \rawcekdecompose
  & : &
  \syndomclos
  &
  \rightarrow
  \syndomval + (\syndompotred \times \syndomcont)
  \\
  \rawcekdecomposep
  & : &
  \syndomclos \times \syndomcont
  &
  \rightarrow
  \syndomval + (\syndompotred \times \syndomcont)
  \\
  \rawcekdecomposepaux
  & : &
  \syndomcont \times \syndomval
  &
  \rightarrow
  \syndomval + (\syndompotred \times \syndomcont)
  \end{array}
\]

\begin{defi}
For any closure $\clo$, value $\val$, and context $\ctxone$,
\[
  \begin{array}{@{}rcl}
  \ecekdecomposep{\clo}{\ctxone}
  & = &
  \left\{
    \begin{array}{l@{\hspace{1cm}}l}
    \cekdecompositionNONE{\val'}
    &
    \mbox{if }
    \cekdecomposeconfq{\clo}{\ctxone}
    \cekdecomposesto^*
    \cekdecompositionNONE{\val'}
    \\
    \cekdecompositionSOME{\varpotred}{\ctxone'}
    &
    \mbox{if }
    \cekdecomposeconfq{\clo}{\ctxone}
    \cekdecomposesto^*
    \cekdecompositionSOME{\varpotred}{\ctxone'}
    \end{array}
  \right.
  \\[4mm]
  \ecekdecomposepaux{\ctxone}{\val}
  & = &
  \left\{
    \begin{array}{l@{\hspace{1cm}}l}
    \cekdecompositionNONE{\val'}
    &
    \mbox{if }
    \cekdecomposeauxconf{\ctxone}{\val}
    \cekdecomposesto^*
    \cekdecompositionNONE{\val'}
    \\
    \cekdecompositionSOME{\varpotred}{\ctxone'}
    &
    \mbox{if }
    \cekdecomposeauxconf{\ctxone}{\val}
    \cekdecomposesto^*
    \cekdecompositionSOME{\varpotred}{\ctxone'}
    \end{array}
  \right.
  \end{array}
\]
and
$\ecekdecompose{\clo}
 =
 \ecekdecomposep{\clo}{\mtctx}$.
\end{defi}

\subsubsection{One-step reduction}
\label{subsubsec:cek-one-step-reduction}

We are now in position to define a total function $\rawreduce$ over
closed closures that
maps a value closure to itself and a non-value closure to the next
closure in the reduction sequence.  This function is defined by
composing the three functions above:

\[
  \ereduce{\clo}
  =
  \begin{array}[t]{@{}l}
  \vcase{\ecekdecompose{\clo}}
        {\cekdecompositionNONE{\val}}
        {\evaluetoclosure{\val}}
        {\cekdecompositionSOME{\varpotred}{\ctxone}}
        {\eplugtwo{\econtractone{\varpotred}}{\ctxone}}
  \end{array}
\]

\noindent
The function $\rawreduce$ is partial because of $\rawcontract$, which is
undefined for stuck closures.

Graphically:
{\let\labelstyle=\textstyle
\spreaddiagramrows{-0.1cm}
\spreaddiagramcolumns{0.33cm}
 \newcommand{\atadless}{\hspace{-2mm}}
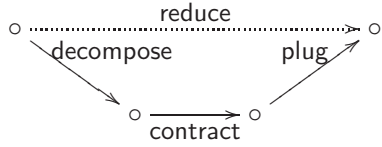
 $$
 \diagram
 \circ
 \drto^{\atadless\text{{\small $\rawdecompose$}}}
 \ar@{..>}[rrr]^{\text{{\small $\rawreduce$}}}
 &
 &
 &
 \circ
 \\
 &
 \circ
 \rto_{\text{{\small $\rawcontract$}}}
 &
 \circ
 \urto^{\text{{\small $\rawplug$}}\atadless}
 &
 \enddiagram
 $$
}

\begin{defi}[One-step reduction]
For any
closure $\clo$,
$ \clo \rightarrow \clo' $
if and only if
$ \ereduce{\clo} = \clo' $.
\end{defi}

\subsubsection{Reduction-based evaluation}
\label{subsubsec:cek-reduction-based-evaluation}

Iterating
$\rawreduce$
defines a reduction-based evaluation function.  The definition below uses
$\rawdecompose$ to distinguish between values and non-values,
and implements iteration (tail-) recursively with the partial function
$\rawiterate$:

\[
  \begin{array}{rcl@{\hspace{1.9cm}}}
  \eevaluate{\clo}
  & = &
  \eiterate{\edecompose{\clo}}
  \end{array}
\]
\noindent
where

\[
  \left\{
  \begin{array}{lcl}
  \eiterate{\cekdecompositionNONE{\val}}
  & = &
  \val
  \\
  \eiterate{\cekdecompositionSOME{\varpotred}{\ctxone}}
  & = &
  \eiterate{\edecompose{\eplugtwo{\econtractone{\varpotred}}{\ctxone}}}
  \end{array}
  \right.
\]

\noindent
The function $\rawevaluate$ is partial because reducing a given closure
might not converge.

Graphically:
{\let\labelstyle=\textstyle
\spreaddiagramrows{-0.1cm}
\spreaddiagramcolumns{0.33cm}
 \newcommand{\atadless}{\hspace{-2mm}}
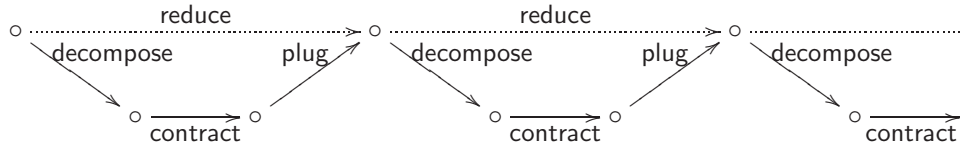
 $$
 \diagram
 \circ
 \drto^{\atadless\text{{\small $\rawdecompose$}}}
 \ar@{..>}[rrr]^{\text{\small $\rawreduce$}}
 &
 &
 &
 \circ
 \drto^{\atadless\text{{\small $\rawdecompose$}}}
 \ar@{..>}[rrr]^{\text{\small $\rawreduce$}}
 &
 &
 &
 \circ
 \drto^{\atadless\text{{\small $\rawdecompose$}}}
 \ar@{..}[rr]
 &
 &
 \\
 &
 \circ
 \rto_{\text{{\small $\rawcontract$}}}
 &
 \circ
 \urto^{\text{{\small $\rawplug$}}\atadless}
 &
 &
 \circ
 \rto_{\text{{\small $\rawcontract$}}}
 &
 \circ
 \urto^{\text{{\small $\rawplug$}}\atadless}
 &
 &
 \circ
 \rto_{\text{{\small $\rawcontract$}}}
 &
 \enddiagram
 $$
}

\begin{defi}[Reduction-based evaluation]
For any
closure $\clo$,
$ \clo \rightarrow^* \val $
if and only if
$ \eevaluate{\clo}$
$= \val $.
\end{defi}

To close, let us adjust the definition of $\rawevaluate$ by exploiting
the fact that for any closure $\clo$, $\eplugtwo{\clo}{\mtctx} =
\clo$:

\[
   \eevaluate{\clo}
   =
   \eiterate{\edecompose{\eplugtwo{\clo}{\mtctx}}}
\]

\noindent
In this adjusted definition, $\rawdecompose$ is always applied to the
result of $\rawplug$.

\subsection{From the reduction semantics for applicative expressions
            to the CEK machine}
\label{subsec:from-red-sem-to-cek-mach}

Deforesting the intermediate terms in the reduction-based evaluation
function of Section~\ref{subsubsec:cek-reduction-based-evaluation} yields a
reduction-free evaluation function in the form of a small-step abstract
machine (Section~\ref{subsubsec:cek-refocusing}).
We simplify this small-step abstract machine by fusing a part of its
driver loop with the contraction function
(Section~\ref{subsubsec:cek-lightweight-fusion}) and compressing its
`corridor' transitions
(Section~\ref{subsubsec:cek-inlining-and-transition-compression}).
Unfolding the recursive data type of closures precisely yields the
caller-save, stackless CEK machine of
Section~\ref{app:subsec:CEK-AM}
(Section~\ref{subsubsec:cek-opening-closures}).

\subsubsection{Refocusing: from reduction-based to reduction-free evaluation}
\label{subsubsec:cek-refocusing}

Following Danvy and Nielsen~\cite{Danvy-Nielsen:RS-04-26}, we deforest
the intermediate closure in the reduction sequence by replacing the
composition of $\rawplug$ and $\rawdecompose$ by a call to a composite
function $\rawrefocus$:

\[
  \begin{array}{@{\hspace{0.5cm}}rcl}
   \eevaluate{\clo}
   & = &
   \eiterate{\ecekrefocus{\clo}{\mtctx}}
  \end{array}
\]
where

\[
  \left\{
  \begin{array}{lcl}
  \eiterate{\cekdecompositionNONE{\val}}
  & = &
  \val
  \\
  \eiterate{\cekdecompositionSOME{\varpotred}{\ctxone}}
  & = &
  \eiterate{\ecekrefocus{\econtractone{\varpotred}}{\ctxone}}
  \end{array}
  \right.
\]

\noindent
and $\rawrefocus$ is optimally defined as continuing the decomposition in
the current reduction context~\cite{Danvy-Nielsen:RS-04-26}:

\[
  \begin{array}{rcl@{\hspace{1.2cm}}}
   \ecekrefocus{\clo}{\ctxone}
   & = &
   \ecekdecomposep{\clo}{\ctxone}
  \end{array}
\]

\noindent
This evaluation function is reduction-free because it no longer
constructs each intermediate closure in the reduction sequence.

Graphically:
{\let\labelstyle=\textstyle
\spreaddiagramrows{-0.1cm}
\spreaddiagramcolumns{0.33cm}
 \newcommand{\atadless}{\hspace{-2mm}}
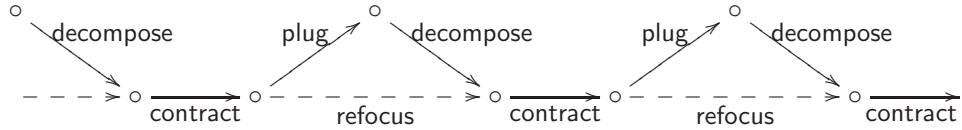
 $$
 \diagram
 \circ
 \drto^{\atadless\text{{\small $\rawdecompose$}}}
 &
 &
 &
 \circ
 \drto^{\atadless\text{{\small $\rawdecompose$}}}
 &
 &
 &
 \circ
 \drto^{\atadless\text{{\small $\rawdecompose$}}}
 &
 &
 \\
 \ar@{-->}[r]
 &
 \circ
 \rto_{\text{{\small $\rawcontract$}}}
 &
 \circ
 \urto^{\text{{\small $\rawplug$}}\atadless}
 \ar@{-->}[rr]_{\text{{\small $\rawrefocus$}}}
 &
 &
 \circ
 \rto_{\text{{\small $\rawcontract$}}}
 &
 \circ
 \urto^{\text{{\small $\rawplug$}}\atadless}
 \ar@{-->}[rr]_{\text{{\small $\rawrefocus$}}}
 &
 &
 \circ
 \rto_{\text{{\small $\rawcontract$}}}
 &
 \enddiagram
 $$
}

\begin{defi}[Reduction-free evaluation]
For any
closure $\clo$,
$ \clo \rightarrow^* \val $
if and only if
$ \eevaluate{\clo} = \val $.
\end{defi}

\newcommand{\cekiterate}[2]{\langle {#1},\,{#2} \rangle_{\mathrm{iter}}}
\newcommand{\cekiteratesto}{\Rightarrow}
\renewcommand{\cekdecomposesto}{\Rightarrow}
\newcommand{\cekrefocusesto}{\Rightarrow}

\newcommand{\cekrefocusconf}[2]{\langle {#1},\,{#2} \rangle_{\mathrm{eval}}}
\newcommand{\cekrefocusconfq}[2]{\langle {#1},\,{#2} \rangle_{\mathrm{eval}}}
\newcommand{\cekrefocusauxconf}[2]{\langle {#1},\,{#2} \rangle_{\mathrm{cont}}}

\subsubsection{Lightweight fusion: making do without driver loop}
\label{subsubsec:cek-lightweight-fusion}

In effect, $\rawiterate$ is as the `driver loop' of a small-step abstract
machine that refocuses and contracts.  Instead, let us fuse
$\rawcontract$ and $\rawiterate$ and express the result with rewriting
rules over a configuration $\cekiterate{\varpotred}{\ctxone}$.  We clone
the rewriting rules for $\rawcekdecomposep$ and
$\rawcekdecomposepaux$ into refocusing rules, indexing their
configurations as $\cekrefocusconf{\clo}{\ctxone}$ and
$\cekrefocusauxconf{\ctxone}{\val}$ instead of as
$\cekdecomposeconf{\clo}{\ctxone}$ and
$\cekdecomposeauxconf{\ctxone}{\val}$, respectively:

\begin{enumerate}[$\bullet$]

\item
instead of rewriting to $\cekdecompositionNONE{\val}$, the cloned
rules rewrite to $\val$;

\item
instead of rewriting to $\cekdecompositionSOME{\varpotred}{\ctxone}$,
the cloned
rules rewrite to $\cekiterate{\varpotred}{\ctxone}$.

\end{enumerate}

\noindent
The result reads as follows:

\[
  \begin{array}{rcl@{\hspace{0.5cm}}l}
  \cekrefocusconf{\esub{x}{\sub}}
                   {\ctxone}
  & \cekrefocusesto &
  \cekiterate{\esub{x}{\sub}}
             {\ctxone}
  \\[1mm]
  \cekrefocusconf{\esub{\synlamp{x}{\tm}}{\sub}}
                   {\ctxone}
  & \cekrefocusesto &
  \cekrefocusauxconf{\ctxone}
                      {\esub{\synlamp{x}{\tm}}{\sub}}
  \\[1mm]
  \cekrefocusconf{\esub{\appp{\tm_0}{\tm_1}}{\sub}}
                   {\ctxone}
  & \cekrefocusesto &
  \cekiterate{\esub{\appp{\tm_0}{\tm_1}}{\sub}}
             {\ctxone}
  \\[1mm]
  \cekrefocusconf{\comp{\clo_0}{\clo_1}}
                   {\ctxone}
  & \cekrefocusesto &
  \cekrefocusconf{\clo_0}
                   {\argctx{\ctxone}{\clo_1}}
  \\[4mm]
  \cekrefocusauxconf{\mtctx}
                      {\val}
  & \cekrefocusesto &
  \val
  \\[1mm]
  \cekrefocusauxconf{\argctx{\ctxone}{\clo_1}}
                      {\val_0}
  & \cekrefocusesto &
  \cekrefocusconf{\clo_1}
                   {\functx{\ctxone}{\val_0}}
  \\[1mm]
  \cekrefocusauxconf{\functx{\ctxone}{\val_0}}
                      {\val_1}
  & \cekrefocusesto &
  \cekiterate{\comp{\val_0}{\val_1}}
             {\ctxone}
  \\[4mm]
  \cekiterate{\esub{x}{\sub}}
             {\ctxone}
  & \cekiteratesto &
  \cekrefocusconf{\val}
                   {\ctxone}
  &
  \textrm{if }lookup (x, \sub) = \val
  \\
  \cekiterate{\comp{\esubp{\synlamp{x}{\tm}}{\sub}}{\val}}
             {\ctxone}
  & \cekiteratesto &
  \cekrefocusconf{\esub{\tm}{\sub'}}
                   {\ctxone}
  &
  \textrm{where }\sub' = extend (x, \val, \sub)
  \\
  \cekiterate{\esub{\appp{\tmone}{\tmtwo}}{\sub}}
             {\ctxone}
  & \cekiteratesto &
  \cekrefocusconf{\comp{\esubp{\tmone}{\sub}}{\esubp{\tmtwo}{\sub}}}
                   {\ctxone}
  \end{array}
\]

\noindent
The following proposition summarizes the situation:

\begin{prop}
For any closure $\clo$,
$ \eevaluate{\clo} = \val $
if and only if
$ \cekrefocusconf{\clo}{\mtctx} \Rightarrow^* \val $.
\end{prop}

\noindent
Proof: straightforward.  The two machines operate in lockstep.
\hfill $\Box$

\subsubsection{Inlining and transition compression}
\label{subsubsec:cek-inlining-and-transition-compression}

The abstract machine of Section~\ref{subsubsec:cek-lightweight-fusion},
while interesting in its own right (it is `staged' in that the
contraction rules are implemented separately from the congruence
rules~\cite{Biernacka-Danvy:TOCL07, Hardin-al:JFP98}), is not minimal: a
number of transitions yield a configuration whose transition is uniquely
determined.  Let us carry out these hereditary, ``corridor'' transitions
once and for all:

\begin{enumerate}[$\bullet$]

\item
$\cekrefocusconf{\esub{x}{\sub}}{\ctxone}
 \cekrefocusesto
 \cekiterate{\esub{x}{\sub}}{\ctxone}
 \cekrefocusesto
 \cekrefocusconf{\val}{\ctxone}
 \cekrefocusesto
 \cekrefocusauxconf{\ctxone}{\val}$
 \hfill
 $\textrm{if }lookup (x, \sub) = \val$

\item
$\cekrefocusconf{\esub{\appp{\tm_0}{\tm_1}}{\sub}}{\ctxone}
 \cekrefocusesto
 \cekiterate{\esub{\appp{\tm_0}{\tm_1}}{\sub}}{\ctxone}
 \cekrefocusesto
 \cekrefocusconf{\comp{\esubp{\tmone}{\sub}}{\esubp{\tmtwo}{\sub}}}{\ctxone}
 \cekrefocusesto
 \cekrefocusconf{\esubp{\tmone}{\sub}}{\argctx{\ctxone}{\esubp{\tmtwo}{\sub}}}$

\item
$\cekrefocusauxconf{\functx{\ctxone}{\esubp{\synlamp{x}{\tm}}{\sub}}}{\val}
 \cekrefocusesto
 \cekiterate{\comp{\esubp{\synlamp{x}{\tm}}{\sub}}{\val}}{\ctxone}
 \cekrefocusesto
 \cekrefocusconf{\esub{\tm}{\sub'}}{\ctxone}$
 \hfill
 $\textrm{where }\sub' = extend (x, \val, \sub)$
\end{enumerate}

\noindent
The result reads as follows:
\[
  \begin{array}{rcl@{\hspace{0.5cm}}l}
  \cekrefocusconf{\esub{x}{\sub}}
                   {\ctxone}
  & \cekrefocusesto &
  \cekrefocusauxconf{\ctxone}{\val}
  &
  \textrm{if }lookup (x, \sub) = \val
  \\[1mm]
  \cekrefocusconf{\esub{\synlamp{x}{\tm}}{\sub}}
                   {\ctxone}
  & \cekrefocusesto &
  \cekrefocusauxconf{\ctxone}
                      {\esub{\synlamp{x}{\tm}}{\sub}}
  \\[1mm]
  \cekrefocusconf{\esub{\appp{\tm_0}{\tm_1}}{\sub}}
                   {\ctxone}
  & \cekrefocusesto &
  \cekrefocusconf{\esubp{\tmone}{\sub}}{\argctx{\ctxone}{\esubp{\tmtwo}{\sub}}}
  \\[4mm]
  \cekrefocusauxconf{\mtctx}
                      {\val}
  & \cekrefocusesto &
  \val
  \\[1mm]
  \cekrefocusauxconf{\argctx{\ctxone}{\clo_1}}
                      {\val_0}
  & \cekrefocusesto &
  \cekrefocusconf{\clo_1}
                   {\functx{\ctxone}{\val_0}}
  \\[1mm]
  \cekrefocusauxconf{\functx{\ctxone}{\esubp{\synlamp{x}{\tm}}{\sub}}}
                      {\val}
  & \cekrefocusesto &
  \cekrefocusconf{\esub{\tm}{\sub'}}{\ctxone}
  &
  \textrm{where }\sub' = extend (x, \val, \sub)
  \end{array}
\]

\noindent
The configuration $\cekiterate{\varpotred}{\ctxone}$ has disappeared and
so is the case for $\comp{\clo_0}{\clo_1}$: they were only transitory.

\begin{prop}
For any closure $\clo$,
$ \eevaluate{\clo} = \val $
if and only if
$ \cekrefocusconf{\clo}{\mtctx} \Rightarrow^* \val $.
\end{prop}

\noindent
Proof: immediate.  We have merely compressed corridor transitions.
\hfill $\Box$

\subsubsection{Opening closures: from explicit substitutions
               to terms and environments}
\label{subsubsec:cek-opening-closures}

The abstract machine above solely operates on ground closures.  If we
open the closures $\esub{\tm}{\sub}$ into pairs $\synpair{\tm}{\sub}$ and
flatten the configuration $\cekrefocusconf{\synpair{\tm}{\sub}}{\ctxone}$
into a triple $\conftripleeval{\tm}{\sub}{\ctxone}$, we obtain an abstract
machine that coincides with the caller-save, stackless CEK machine of
Section~\ref{app:subsec:CEK-AM}.

\subsection{Conclusion and perspectives}

Appendix~\ref{app:caller-save-stackless-evaluator} illustrated the
functional correspondence between the functional implementation of a
denotational or natural semantics and of an abstract machine, the CEK
machine, for the $\lambda$-calculus with left-to-right applicative order.
The present
appendix illustrates the syntactic correspondence between the
functional implementation of a reduction semantics and of an abstract
machine, again the CEK machine, for the $\lambda$-calculus with
left-to-right applicative order.  Together, the functional correspondence
and the syntactic correspondence therefore demonstrate the natural fit of
the CEK machine in the semantic spectrum of the $\lambda$-calculus with
explicit substitutions and left-to-right applicative order.

\bibliography{danvy-millikin-LMCS}
\bibliographystyle{plain}

\end{document}